\documentclass{ws-ijqi}
\usepackage{mathrsfs}
\usepackage{upmath}
\usepackage{yfonts}
\usepackage{dsfont}
\usepackage[all]{xy}
\usepackage{tikz}
\begin{document}
%%%%%%%%%%%%%%%%%%%%%%%%%%%%%%%%%%%%%%%%%%%%%%%%%%%%%%%%%%%%%%%%%%%%%%%%%%%%%%%%%%%%%%%%%%%%%%%%%%%%%%%%%%%%%%%%%%%%%%%%%%%%%%%%%%%%%%%%%%%%%%%%%%%%%%%%
\newcommand{\be}{\begin{equation}}
\newcommand{\ee}{\end{equation}}
\newcommand{\brr}{\begin{eqnarray}}
\newcommand{\err}{\end{eqnarray}}
\newcommand{\nn}{\nonumber}
\newcommand{\bd}{\begin{displaymath}}
\newcommand{\ed}{\end{displaymath}}
\newcommand{\ovl}{\overline}
\newcommand{\bib}{\bibitem}
\newcommand{\bfig}{\begin{figure}}
\newcommand{\efig}{\end{figure}}
\newcommand{\ie}{i.e.}
\newcommand{\eg}{e.g.}
%%%%%%%%%%%%%%%%%%%%%%%%%%%%%%%%%%%%%%%%%%%%%%%%%%%%%%%%%%%%%%%%%%%%%%%%%%%%%%%%%%%%%%%%%%%%%%%%%%%%%%%%%%%%%%%%%%%%%%%%%%%%%%%%%%%%%%%%%%%%%%%%%%%%%%%%
\hyphenation{inhe-rent trans-for-ma-tions cha-rac-te-ri-zes ope-ra-tors des-cri-bing ope-ra-tor Pu-bli-shing Chi-ches-ter me-tho-ds appro-xi-ma-tion}
\hyphenation{des-cri-bes un-cer-tain-ty li-ne-ar ma-the-ma-ti-cal es-ta-bli-shes fur-ther phy-si-cal in-tri-ca-te ge-ne-ra-li-za-tion lo-ca-tion}
\hyphenation{po-si-ti-ve mar-gi-nal mo-di-fied Ha-mil-to-nian o-pe-ra-tion mo-di-fy par-ti-cu-lar}
%%%%%%%%%%%%%%%%%%%%%%%%%%%%%%%%%%%%%%%%%%%%%%%%%%%%%%%%%%%%%%%%%%%%%%%%%%%%%%%%%%%%%%%%%%%%%%%%%%%%%%%%%%%%%%%%%%%%%%%%%%%%%%%%%%%%%%%%%%%%%%%%%%%%%%%%
\def\alf{\alpha}
\def\bet{\beta}
\def\gam{\gamma}
\def\th{\theta}
\def\lam{\lambda}
\def\om{\omega}
\def\eps{\epsilon}
\def\rpar{\right)}
\def\lpar{\left(}
\def\rbk{\right]}
\def\lbk{\left[}
\def\rbr{\right\}}
\def\lbr{\left\{}
\def\lb{\label}
\def\ro{\mbox{\boldmath $\rho$}}
\def\sig{\mbox{\boldmath $\sigma$}}
\def\mpi{\mbox{\boldmath $\Pi$}}
\def\bop{\mbox{\boldmath $\Delta$}}
\def\hei{\mbox{\tiny ${\rm H}$}}
\def\dim{\mbox{\tiny ${\rm N}$}}
\def\indb{\mbox{\tiny ${\rm B}$}}
\def\opj{\mbox{\tiny ${\rm J}$}}
\def\opx{\mbox{\scriptsize ${\rm X}$}}
\def\opy{\mbox{\scriptsize ${\rm Y}$}}
\def\opz{\mbox{\scriptsize ${\rm Z}$}}
\def\opxy{\mbox{\scriptsize ${\rm XY}$}}
\def\opxz{\mbox{\scriptsize ${\rm XZ}$}}
\def\opyz{\mbox{\scriptsize ${\rm YZ}$}}
\def\tr{\mbox{${\rm Tr}$}}
\def\ima{\mbox{${\rm Im}$}}
\def\re{\mbox{${\rm Re}$}}
\def\rg{\rangle}
\def\lg{\langle}
\def\nc{\mathrm{i}}
\def\coloneq{\mathrel{\mathop:}=}
\def\ns{\mbox{\scriptsize $N$}}
\def\half{\frac{1}{2}}
\def\fa{\mathfrak{a}}
\def\hp{\hspace{\parindent}}
\def\entellipse{(1.95,0) ellipse (5cm and 1.6cm)}
\def\corellipse{(0,0) ellipse (3cm and 0.8cm)}
\def\sqzellipse{(2.25,-0.8) ellipse (1.5cm and 0.6cm)}
%%%%%%%%%%%%%%%%%%%%%%%%%%%%%%%%%%%%%%%%%%%%%%%%%%%%%%%%%%%%%%%%%%%%%%%%%%%%%%%%%%%%%%%%%%%%%%%%%%%%%%%%%%%%%%%%%%%%%%%%%%%%%%%%%%%%%%%%%%%%%%%%%%%%%%%%
\markboth{Marchiolli, Galetti, and Debarba}
{Spin squeezing and entanglement via finite-dimensional discrete phase space description}
%%%%%%%%%%%%%%%%%%%%%%%%%%%%%%%%%%%%%%%%%%%%%%%%%%%%%%%% Publisher's Area please ignore %%%%%%%%%%%%%%%%%%%%%%%%%%%%%%%%%%%%%%%%%%%%%%%%%%%%%%%%%%%%%%%%
\catchline{}{}{}{}{}
%%%%%%%%%%%%%%%%%%%%%%%%%%%%%%%%%%%%%%%%%%%%%%%%%%%%%%%%%%%%%%%%%%%%%%%%%%%%%%%%%%%%%%%%%%%%%%%%%%%%%%%%%%%%%%%%%%%%%%%%%%%%%%%%%%%%%%%%%%%%%%%%%%%%%%%%
\title{SPIN SQUEEZING AND ENTANGLEMENT VIA FINITE-DIMENSIONAL DISCRETE PHASE-SPACE DESCRIPTION}
%%%%%%%%%%%%%%%%%%%%%%%%%%%%%%%%%%%%%%%%%%%%%%%%%%%%%%%%%%%%%%%%%%%%%%%%%%%%%%%%%%%%%%%%%%%%%%%%%%%%%%%%%%%%%%%%%%%%%%%%%%%%%%%%%%%%%%%%%%%%%%%%%%%%%%%%
\author{Marcelo A. Marchiolli}
\address{Avenida General Os\'{o}rio 414, centro, Jaboticabal, SP 14870-100, Brazil \\
         marcelo\_march@bol.com.br} 
\author{Di\'{o}genes Galetti}
\address{Instituto de F\'{\i}sica Te\'{o}rica, Universidade Estadual Paulista, \\
         Rua Dr Bento Teobaldo Ferraz 271, Bloco II, Barra Funda, S\~{a}o Paulo, SP 01140-070, Brazil \\
         galetti@ift.unesp.br}
\author{Tiago Debarba}
\address{Instituto de Ci\^{e}ncias Exatas, Universidade Federal de Minas Gerais, Departamento de F\'{\i}sica, \\ 
         Avenida Ant\^{o}nio Carlos 6627, Belo Horizonte, MG 31270-901, Brazil \\
         debarba@fisica.ufmg.br}
\maketitle
%%%%%%%%%%%%%%%%%%%%%%%%%%%%%%%%%%%%%%%%%%%%%%%%%%%%%%%%%%%%%%%%%%%%%%%%%%%%%%%%%%%%%%%%%%%%%%%%%%%%%%%%%%%%%%%%%%%%%%%%%%%%%%%%%%%%%%%%%%%%%%%%%%%%%%%%
%\begin{history}
%\received{Day Month Year}
%\revised{Day Month Year}
%\accepted{(Day Month Year)}
%\comby{(xxxxxxxxxx)}
%\end{history}
%%%%%%%%%%%%%%%%%%%%%%%%%%%%%%%%%%%%%%%%%%%%%%%%%%%%%%%%%%%%%%%%%%%%%%%%%%%%%%%%%%%%%%%%%%%%%%%%%%%%%%%%%%%%%%%%%%%%%%%%%%%%%%%%%%%%%%%%%%%%%%%%%%%%%%%%
\begin{abstract}
We show how mapping techniques inherent to ${\rm N}^{2}$-dimensional discrete phase spaces can be used to treat a wide family of spin systems which 
exhibits squeezing and entanglement effects. This algebraic framework is then applied to the modified Lipkin-Meshkov-Glick (LMG) model in order to 
obtain the time evolution of certain special parameters related to the Robertson-Schr\"{o}dinger (RS) uncertainty principle and some particular proposals 
of entanglement measure based on collective angular-momentum generators. Our results reinforce the connection between both the squeezing and 
entanglement effects, as well as allow to investigate the basic role of spin correlations through the discrete representatives of quasiprobability 
distribution functions. Entropy functionals are also discussed in this context. The main sequence {\it correlations} $\mapsto$ {\it entanglement}
$\mapsto$ {\it squeezing} of quantum effects embraces a new set of insights and interpretations in this framework, which represents an effective 
gain for future researches in different spin systems. 
\end{abstract}

\keywords{Spin squeezing; entanglement; finite-dimensional discrete phase spaces.}

%%%%%%%%%%%%%%%%%%%%%%%%%%%%%%%%%%%%%%%%%%%%%%%%%%%%%%%%%%%%%%%%%%%%%%%%%%%%%%%%%%%%%%%%%%%%%%%%%%%%%%%%%%%%%%%%%%%%%%%%%%%%%%%%%%%%%%%%%%%%%%%%%%%%%%%%
\section{Introduction}
%%%%%%%%%%%%%%%%%%%%%%%%%%%%%%%%%%%%%%%%%%%%%%%%%%%%%%%%%%%%%%%%%%%%%%%%%%%%%%%%%%%%%%%%%%%%%%%%%%%%%%%%%%%%%%%%%%%%%%%%%%%%%%%%%%%%%%%%%%%%%%%%%%%%%%%%

Originally introduced by E. Schr\"{o}dinger in his seminal work on probability relations between separated systems,\cite{ES} ``entanglement" indeed 
corresponds to a fundamental concept in physics that lies at the heart of many conceptual problems in quantum mechanics.\cite{EPR,Peres} The mere
comprehension of this abstract concept and its respective changing of status, in the recent past, for experimental measure, certainly represents a 
concatenation of efforts with significant progress in theoretical and experimental physics. It is important to stress that `quantum entanglement' plays 
an essential role in multipartite quantum systems,\cite{Horodecki} since its underlying physical properties can be used as a specific resource for 
determined quantum-information tasks\cite{Winter} --- recently, different proposals of quantumness correlations have appeared in current literature,\cite{Ollivier} 
emphasizing the possible connections with quantum entanglement.\cite{Modi} Therefore, understanding, identifying, measuring and, consequently, exploiting 
genuine quantum effects (or ``quantumness"\cite{Zurek}) in multipartite systems constitute a set of obligatory prerequisites for grasping the fundamental 
implications of any quantum theory.\cite{Adesso}

Recent theoretical proposals\cite{SS1} and experiments\cite{Wine,Hald} involving measurements upon collective angular-momentum generators in different physical 
systems fulfil, in part, the aforementioned prerequisites, as well as corroborate the subtle match between spin-squeezing\cite{Kitagawa} and entanglement effects. 
However, some necessary questions concerning the correlations among different spin components as chief agents responsible for spin-squeezing effects deserve be 
properly answered. Then, it seems reasonable to adopt a particular algebraic framework that, within all the inherent mathematical virtues, allows to: (i) map the 
kinematical and dynamical contents of a given spin system with a finite space of states into a ${\rm N}^{2}$-dimensional discrete phase space; (ii) comprehend the 
role of correlations through discrete Wigner functions and their connections with spin-squeezing effects; and finally, (iii) gain new insights on the effects 
exhibited in the Venn diagram below which lead us to take advantage operationally of their quantumness in many-particle experiments. Notwithstanding the appreciable 
number of papers in current literature proposing similar theoretical frameworks with different intrinsic mathematical properties,\cite{Vourdas} let us focus our 
attention upon the formalism developed in Refs.~\refcite{GP1}--\refcite{MR} which complies such demands.
\begin{figure}
\begin{center}
\begin{tikzpicture}[line width=0.25pt][h]
\begin{scope}
\fill[gray] \entellipse;
\fill[red] \corellipse;
\fill[yellow] \sqzellipse;
\end{scope}
\draw(5.0cm,0) node {{\bf Correlations}};
\draw(-1.3cm,0) node {{\bf Entanglement}};
\draw(2.22cm,-0.8cm) node {{\bf Spin Squeezing}};
\end{tikzpicture}
\end{center}
\caption{This particular Venn diagram depicts, in a general form, the interplay among correlations, entanglement, and spin-squeezing effects in finite-size
quantum spin systems.}
\end{figure}
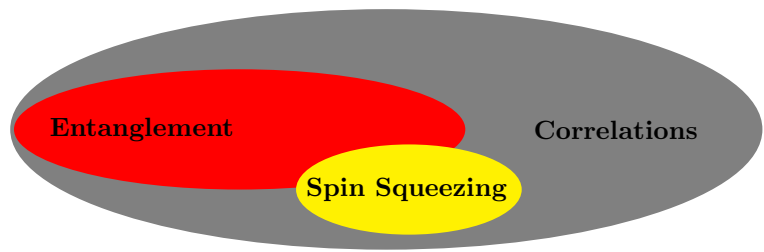

Based on the technique of constructing unitary operator bases initially formulated by Schwinger,\cite{Schwinger} this particular discrete quantum 
phase-space approach embraces a well-defined algebraic structure where the ${\rm N}$-dimensional pre-Hilbert spaces equipped with a Hilbert-Schmidt inner
product\cite{Prugo} endorse the finite space of states. Moreover, it leads us (i) to exhibit and handle the pair of complementary variables related to a 
specific degree of freedom we are dealing with (as well as to recognize the quantum correlations between them), and also (ii) to obtain additional quantum
information about the physical system through the analysis of the corresponding discrete quasiprobability distribution functions. Henceforth, the basic idea 
consists in exploring this mathematical tool in order to study the role of those correlations in connection with spin-squeezing and entanglement processes.

Initially focused on some mathematically appealing features inherent to the collective angular-momentum generators, the first part of this paper basically
discuss the spin coherent states and their relations with unitary transformations through a constructive point of view. Next, we establish a 
$\mbox{mod}({\rm N})$-invariant operator basis which can be immediately employed in the mapping of quantum operators (acting into that particular 
${\rm N}$-dimensional space of states) onto well-behaved functions of discrete variables by means of a trace operation.\cite{Fano} These functions 
correspond to the representatives of the operators in a ${\rm N}^{2}$-dimensional phase space labeled by a pair of discrete variables for each degree of freedom 
of the physical system under investigation. Thus, all the necessary quantities for describing its kinematical and dynamical contents can now be promptly mapped 
one-to-one on such phase spaces. For completeness sake, we also present two different prescriptions that essentially determine the time evolution for both 
the discrete Wigner and Weyl functions.

These results are then applied to an extended version of the LMG model,\cite{Lipkin} which was initially introduced by Vidal and coworkers\cite{Dusuel}
with the aim of investigating the statistical-mechanical properties and entanglement effects of a particular interacting spin system. Summarizing, the
modified LMG model describes a finite set of spins $\half$ with a mutual anisotropic $(XY)$ ferromagnetic interaction, and also subjected to a transverse
magnetic field. The second part of this paper shows how the quantum correlations (here controlled by parameters related to the transverse magnetic field 
and anisotropy) affect the connection between the spin-squeezing and entanglement effects. To develop this specific task, we first employ the RS uncertainty 
principle\cite{Dodonov} in order to determine a spin-squeezing measure which incorporates, in its definition, the covariance function --- being such a function 
responsible for introducing, within the Heisenberg uncertainty principle, important contributions associated with the anticommutation relations of the 
angular-momentum generators. The subsequent comparison with the entanglement measure proposed in Ref.~\refcite{Hald} establishes, in this way, the aforementioned 
link and reinforces the fundamental role of correlations in our initial analysis. Moreover, we also study the behaviour of the discrete Wigner function for certain 
particular values of time where occur an `almost perfect match' between both the spin-squeezing and entanglement measurements (such procedure leads us to 
comprehend the intricate role of correlations in a finite-dimensional phase space). We have finalized this work with the analysis of some entropy functionals.\cite{Mizrahi} 
For completeness sake, it is worth stressing that we have worked with a number of spins which preserves basic quantum effects, contrary to the thermodynamic limit.

This paper is structured as follows. In Section 2, we present an important set of essential mathematical tools related to the angular-momentum generators 
and spin coherent states, as well as an interesting discussion on unitary transformations and their corresponding geometric interpretations. In Section 3,
we introduce a mapping kernel for finite-dimensional discrete phase spaces whose inherent properties allow, within other important virtues, to describe the 
kinematical and dynamical contents of a given physical system with a finite space of states. In Section 4, we apply the quantum-algebraic approach developed
in the previous section to the modified LMG model, with the aim of analysing the influence of quantum correlations on the spin-squeezing and entanglement
effects. Section 5 summarizes the main results obtained in this paper and discuss some possible perspectives for future research. We have added three appendixes 
related to the calculational details of certain topics and expressions used in the previous sections: Appendix A shows how the unitary transformations associated 
with angular-momentum generators affect the expressions for variances, covariances, and RS uncertainty relation; Appendix B discuss the Kitagawa-Ueda model\cite{Kitagawa}
and the inherent spin-squeezing and entanglement effects; while Appendix C exhibits an important set of specific numerical computations which leads us to establish a 
validity domain for the entanglement criteria studied in this work.       

%%%%%%%%%%%%%%%%%%%%%%%%%%%%%%%%%%%%%%%%%%%%%%%%%%%%%%%%%%%%%%%%%%%%%%%%%%%%%%%%%%%%%%%%%%%%%%%%%%%%%%%%%%%%%%%%%%%%%%%%%%%%%%%%%%%%%%%%%%%%%%%%%%%%%%%%
\section{Definitions and background for spin coherent states}
%%%%%%%%%%%%%%%%%%%%%%%%%%%%%%%%%%%%%%%%%%%%%%%%%%%%%%%%%%%%%%%%%%%%%%%%%%%%%%%%%%%%%%%%%%%%%%%%%%%%%%%%%%%%%%%%%%%%%%%%%%%%%%%%%%%%%%%%%%%%%%%%%%%%%%%%

In this section, we will briefly survey some mathematically appealing features inherent to the $\mathfrak{su}(2)$ Lie algebra with emphasis on a
specific group of unitary transformations involving the generators of this algebra which leads us to properly define the spin coherent states. For 
this task, let us initially consider the standard angular-momentum generators $\lbr {\bf J}_{x},{\bf J}_{y},{\bf J}_{z} \rbr$ which act on a
finite-dimensional Hilbert space $\mathcal{H}_{\dim}$. The familiar commutation relation $\lbk {\bf J}_{a},{\bf J}_{b} \rbk = \nc \eps_{abc} 
{\bf J}_{c}$ for $a,b,c = x,y,z$ (letting $\hbar = 1$ for convenience), where $\eps_{abc}$ denotes the Levi-Civita symbol associated with the three
orthogonal directions, permits us to define the raising and lowering operators through the particular decomposition ${\bf J}_{\pm} \coloneq {\bf J}_{x}
\pm \nc {\bf J}_{y}$ with $\lbk {\bf J}_{+},{\bf J}_{-} \rbk = 2 {\bf J}_{z}$ and $\lbk {\bf J}_{z},{\bf J}_{\pm} \rbk = \pm {\bf J}_{\pm}$. Moreover,
$\vec{{\bf J}}^{2} \coloneq {\bf J}_{x}^{2}+{\bf J}_{y}^{2}+{\bf J}_{z}^{2} = {\bf J}_{z}^{2} + \half \lbr {\bf J}_{+},{\bf J}_{-} \rbr$ characterizes
the total spin operator and constitutes an important element of this algebra since $\vec{{\bf J}}^{2}$ (also known as a Casimir operator) commutes with all 
the generators ${\bf J}_{a}$. In particular, these results allow to construct, for example, a set of simultaneous eigenstates of $\vec{{\bf J}}^{2}$
and ${\bf J}_{z}$, that is, $\lbr | j,m \rg : |m| \leq j \; , \; j = 0,\half,1,\frac{3}{2},\ldots \rbr$, with well-known mathematical properties.\cite{Gilmore}
\begin{romanlist}[(ii)]
\item It is interesting to note that each particular member of this representation (here characterized by a specific eigenvalue $j$) is analogous to the Fock 
state\cite{Prugo} of the electromagnetic field, since it can be created by the repeated application of the raising operator (remembering that 
${\bf J}_{\pm}^{2j+1} = 0$) on the `vaccum state', 
\bd
| j,m \rg \coloneq \mathscr{C}_{2j,j+m}^{- \half} \frac{{\bf J}_{+}^{j+m}}{(j+m)!} | j,-j \rg \quad \mbox{for} \quad \mathscr{C}_{r,s} \equiv
\frac{r!}{s! (r-s)!} .
\ed
Thus, the set $\lbr | j,m \rg \rbr$ corresponds to a discrete, orthonormal and complete basis for the $(2j+1)$-dimensional vector space $\mathbb{C}^{2j+1}$
of angular-momentum states,\cite{Nou} whose completeness and orthonormality relations are expressed as follow: 
\bd
\sum_{m=-j}^{j} | j,m \rg \lg j,m | = {\bf 1} \quad \mbox{and} \quad \lg j^{\prime},m^{\prime} | j,m \rg = \delta_{j^{\prime},j} \,
\delta_{m^{\prime},m} .
\ed
Consequently, the expansion of any quantum state $| \Psi \rg$ belonging to this finite-dimensional space can now be prompty obtained,
\bd
| \Psi \rg = \sum_{m=-j}^{j} c_{j,m} | j,m \rg
\ed
where $c_{j,m} \equiv \lg j,m | \Psi \rg$ denotes the coefficients of such a discrete expansion.

\item The nondiagonal matrix elements $\lg j^{\prime},m^{\prime} | {\bf J}_{\pm}^{k} | j,m \rg$ associated with the moments ${\bf J}_{\pm}^{k}$ 
show a dependence on the binomial coefficients $\mathscr{C}_{r,s}$, that is
\bd
\lg j^{\prime},m^{\prime} | {\bf J}_{\pm}^{k} | j,m \rg = \frac{(j \pm m^{\prime})!}{(j \pm m)!} \lpar \frac{\mathscr{C}_{2j,j+m^{\prime}}}
{\mathscr{C}_{2j,j+m}} \rpar^{\half} \delta_{j^{\prime},j} \, \delta_{m^{\prime},m \pm k} \qquad (0 \leq k \leq 2j),
\ed
while $\lg j^{\prime},m^{\prime} | {\bf J}_{z}^{k} | j,m \rg = m^{k} \delta_{j^{\prime},j} \, \delta_{m^{\prime},m}$ demonstrates an explicit connection
with the discrete component $m$. These results are extremely useful when applied to the evaluation of nondiagonal matrix elements related to well-known 
families of unitary transformations involving the linear combination of spin operators,\cite{Gilmore,Nou} and also in the study of spin coherent 
states.\cite{Perelomov,Inomata,Puri}
\end{romanlist}
%
%%%%%%%%%%%%%%%%%%%%%%%%%%%%%%%%%%%%%%%%%%%%%%%%%%%%%%%%%%%%%%%%%%%%%%%%%%%%%%%%%%%%%%%%%%%%%%%%%%%%%%%%%%%%%%%%%%%%%%%%%%%%%%%%%%%%%%%%%%%%%%%%%%%%%%%%
\subsection{Unitary transformations}
%%%%%%%%%%%%%%%%%%%%%%%%%%%%%%%%%%%%%%%%%%%%%%%%%%%%%%%%%%%%%%%%%%%%%%%%%%%%%%%%%%%%%%%%%%%%%%%%%%%%%%%%%%%%%%%%%%%%%%%%%%%%%%%%%%%%%%%%%%%%%%%%%%%%%%%%

Let ${\bf T}(\Omega_{\pm},\Omega_{z}) \coloneq \exp \lpar \Omega_{+} {\bf J}_{+} + \Omega_{z} {\bf J}_{z} + \Omega_{-} {\bf J}_{-} \rpar$ denote a
kind of general abstract operator expressed in terms of generators of the $\mathfrak{su}(2)$ Lie algebra and arbitrary c-number parameters 
$\Omega_{\pm}$ and $\Omega_{z}$. For $\Omega_{+} = \xi$, $\Omega_{-} = - \xi^{\ast}$ and $\Omega_{z} = \nc \om$, with $\xi \in \mathbb{C}$ and
$\om \in \mathbb{R}$, such an abstract operator represents a generator of unitary transformations whose respective generalized `normal'- and
`antinormal'-order decomposition formulae display the following expressions:\cite{Eberly,Ban} 
\brr
\lb{e1}
\!\!\!\!\! {\bf T}(\xi,\om) &=& \exp \lpar \Lambda_{+} {\bf J}_{+} \rpar \exp \lbk \ln \lpar \Lambda_{z} \rpar {\bf J}_{z} \rbk \exp \lpar \Lambda_{-} 
{\bf J}_{-} \rpar \nn \\
&=& \exp \lpar - \Lambda_{+}^{\ast} {\bf J}_{-} \rpar \exp \lbk - \ln \lpar \Lambda_{z}^{\ast} \rpar {\bf J}_{z} \rbk \exp \lpar - \Lambda_{-}^{\ast} 
{\bf J}_{+} \rpar
\err
with
\brr
\Lambda_{+} &=& \frac{\lpar \xi / \phi \rpar \sin (\phi)}{\cos (\phi) - \nc \lpar \om / 2 \phi \rpar \sin (\phi)} , \qquad
\Lambda_{-} = - \frac{\lpar \xi^{\ast} / \phi \rpar \sin (\phi)}{\cos (\phi) - \nc \lpar \om / 2 \phi \rpar \sin (\phi)} , \nn \\
\Lambda_{z} &=& \lbk \cos (\phi) - \nc \lpar \om / 2 \phi \rpar \sin (\phi) \rbk^{-2} \quad \mbox{and} \quad \phi = \lbk | \xi |^{2} + \lpar \om / 2
\rpar^{2} \rbk^{\half} . \nn
\err
So, the action of ${\bf T}(\xi,\om)$ on the generators $\{ {\bf J}_{a} \}$ --- here defined through the relation $\overline{{\bf J}}_{a} \coloneq 
{\bf T}^{\dagger}(\xi,\om) {\bf J}_{a} {\bf T}(\xi,\om)$ for $a=x,y,z$ --- produces effectively a new set of angular-momentum operators
$\{ \overline{{\bf J}}_{a} \}$ expressed in terms of the old ones that, by their turn, are multiplied by determined coefficients which depend on the
parameters $\xi$ and $\om$, namely,
\brr
\lb{e2}
\overline{{\bf J}}_{x} &=& \lbk \cos (2 \phi) + 2 \ima^{2} (\xi) \frac{\sin^{2} (\phi)}{\phi^{2}} \rbk {\bf J}_{x} + \lbk \om \, 
\frac{\sin (2 \phi)}{2 \phi} + 2 \re (\xi) \ima (\xi) \frac{\sin^{2} (\phi)}{\phi^{2}} \rbk {\bf J}_{y} \nn \\
& & + \lbk \om \ima (\xi) \frac{\sin^{2} (\phi)}{\phi^{2}} - \re (\xi) \frac{\sin (2 \phi)}{\phi} \rbk {\bf J}_{z} , \\
\lb{e3}
\overline{{\bf J}}_{y} &=& \lbk - \om \, \frac{\sin (2 \phi)}{2 \phi} + 2 \re (\xi) \ima (\xi) \frac{\sin^{2} (\phi)}{\phi^{2}} \rbk {\bf J}_{x} +
\lbk \cos (2 \phi) + 2 \re^{2} (\xi) \frac{\sin^{2} (\phi)}{\phi^{2}} \rbk {\bf J}_{y} \nn \\
& & + \lbk \om \re (\xi) \frac{\sin^{2} (\phi)}{\phi^{2}} + \ima (\xi) \frac{\sin (2 \phi)}{\phi} \rbk {\bf J}_{z} , \\
\lb{e4}
\overline{{\bf J}}_{z} &=& \lbk \om \ima (\xi) \frac{\sin^{2} (\phi)}{\phi^{2}} + \re (\xi) \frac{\sin (2 \phi)}{\phi} \rbk {\bf J}_{x} +
\lbk \om \re (\xi) \frac{\sin^{2} (\phi)}{\phi^{2}} - \ima (\xi) \frac{\sin (2 \phi)}{\phi} \rbk {\bf J}_{y} \nn \\
& & + \lbk \cos (2 \phi) + \frac{\om^{2}}{2} \frac{\sin^{2} (\phi)}{\phi^{2}} \rbk {\bf J}_{z} .
\err
Note that $\vec{{\bf J}}^{2}$ remains invariant under the unitary transformation (\ref{e1}), which implies in the identity $\vec{\overline{{\bf J}}}^{2} 
\equiv \vec{{\bf J}}^{2}$; besides, the geometric counterpart of this particular result is directly associated with arbitrary rotations on the surface of 
a sphere of radius $j(j+1)$. Indeed, let ${\bf T}(\th,\varphi)$ denote a particular case of ${\bf T}(\xi,\om)$ when $\xi = \frac{\th}{2} \exp (- \nc \varphi)$ 
and $\om = 0$, which implies in specific decomposition formulae characterized by $\Lambda_{\pm} = \pm \tan (\th/2) \exp ( \mp \nc \varphi )$ and $\Lambda_{z} = 
\sec^{2} (\th/2)$. The connection between $\xi$ and $\Lambda_{+}$ reflects the stereographic projection of a two-dimensional sphere $S^{2}$ on the complex plane
$\mathbb{C}$ with one-point compactification (in this situation, the infinite point corresponds to the north pole of $S^{2} = \mathbb{C} \cup \{ \infty \}$).
 
As a consequence, the action of ${\bf T}(\th,\varphi)$ on the vacuum state $| j,-j \rg$ will define our next object of study: the spin coherent states --- also 
known in current literature as $\mathrm{SU}(2)$ coherent states or spin coherent states.\cite{Puri} In fact, we will establish some few important mathematical 
results connected with the spin coherent states which constitute the first basic tools for discussing the subtle link between spin squeezing and entanglement.

%%%%%%%%%%%%%%%%%%%%%%%%%%%%%%%%%%%%%%%%%%%%%%%%%%%%%%%%%%%%%%%%%%%%%%%%%%%%%%%%%%%%%%%%%%%%%%%%%%%%%%%%%%%%%%%%%%%%%%%%%%%%%%%%%%%%%%%%%%%%%%%%%%%%%%%%
\subsection{Spin coherent states}
%%%%%%%%%%%%%%%%%%%%%%%%%%%%%%%%%%%%%%%%%%%%%%%%%%%%%%%%%%%%%%%%%%%%%%%%%%%%%%%%%%%%%%%%%%%%%%%%%%%%%%%%%%%%%%%%%%%%%%%%%%%%%%%%%%%%%%%%%%%%%%%%%%%%%%%%

In general, the spin coherent states can be defined by means of two different mathematical procedures: the first one\cite{Atkins} basically follows the Schwinger's 
prescription of angular momentum (note that LMG model\cite{Ring} represents a typical example of this prescription) in order to produce such states, while the second 
one\cite{Thomas} pursues an algebraic framework analogous to that adopted for the field coherent states which allows us to describe physical systems consisting of 
$N$ two-level atoms\footnote{See Refs.~\refcite{N1}--\refcite{Ag2} for a detailed discussion on the atomic coherent-state representation and its practical applications 
in multitime-correlation functions and phase-space quasidistributions.} (for instance, see the Dicke model\cite{Dicke}). Here, we follow closely the last way that 
consists in defining the spin coherent states through the mathematical relation $| \th,\varphi \rg \coloneq {\bf T}(\th,\varphi) | j,-j \rg$, with the vacuum state 
$| j,-j \rg$ written in terms of the previously mentioned basis $\{ | j,m \rg \}_{-j \leq m \leq j}$. Thus, after some algebra, it is easy to show that
\be
\lb{e5}
\!\!\!\!\! | \th,\varphi \rg = \sum_{k=0}^{2j} \lbk \mathscr{C}_{2j,k} \sin^{2k} (\th/2) \cos^{2(2j-k)} (\th/2) \rbk^{\half} \exp (- \nc k \varphi) 
| j,k-j \rg
\ee
presents a bijective mapping with the $N$-photon generalized binomial states related to the electromagnetic field, being this correspondence properly
explored in Ref.~\refcite{Messina}.

Next, we focus our attention on an important set of mathematical properties inherent to the spin coherent states (\ref{e5}) that leads us to discuss the 
squeezing effects in determined physical systems under different circumstances. Note that a complete list of such properties can be promptly found in 
Refs.~\refcite{Thomas} and \refcite{Zhang}.
\begin{description}
\item[Non-orthogonality.] The inner product of two distinct spin coherent states can be evaluated through the completeness relation associated with 
$\{ | j,m \rg \}$, yielding the well-known result 
\bd
\lg \th^{\prime},\varphi^{\prime} | \th,\varphi \rg = \lbr \cos (\th^{\prime}) \cos (\th) + \sin (\th^{\prime}) \sin (\th) 
\exp [ \nc (\varphi^{\prime} - \varphi) ] \rbr^{2j} .
\ed
For $\th^{\prime} = \th$ and $\varphi^{\prime} = \varphi$, we achieve $\lg \th,\varphi | \th,\varphi \rg = 1$ (normalization condition). Now, excepting 
for the antipodal points, the spin coherent states are in general not orthogonal. Hence, the Majorana-Bloch sphere\cite{Majorana} of unit radius represents 
the ideal geometric element to describe such states, since its respective north $(\th = \pi)$ and south $(\th = 0)$ poles correspond to the highest/lowest 
states $| j,\pm j \rg$. For completeness sake, let us briefly mention that $| \lg \th^{\prime},\varphi^{\prime} | \th,\varphi \rg |^{2} = \cos^{4j} ( \Theta/2 )$ 
(overlap probability) with $\cos (\Theta) \equiv \cos (\th^{\prime}) \cos (\th) + \sin (\th^{\prime}) \sin (\th) \cos (\varphi^{\prime} - \varphi)$ is limited 
to the closed interval $[0,1]$, which reinforces the overcomplete character of the states under consideration (they do not form an orthonormal set).

\item[Completeness relation.] Let ${\bf P}(\th,\varphi) = | \th,\varphi \rg \lg \th,\varphi |$ denote the diagonal projector operator related to the spin
coherent states which satisfies the property\cite{Perelomov}
\bd
\int \mathrm{d} \Omega (\th,\varphi) {\bf P}(\th,\varphi) = {\bf 1} \quad \mbox{with} \quad \mathrm{d} \Omega (\th,\varphi) = \frac{2j+1}{4 \pi} 
\sin (\th) \mathrm{d} \th \mathrm{d} \varphi .
\ed
In particular, such a completeness relation asserts that any quantum state $| \Psi \rg$ belonging to $\mathbb{C}^{2j+1}$ can be properly expanded in this 
overcomplete basis, namely
\bd
| \Psi \rg = \int \mathrm{d} \Omega (\th,\varphi) \Psi (\th,\varphi) | \th,\varphi \rg .
\ed
In this case, $\Psi (\th,\varphi)$ represents a polynomial function of degree $2j$, whose analytical expression is given by
\bd
\Psi (\th,\varphi) = \sum_{k=0}^{2j} \lbk \mathscr{C}_{2j,k} \sin^{2k} (\th/2) \cos^{2(2j-k)} (\th/2) \rbk^{\half} \exp ( \nc k \varphi) \,
\lg j,k-j | \Psi \rg .
\ed 
It is worth mentioning that operators acting on this Hilbert space also admit expansions in both the nondiagonal and diagonal forms, the diagonal 
representation being of particular interest to deal with statistical operators for atoms.\cite{Thomas}

\item[Minimum uncertainty states.] Note that spin coherent states constitute an important class of minimum uncertainty states. To demonstrate this assertion, 
let us initially consider the RS uncertainty principle\cite{Dodonov} for ${\bf J}_{x}$ and ${\bf J}_{y}$, \ie,
\be
\lb{e6}
\mathscr{V}_{\mathrm{J}_{x}} \mathscr{V}_{\mathrm{J}_{y}} - \lpar \mathscr{V}_{\mathrm{J}_{x} \mathrm{J}_{y}} \rpar^{2} \geq \frac{1}{4} \left| 
\lg \lbk {\bf J}_{x},{\bf J}_{y} \rbk \rg \right|^{2} = \frac{1}{4} | \lg {\bf J}_{z} \rg |^{2}
\ee
where $\mathscr{V}_{\mathrm{J}_{x} \mathrm{J}_{y}} \equiv \lg \half \lbr {\bf J}_{x},{\bf J}_{y} \rbr \rg - \lg {\bf J}_{x} \rg \lg {\bf J}_{y} \rg$
represents the covariance function, and $\mathscr{V}_{\mathrm{J}_{a}} \equiv \lg {\bf J}_{a}^{2} \rg - \lg {\bf J}_{a} \rg^{2}$ stands for usual
variance when $a=x,y$. So, if one considers the previous mean values evaluated for the spin coherent states, it is immediate to verify the equality
sign in this equation since both the expressions are equal to $\frac{1}{4} j^{2} \cos^{2} (\th)$. In fact, this particular result can be extended in
order to include any non-parametrized complex variable --- see Appendix A for further results.
\end{description}

%%%%%%%%%%%%%%%%%%%%%%%%%%%%%%%%%%%%%%%%%%%%%%%%%%%%%%%%%%%%%%%%%%%%%%%%%%%%%%%%%%%%%%%%%%%%%%%%%%%%%%%%%%%%%%%%%%%%%%%%%%%%%%%%%%%%%%%%%%%%%%%%%%%%%%%%
\section{Mappings via finite-dimensional discrete phase spaces}
%%%%%%%%%%%%%%%%%%%%%%%%%%%%%%%%%%%%%%%%%%%%%%%%%%%%%%%%%%%%%%%%%%%%%%%%%%%%%%%%%%%%%%%%%%%%%%%%%%%%%%%%%%%%%%%%%%%%%%%%%%%%%%%%%%%%%%%%%%%%%%%%%%%%%%%%

In this section, we will introduce certain basic mathematical tools constituents of the pioneering formalism initially developed in Refs.~\refcite{GP1}
and \refcite{GP2} for physical systems with finite-dimensional space of states $\mathcal{H}_{\dim}$. It is worth stressing that such tools will represent 
our guidelines for the fundamentals of the formal description of finite-dimensional phase spaces by means of discrete variables. For this specific task, 
let us first establish the mod(N)-invariant operator basis\cite{GP2}
\be
\lb{e7}
{\bf G}(\mu,\nu) \coloneq \frac{1}{\sqrt{{\rm N}}} \sum_{\eta,\xi = -\ell}^{\ell} \exp \lbk - \frac{2 \pi \nc}{{\rm N}} \lpar \eta \mu + \xi \nu \rpar 
\rbk {\bf S}(\eta,\xi)
\ee
expressed in terms of a discrete double Fourier transform of the symmetrized unitary operator basis\cite{GP1}
\bd
{\bf S}(\eta,\xi) \coloneq \frac{1}{\sqrt{{\rm N}}} \exp \lpar \frac{\pi \nc}{{\rm N}} \eta \xi \rpar {\bf U}^{\eta} {\bf V}^{\xi} ,
\ed
where the labels $\eta$ and $\xi$ are associated with the dual momentum and coordinatelike variables of an ${\rm N}^{2}$-dimensional discrete phase space
here endorsed by an underlying presymplectic structure of geometric origin.\cite{AD} Note that these discrete labels obey the arithmetic modulo N and
assume integer values in the symmetric interval $[ -\ell,\ell ]$ for $\ell = \frac{{\rm N}-1}{2}$ fixed.\footnote{Henceforth, for convenience, we assume 
${\rm N}$ odd throughout this paper. However, it is important to stress that even dimensionalities can also be dealt with simply by working on
non-symmetrized intervals.} A comprehensive and useful compilation of results and properties of the unitary operators ${\bf U}$ and ${\bf V}$ can be found in 
Ref.~\refcite{DM1}, since the primary focus of our attention is the essential features exhibited by $\{ {\bf G}(\mu,\nu) \}_{\mu,\nu = -\ell, \ldots, \ell}$.

The set of ${\rm N}^{2}$ operators $\{ {\bf G}(\mu,\nu) \}$ allows, for instance, to decompose any linear operator ${\bf O}$ acting on $\mathcal{H}_{\dim}$
by means of the expansion
\be
\lb{e8}
{\bf O} = \frac{1}{{\rm N}} \sum_{\mu,\nu = -\ell}^{\ell} \mathscr{O}(\mu,\nu) {\bf G}(\mu,\nu) ,
\ee
where the coefficients $\mathscr{O}(\mu,\nu) \equiv \tr \lbk {\bf G}^{\dagger}(\mu,\nu) {\bf O} \rbk$ are evaluated through trace operation and correspond, 
in this context, to a one-to-one mapping between operators and functions embedded in a finite phase-space characterized by the discrete variables $\mu$ and 
$\nu$.\cite{Fano} So, if one considers the density operator $\ro$ in such a case, we verify that
\be
\lb{e9}
\ro = \frac{1}{{\rm N}} \sum_{\mu,\nu = -\ell}^{\ell} \mathscr{W}_{\rho}(\mu,\nu) {\bf G}(\mu,\nu)
\ee
admits a plausible expansion with intrinsic mathematical properties since the coefficients result in the discrete Wigner function $\mathscr{W}_{\rho}(\mu,\nu) 
\coloneq \tr \lbk {\bf G}^{\dagger}(\mu,\nu) \ro \rbk$. The first practical consequence of this decomposition yields the mean value
\be
\lb{e10}
\lg {\bf O} \rg \coloneq \tr [ {\bf O} \ro ] = \frac{1}{{\rm N}} \sum_{\mu,\nu = -\ell}^{\ell} \mathscr{O}(\mu,\nu) \mathscr{W}_{\rho}(\mu,\nu) ,
\ee
whose formal expression depends explicitly on the product of both mapped forms related to ${\bf O}$ and $\ro$. Consequently, the moments $\lg {\bf J}_{a}^{k} \rg$
for $a=x,y,z$ and $0 \leq k \leq 2j$ can now be promptly obtained from this formalism for any finite quantum state belonging to $\mathcal{H}_{2j+1}$.
Next, we apply such a mapping technique for the spin coherent states in order to establish an important link between angular-momentum operators and
$(2j+1)^{2}$-dimensional discrete phase spaces.

%%%%%%%%%%%%%%%%%%%%%%%%%%%%%%%%%%%%%%%%%%%%%%%%%%%%%%%%%%%%%%%%%%%%%%%%%%%%%%%%%%%%%%%%%%%%%%%%%%%%%%%%%%%%%%%%%%%%%%%%%%%%%%%%%%%%%%%%%%%%%%%%%%%%%%%%
\subsection{Applications}
%%%%%%%%%%%%%%%%%%%%%%%%%%%%%%%%%%%%%%%%%%%%%%%%%%%%%%%%%%%%%%%%%%%%%%%%%%%%%%%%%%%%%%%%%%%%%%%%%%%%%%%%%%%%%%%%%%%%%%%%%%%%%%%%%%%%%%%%%%%%%%%%%%%%%%%%

According to a prescription adopted in Ref.~\refcite{GP3}, let $\{ | j,m \rg \}_{-j \leq m \leq j}$ denote the eigenstates of ${\bf U}$ with eigenvalues
$\{ \om^{m} \}$ for $\om = \exp \lpar \frac{2 \pi \nc}{2j+1} \rpar$ fixed. This assumption leads us to establish the general properties ${\bf U}^{\eta} 
| j,m \rg = \om^{m \eta} | j,m \rg$ and ${\bf V}^{\xi} | j,m \rg = | j,m - \xi \rg$ (together with the relations ${\bf U}^{2j+1} = {\bf 1}$, 
${\bf V}^{2j+1} = {\bf 1}$, and ${\bf V}^{\xi} {\bf U}^{\eta} = \om^{\eta \xi} {\bf U}^{\eta} {\bf V}^{\xi}$), which represent a satisfactory mathematical
connection for the present purpose. Next, we determine some useful results related to the angular-momentum operators and spin coherent states, remembering
that ${\bf G}^{\dagger}(\mu,\nu) = {\bf G}(\mu,\nu)$ for $-\ell \leq \mu,\nu \leq \ell$.

As a first and pertinent application, let us consider the mapped expressions
\brr
\lpar {\bf J}_{z}^{k} \rpar (\mu,\nu) &=& \sum_{m=-j}^{j} m^{k} \underbrace{\lg j,m | {\bf G}(\mu,\nu) | j,m \rg}_{\mathscr{G}_{j,m|j,m}(\mu,\nu)} \nn \\
\lpar {\bf J}_{+}^{k} \rpar (\mu,\nu) &=& \sum_{m=-j}^{j-k} \frac{(j+m+k)!}{(j+m)!} \lpar \frac{\mathscr{C}_{2j,j+m+k}}{\mathscr{C}_{2j,j+m}} \rpar^{\half} 
\underbrace{\lg j,m | {\bf G}(\mu,\nu) | j,m+k \rg}_{\mathscr{G}_{j,m|j,m+k}(\mu,\nu)} \nn \\
\lpar {\bf J}_{-}^{k} \rpar (\mu,\nu) &=& \sum_{m=-j+k}^{j} \frac{(j-m+k)!}{(j-m)!} \lpar \frac{\mathscr{C}_{2j,j+m-k}}{\mathscr{C}_{2j,j+m}} \rpar^{\half}
\underbrace{\lg j,m | {\bf G}(\mu,\nu) | j,m-k \rg}_{\mathscr{G}_{j,m|j,m-k}(\mu,\nu)} , \nn
\err
which depend on the nondiagonal matrix elements
\bd
\mathscr{G}_{j,m|j,m^{\prime}}(\mu,\nu) = \frac{1}{2j+1} \sum_{\bet = -j}^{j} \exp \lbr - \frac{2 \pi \nc}{2j+1} \lbk \bet ( \mu - m^{\prime} ) +
(m^{\prime} - m) \lpar \nu + \frac{\bet}{2} \rpar \rbk \rbr
\ed
associated with the discrete mapping kernel ${\bf G}(\mu,\nu)$ for $m^{\prime} = m \pm k$ and $0 \leq k \leq 2j$. Since the diagonal matrix element
$\mathscr{G}_{j,m|j,m}(\mu,\nu) = \delta_{m,\mu}^{[2j+1]}$ gives the Kronecker delta function (in such a case, the superscript $[2j+1]$ denotes that this
function is different from zero when its labels are congruent modulo $2j+1$), it seems reasonable to obtain a simple expression in the first example, that
is $\lpar {\bf J}_{z}^{k} \rpar (\mu,\nu) = \mu^{k}$; however, this fact is not verified for $\lpar {\bf J}_{\pm}^{k} \rpar (\mu,\nu)$. From the kinematical 
and dynamical point of view, these mapped expressions represent an advantageous set of mathematical tools which allows us to comprehend certain intriguing 
problems related to discrete symmetries of spin systems.\cite{GP3}

Basically, the second application consists in evaluating the discrete Wigner function $\mathscr{W}_{\th,\varphi}(\mu,\nu)$ present in the expansion of the
projector ${\bf P}(\th,\varphi)$ via Eq. (\ref{e9}), namely,
\bd
{\bf P}(\th,\varphi) = \frac{1}{2j+1} \sum_{\mu,\nu = -j}^{j} \mathscr{W}_{\th,\varphi}(\mu,\nu) {\bf G}(\mu,\nu) .
\ed
In general, this particular quasidistribution can be written in terms of its respective discrete Weyl function $\widetilde{\mathscr{W}}_{\th,\varphi} 
(\eta,\xi) \coloneq \tr [ {\bf S}(\eta,\xi) {\bf P}(\th,\varphi) ]$ as follows:
\be
\lb{e11}
\!\!\!\!\! \mathscr{W}_{\th,\varphi}(\mu,\nu) = \frac{1}{\sqrt{2j+1}} \sum_{\mu,\nu = -j}^{j} \exp \lbk - \frac{2 \pi \nc}{2j+1} ( \eta \mu + \xi \nu ) 
\rbk \widetilde{\mathscr{W}}_{\th,\varphi} (\eta,\xi) .
\ee
So, after some lengthy calculations, the analytical expression for $\widetilde{\mathscr{W}}_{\th,\varphi} (\eta,\xi)$ assumes the exact form
\bd
\widetilde{\mathscr{W}}_{\th,\varphi} (\eta,\xi) = \frac{1}{\sqrt{2j+1}} \sum_{m=-j}^{j} \exp \lbk \frac{2 \pi \nc}{2j+1} \eta \lpar m - \frac{\xi}{2} 
\rpar \rbk \lg \th,\varphi | j,m- \xi \rg \lg j,m | \th,\varphi \rg , 
\ed
where $\lg j,m | \th,\varphi \rg$ can be promptly obtained from Eq. (\ref{e5}). It is interesting to stress that both the north $(\th = \pi)$ and south
$(\th = 0)$ poles of the Majorana-Bloch sphere correspond to the limit cases $\mathscr{W}_{\pi,\varphi}(\mu,\nu) = \delta_{\mu,j}^{[2j+1]}$ and
$\mathscr{W}_{0,\varphi}(\mu,\nu) = \delta_{\mu,-j}^{[2j+1]}$, which reflect directly the highest/lowest states $| j, \pm j \rg$ for any $\varphi \in
[0,2 \pi)$. Furthermore, numerical calculations related to Eq. (\ref{e11}) suggest that some relevant contributions are connected with discrete 
values of the angles $\th$ and $\varphi$ --- in particular, integer multiples of $\frac{\pi}{2j+1}$. This evidence supports the ideas of Buniy and
coworkers\cite{Zee} about a `possible discretization' of the Majorana-Bloch sphere, as well as reveals some important features --- and not yet properly 
explored --- on the discrete nature of quantum states in finite-dimensional Hilbert spaces.\footnote{In the quantum-gravity scope,\cite{Calmet} it is 
worth mentioning that certain theoretical approaches to the generalized uncertainty principle (GUP) also suggest the breakdown of the spacetime
continuum picture near to Planck scale.\cite{Vagenas} From an experimental point of view, it seems reasonable to argue that `discreteness is actually 
less speculative than absolute continuity'.} 

%%%%%%%%%%%%%%%%%%%%%%%%%%%%%%%%%%%%%%%%%%%%%%%%%%%%%%%%%%%%%%%%%%%%%%%%%%%%%%%%%%%%%%%%%%%%%%%%%%%%%%%%%%%%%%%%%%%%%%%%%%%%%%%%%%%%%%%%%%%%%%%%%%%%%%%%
\subsection{Time evolution}
%%%%%%%%%%%%%%%%%%%%%%%%%%%%%%%%%%%%%%%%%%%%%%%%%%%%%%%%%%%%%%%%%%%%%%%%%%%%%%%%%%%%%%%%%%%%%%%%%%%%%%%%%%%%%%%%%%%%%%%%%%%%%%%%%%%%%%%%%%%%%%%%%%%%%%%%

Here, we establish a mathematical recipe that permits to investigate the dynamics of a particular quantum system characterized by a finite space of 
(discrete) states. For this task, let $\ro(t)$ describe the state of this physical system whose interaction with any dissipative environment is, in 
principle, automatically discarded. Besides, let us consider (for convenience) only time-independent Hamiltonians; consequently, the time-evolution of
that density operator will be governed, in such a case, by the well-known von Neumann-Liouville equation $\nc \hbar \partial_{t} \ro(t) = [ {\bf H},\ro(t) ]$. 
So, those initial premises represent the constituting blocks for the aforementioned recipe.

As a first proposal in our prescription, we determine a mapped expression for the von Neumann-Liouville equation which describes the time evolution of
the discrete Wigner function $\mathscr{W}_{\rho}(\mu,\nu;t)$. In this sense, the mapping technique sketched in this section yields the differential 
equation
\be
\lb{e12}
\nc \hbar \partial_{t} \mathscr{W}_{\rho}(\mu,\nu;t) = \sum_{\mu^{\prime},\nu^{\prime} = -\ell}^{\ell} \mathscr{L}_{\hei}(\mu,\nu,\mu^{\prime},\nu^{\prime})
\mathscr{W}_{\rho}(\mu^{\prime},\nu^{\prime};t) ,
\ee
where $\mathscr{L}_{\hei}(\mu,\nu,\mu^{\prime},\nu^{\prime})$ represents the mapped form of the Liouville operator written in terms of 
$\mathit{H}(\mu^{\prime \prime},\nu^{\prime \prime}) = \tr \lbk {\bf G}(\mu^{\prime \prime},\nu^{\prime \prime}) {\bf H} \rbk$, that is,
\brr
\mathscr{L}_{\hei}(\mu,\nu,\mu^{\prime},\nu^{\prime}) &=& \frac{2 \nc}{{\rm N}^{4}} \sum_{\Delta} \sin \lbk \frac{\pi}{{\rm N}} ( \alf \bet^{\prime} -
\alf^{\prime} \bet ) \rbk \exp \lbr \frac{2 \pi \nc}{{\rm N}} \lbk \alf (\mu - \mu^{\prime}) + \bet (\nu - \nu^{\prime}) \rbk \rbr \nn \\
& & \times \exp \lbr \frac{2 \pi \nc}{{\rm N}} \lbk \alf^{\prime} (\mu - \mu^{\prime \prime}) + \bet^{\prime} (\nu - \nu^{\prime \prime}) \rbk \rbr
\mathit{H}(\mu^{\prime \prime},\nu^{\prime \prime}) . \nn
\err
Note that $\Delta$ denotes the set $\{ \alf,\bet,\alf^{\prime},\bet^{\prime},\mu^{\prime \prime},\nu^{\prime \prime} \} \in [-\ell,\ell]$ in this expression. 
In the following, let us mention some few words about the formal solution of Eq. (\ref{e12}): it can be expressed analytically in terms of the series\cite{GR1}
\be
\lb{e13}
\mathscr{W}_{\rho}(\mu,\nu;t) = \sum_{\kappa,\tau = -\ell}^{\ell} \mathscr{P}( \mu,\nu;t | \kappa,\tau;t_{0} ) \mathscr{W}_{\rho}(\kappa,\tau;t_{0}) ,
\ee
whose ${\rm N}^{2}$-dimensional discrete phase-space propagator admits the expansion 
\brr
\mathscr{P}( \mu,\nu;t | \kappa,\tau;t_{0} ) &=& \delta_{\kappa,\mu}^{[\dim]} \delta_{\tau,\nu}^{[\dim]} + \frac{\nc}{1! \hbar} (t-t_{0})
\mathscr{L}_{\hei}(\mu,\nu,\kappa,\tau) \nn \\
& & + \frac{\nc^{2}}{2! \hbar^{2}} (t-t_{0})^{2} \sum_{\kappa^{\prime},\tau^{\prime} = -\ell}^{\ell} \mathscr{L}_{\hei}(\mu,\nu,\kappa^{\prime},\tau^{\prime})
\mathscr{L}_{\hei}(\kappa^{\prime},\tau^{\prime},\kappa,\tau) + \cdots \nn
\err
which allows to evaluate directly the time evolution of $\mathscr{W}_{\rho}(\mu,\nu;t)$ by using the series related to the iterated application of the
mapped Liouville operator. The advantages and/or disadvantages from this particular formal solution were adequately discussed in Ref.~\refcite{GR1}, and 
subsequently applied with great success in the discrete Husimi-function context for the LMG model.\cite{MED}

The alternative proposal establishes a differential equation for the discrete Weyl function analogous to Eq. (\ref{e12}). Then, after some calculations related 
to the product of three symmetrized unitary operator bases and its respective trace operation,\cite{DM1} it is immediate to show that such an equation can be 
written as
\be
\lb{e14}
\nc \hbar \partial_{t} \widetilde{\mathscr{W}}_{\rho} (\eta,\xi;t) = \sum_{\eta^{\prime},\xi^{\prime} = - \ell}^{\ell} \mathcal{L}_{\hei}
(\eta,\xi,\eta^{\prime},\xi^{\prime}) \widetilde{\mathscr{W}}_{\rho} (\eta^{\prime},\xi^{\prime};t) ,
\ee
where $\widetilde{\mathscr{W}}_{\rho} (\eta,\xi;t) \coloneq \tr \lbk {\bf S}(\eta,\xi) \ro(t) \rbk$ corresponds to our object of study and
\bd
\mathcal{L}_{\hei}(\eta,\xi,\eta^{\prime},\xi^{\prime}) = \frac{2 \nc}{\sqrt{N}} \sin \lbk \frac{\pi}{N} \lpar \eta^{\prime} \xi - \xi^{\prime} \eta \rpar
\rbk \underbrace{\mathcal{H}(\eta - \eta^{\prime}, \xi - \xi^{\prime})}_{\mathrm{Tr} \lbk {\bf S}(\eta - \eta^{\prime}, \xi - \xi^{\prime}) {\bf H} \rbk}
\ed
yields the mapped forms of the Liouville and Hamiltonian operators. As expected, Eq. (\ref{e14}) describes the time evolution of the discrete Weyl function 
whose formal solution is given by
\be
\lb{e15}
\widetilde{\mathscr{W}}_{\rho} (\eta,\xi;t) = \sum_{\eta^{\prime},\xi^{\prime} = - \ell}^{\ell} \mathcal{P}(\eta,\xi;t | \eta^{\prime},\xi^{\prime};t_{0})
\widetilde{\mathscr{W}}_{\rho} (\eta^{\prime},\xi^{\prime};t_{0}) ,
\ee
with $\mathcal{P}(\eta,\xi;t | \eta^{\prime},\xi^{\prime};t_{0})$ being the discrete dual phase-space propagator which admits a time expansion similar
to $\mathscr{P}( \mu,\nu;t | \kappa,\tau;t_{0} )$, that is
\brr
\mathcal{P}( \eta,\xi;t | \eta^{\prime},\xi^{\prime};t_{0} ) &=& \delta_{\eta^{\prime},\eta}^{[\dim]} \delta_{\xi^{\prime},\xi}^{[\dim]} + 
\frac{\nc}{1! \hbar} (t-t_{0}) \mathcal{L}_{\hei}(\eta,\xi,\eta^{\prime},\xi^{\prime}) \nn \\
& & + \frac{\nc^{2}}{2! \hbar^{2}} (t-t_{0})^{2} \sum_{\eta^{\prime \prime},\xi^{\prime \prime} = -\ell}^{\ell} \mathcal{L}_{\hei}
(\eta,\xi,\eta^{\prime \prime},\xi^{\prime \prime}) \mathcal{L}_{\hei}(\eta^{\prime \prime},\xi^{\prime \prime},\eta^{\prime},\xi^{\prime}) + \cdots . \nn
\err
Since the discrete Wigner and Weyl functions are connected by means of a double Fourier transform, the low operational costs involved in this mathematical
recipe are advantageous --- from a computational point of view --- if compared with the previous one. Notwithstanding the apparent advantage, it is worth
remembering that $\widetilde{\mathscr{W}}_{\rho} (\eta,\xi;t)$ is a complex function and hence it represents an intermediate step for the main goal, namely,
the time evolution of the discrete Wigner function --- and consequently, the time evolution of certain mean values via Eq. (\ref{e10}). The schematic
diagram shown below
\brr
\xymatrix@1{
\mathit{H}(\mu^{\prime \prime},\nu^{\prime \prime}) \ar[d] & \mathcal{H}(\eta - \eta^{\prime}, \xi - \xi^{\prime}) \ar[d] \\
\mathscr{L}_{\hei}(\mu,\nu,\mu^{\prime},\nu^{\prime}) \ar[d] & \mathcal{L}_{\hei}(\eta,\xi,\eta^{\prime},\xi^{\prime}) \ar[d] \\
\mathscr{P}( \mu,\nu;t | \kappa,\tau;t_{0} ) \ar[d] & \mathcal{P}(\eta,\xi;t | \eta^{\prime},\xi^{\prime};t_{0}) \ar[d] \\
\mathscr{W}_{\rho}(\mu,\nu;t) & \widetilde{\mathscr{W}}_{\rho} (\eta,\xi;t) \ar[l]^{\mathrm{FT}} } \nn
\err
represents a summary of the previous proposals for describing the dynamics of a finite quantum system represented in a ${\rm N}^{2}$-dimensional discrete 
phase space.\cite{GR1} 

%%%%%%%%%%%%%%%%%%%%%%%%%%%%%%%%%%%%%%%%%%%%%%%%%%%%%%%%%%%%%%%%%%%%%%%%%%%%%%%%%%%%%%%%%%%%%%%%%%%%%%%%%%%%%%%%%%%%%%%%%%%%%%%%%%%%%%%%%%%%%%%%%%%%%%%%
\section{Spin squeezing and entanglement}
%%%%%%%%%%%%%%%%%%%%%%%%%%%%%%%%%%%%%%%%%%%%%%%%%%%%%%%%%%%%%%%%%%%%%%%%%%%%%%%%%%%%%%%%%%%%%%%%%%%%%%%%%%%%%%%%%%%%%%%%%%%%%%%%%%%%%%%%%%%%%%%%%%%%%%%%

This section will illustrate how the mapping techniques leading to finite-dimensional discrete phase spaces can be effectively used in the study of 
important quantum effects, such as spin squeezing and entanglement. Basically, we will invoke the LMG model already discussed in current literature 
(\eg, see Ref.~\refcite{Ring}) which exhibits not only strong mathematical and physical appeals,\cite{MED} but also a fundamental physical 
property: quantum correlations due to different types of interactions and mediated by a transverse magnetic field. Since quantum correlations are 
responsible for the aforementioned effects, it is natural to investigate how the interactions involved in this model affect the standard quantum 
noise related to the spin coherent states.\footnote{In Appendix B, we investigate a soluble spin model with lowest-order nonlinear interaction in 
${\bf J}_{z}$,\cite{Kitagawa} whose algebraic features permit to attain a particular set of analytical results very useful in the study of spin squeezing 
and entanglement effects via quantum correlations.\cite{Wine}}

%%%%%%%%%%%%%%%%%%%%%%%%%%%%%%%%%%%%%%%%%%%%%%%%%%%%%%%%%%%%%%%%%%%%%%%%%%%%%%%%%%%%%%%%%%%%%%%%%%%%%%%%%%%%%%%%%%%%%%%%%%%%%%%%%%%%%%%%%%%%%%%%%%%%%%%%
\subsection{The modified Lipkin-Meshkov-Glick model}
%%%%%%%%%%%%%%%%%%%%%%%%%%%%%%%%%%%%%%%%%%%%%%%%%%%%%%%%%%%%%%%%%%%%%%%%%%%%%%%%%%%%%%%%%%%%%%%%%%%%%%%%%%%%%%%%%%%%%%%%%%%%%%%%%%%%%%%%%%%%%%%%%%%%%%%%

Here, we adopt the prescription established by Vidal and coworkers\cite{Dusuel} for the modified LMG model through the Hamiltonian operator 
(written in the spin language)
\be
\lb{e16}
{\bf H} = -2h {\bf J}_{z} - 2 \wp_{+} \lbk \vec{{\bf J}}^{2} - {\bf J}_{z}^{2} - (N/2) {\bf I} \rbk - \wp_{-} \lpar {\bf J}_{+}^{2} + {\bf J}_{-}^{2} \rpar ,
\ee
which describes a set of $N$ spins half mutually interacting in the $xy$ plane subjected to a transverse magnetic field $h$. The coefficients 
$\wp_{\pm} = \frac{\lam}{2N} (1 \pm \gam)$ have an important role in this description: they allow to (i) investigate the different anti-ferromagnetic 
$( \lam < 0 )$ and ferromagnetic $( \lam > 0 )$ cases inherent to the model for any anisotropy parameter $| \gam | \leq 1$ ($\gam = 1$ refers to the 
isotropic case), and also (ii) ensure that the free energy per spin is finite in the thermodynamical limit. In particular, such a model presents a 
second-order quantum phase transition at $\lam = | h |$ for $\lam > 0$ fixed, whose symmetric $( \lam < |h| )$ and broken $( \lam > |h| )$ phases are
well-defined within the mean-field approach; furthermore, its ground-state entanglement properties exhibit a rich structure which reflects the internal
symmetries of the Hamiltonian operator ${\bf H}$.\cite{Dusuel,Vidal} Indeed, such an operator preserves the magnitude of the total spin operator since
$[ {\bf H},\vec{{\bf J}}^{2} ] = 0$ for all $\gam$, and does not couple states having a different number of spins pointing in the field direction, namely,
\bd
\biggl[ {\bf H}, \prod_{j=1}^{N} \sig_{j,z} \biggr] = 0 \qquad (\mbox{spin-flip symmetry}) .
\ed
Consequently, it is immediate to verify that Eq. (\ref{e16}) can be diagonalized within each $(2j+1)$-dimensional multiplet labelled by the eigenvalues
of $\vec{{\bf J}}^{2}$ and ${\bf J}_{z}$, this fact being responsible for the soluble character of the associated spin model.\footnote{It is important to
stress that the prescritpion here adopted does not coincide with the original Lipkin-Meshkov-Glick model\cite{Lipkin}, which was initially introduced 
over 40 years ago in nuclear physics\cite{Ring} for treating certain fermionic systems. In fact, this new modified version brings to scene the modern 
statistical mechanics point of view, where the collective properties of spin systems can be worked out with great success.\cite{Dusuel,Vidal} The extended
original version of the LMG model, which explores the parity symmetry via finite-dimensional discrete phase spaces, can be found in Ref.~\refcite{GP3}.}

Next, let us mention some few words about two discrete conserved quantities inherent to the spin model under investigation which reflect certain additional
symmetry properties for $\gam = 1$ fixed. In this case, the simplest operator commuting with ${\bf H}$, therefore giving a constant of motion, is the parity 
operator $\mpi$ --- here defined as $\mpi \coloneq {\bf R}_{z} (\pi) \equiv \exp ( \nc \pi {\bf J}_{z} )$. This result tells us that the Hamiltonian matrix, in 
the ${\bf J}_{z}$ representation, breaks into two disjoint blocks involving only even and odd eigenvalues of ${\bf J}_{z}$, respectively. The second interesting 
quantity comes from the anticommutation relation $\{ {\bf H},{\bf R} \} = 0$ for ${\bf R}(\pi,0,\pi/2) \equiv {\bf R}_{x}(\pi) {\bf R}_{z}(\pi/2)$: it corresponds 
to a particular rotation of the angular-momentum quantization frame by the Euler angles $(\pi,0,\pi/2)$, transforming, in this way, ${\bf H} \rightarrow - {\bf H}$ 
for the isotropic case. Thus, if $| E_{j} \rg$ is an energy eigenstate with eigenvalue $E_{j}$, then ${\bf R} | E_{j} \rg$ is also an eigenstate of ${\bf H}$ 
with eigenvalue $-E_{j}$. This specific symmetry property gives rise to an energy spectrum that is symmetric about zero.\cite{GP3,MED,Dusuel}

After this condensed review, we establish below a sequence of steps that allows us to calculate the time evolution of several quantities necessary for investigating 
both the squeezing and entanglement effects associated with the aforementioned spin model in terms of the parameters $h$ and $\wp_{\pm}$. The first one consists in 
adopting the theoretical framework described in the previous section for the time-dependent discrete Wigner function $\mathscr{W}_{\rho}(\mu,\nu;t)$ defined upon a 
$(2j+1)^{2}$-dimensional discrete phase space labeled by the angular-momentum and angle pair $(\mu,\nu) \in [-j,j]$ --- in particular, we adopt the theoretical
prescription established in Ref.~\refcite{GR1} for the angle variable, that is $\th_{\nu} = \frac{2 \pi}{2j+1} \nu$. Since this approach depends on the Wigner function 
evaluated at time $t_{0}=0$, the second step consists in fixing the spin coherent states as initial state --- see Eq. (\ref{e11}). The last step refers to the numerical 
calculation of the moments
\be
\lb{e17}
\lg {\bf J}_{a}^{k} \rg_{\th,\varphi}(t) = \frac{1}{2j+1} \sum_{\mu,\nu = -j}^{j} \lpar {\bf J}_{a}^{k} \rpar (\mu,\nu) \mathscr{W}_{\th,\varphi}(\mu,\nu;t)
\ee
and covariance functions in the Schr\"{o}dinger picture fixing the initial state in $\th = \frac{\pi}{2}$ and $\varphi = 0$ (in this way, the uncertainties 
are redistributed between the orthogonal components in the $yz$-plane), and also considering (for convenience, not necessity) only the ferromagnetic case 
$\lam = 1$. Henceforth, Eq. (\ref{e16}) will assume the simplified form ${\bf H}^{\prime} = - h {\bf J}_{z} - \frac{1}{N} \lpar {\bf J}_{x}^{2} + \gam
{\bf J}_{y}^{2} \rpar$, where the constant term $\frac{1}{4} (1 + \gam) {\bf I}$ was suppressed at this initial stage since it represents only a phase for 
the time evolution operator with null contribution (as expected) within our computational approach.

%%%%%%%%%%%%%%%%%%%%%%%%%%%%%%%%%%%%%%%%%%%%%%%%%%%%%%%%%%%%%%%%%%%%%%%%%%%%%%%%%%%%%%%%%%%%%%%%%%%%%%%%%%%%%%%%%%%%%%%%%%%%%%%%%%%%%%%%%%%%%%%%%%%%%%%%%
\subsection{The match between squeezing and entanglement effects}
%%%%%%%%%%%%%%%%%%%%%%%%%%%%%%%%%%%%%%%%%%%%%%%%%%%%%%%%%%%%%%%%%%%%%%%%%%%%%%%%%%%%%%%%%%%%%%%%%%%%%%%%%%%%%%%%%%%%%%%%%%%%%%%%%%%%%%%%%%%%%%%%%%%%%%%%%

In this subsection, we adopt the theoretical framework established for the Kitagawa-Ueda model (see Appendix B) concerning the match between squeezing 
and entanglement effects. In particular, let us initially consider the results obtained in Table \ref{tab1} for the RS uncertainty principle related to
the angular-momentum generators and its connections with the $\mathcal{S}$-inequalities. Since the covariance function $\mathscr{V}_{\mathrm{J}_{a} 
\mathrm{J}_{b}}(t)$ has a central role in this algebraic approach and requires only numerical computations of $\lg {\bf J}_{a}^{k} \rg_{\th,\varphi}(t)$
and $\lg \{ {\bf J}_{a},{\bf J}_{b} \} \rg_{\th,\varphi}(t)$, it is convenient to clarify certain subtle steps that are inherent to the exact analytical
calculation of the last quantity. According to the mapping technique discussed in Section 3, the expression for
\be
\lb{e18}
\!\!\!\!\! \lg \{ {\bf J}_{a},{\bf J}_{b} \} \rg_{\th,\varphi}(t) = \frac{1}{2j+1} \sum_{\mu,\nu = -j}^{j} \lpar \{ {\bf J}_{a},{\bf J}_{b} \} \rpar (\mu,\nu) 
\mathscr{W}_{\th,\varphi}(\mu,\nu;t)
\ee
basically depends on the mapped form of the anticommutation relation, that is
\bd
\lpar \{ {\bf J}_{a},{\bf J}_{b} \} \rpar (\mu,\nu) = \frac{1}{(2j+1)^{2}} \sum_{\Omega} \Gamma_{{\rm A}}(\mu,\nu | \mu^{\prime},\nu^{\prime},
\mu^{\prime \prime},\nu^{\prime \prime}) \lpar {\bf J}_{a} \rpar (\mu^{\prime},\nu^{\prime}) \lpar {\bf J}_{b} \rpar (\mu^{\prime \prime},\nu^{\prime \prime}) 
\ed
where $\Omega$ stands for $\{ \mu^{\prime},\nu^{\prime},\mu^{\prime \prime},\nu^{\prime \prime} \} \in [-j,j]$. It is worth stressing that\cite{DM1}
\brr
& & \Gamma_{{\rm A}}(\mu,\nu | \mu^{\prime},\nu^{\prime},\mu^{\prime \prime},\nu^{\prime \prime}) = \frac{2}{(2j+1)^{2}} \sum_{\Omega^{\prime}} \exp \lbr 
\frac{2 \pi \nc}{2j+1} \lbk \eta^{\prime} ( \mu - \mu^{\prime} ) + \xi^{\prime} ( \nu - \nu^{\prime} ) \rbk \rbr \nn \\
& & \qquad \times \exp \lbr \frac{2 \pi \nc}{2j+1} \lbk \eta^{\prime \prime} ( \mu - \mu^{\prime \prime} ) + \xi^{\prime \prime} ( \nu - \nu^{\prime \prime} ) 
\rbk \rbr \cos \lbk \frac{\pi}{2j+1} ( \eta^{\prime} \xi^{\prime \prime} - \xi^{\prime} \eta^{\prime \prime} ) \rbk \nn
\err
shows explicitly the embryonic structure of the continuous cosine function presents in the well-known Weyl-Wigner-Moyal phase space approach\cite{Zachos}
--- here, $\Omega^{\prime}$ denotes the set $\{ \eta^{\prime},\xi^{\prime},\eta^{\prime \prime},\xi^{\prime \prime} \} \in [-j,j]$. Thus, the
$\mathcal{S}$-inequalities can be properly estimated for the modified LMG model, which lead us to investigate the squeezing effects. For completeness sake,
let us briefly mention that $\lg [ {\bf J}_{a},{\bf J}_{b} ] \rg_{\th,\varphi}(t)$ yields an expression analogous to Eq. (\ref{e18}), but with two minor
modifications: the function $\cos (z)$ in $\Gamma_{{\rm A}}$ should be replaced by $\nc \sin (z)$ (keeping constant the argument of the trigonometric
functions), which implies in the change $\Gamma_{{\rm A}} \rightarrow \Gamma_{{\rm C}}$.
%%%%%%%%%%%%%%%%%%%%%%%%%%%%%%%%%%%%%%%%%%%%%%%%%%%%%%%%%%%%%%%%%%%%%%%%%%%%%%%%%%%%%%%%%%%%%%%%%%%%%%%%%%%%%%%%%%%%%%%%%%%%%%%%%%%%%%%%%%%%%%%%%%%%%%%%%%
\begin{table}[!t]
\tbl{The $\mathcal{S}$-inequalities exhibit unique mathematical virtues since they yield a direct connection with the RS uncertainty principle related to
the non-commuting pair $\{ {\bf J}_{a},{\bf J}_{b} \}_{a,b=x,y,z}$ of angular-momentum operators, where the covariance functions have an important role.
Indeed, such inequalities lead us to investigate the squeezing effects through a well-established criterion in literature: for $\mathcal{S}_{a}^{(c)} < 1$ 
(squeezing condition) and $\mathcal{S}_{b}^{(c)} > 1$, the inequality $\mathcal{S}_{a}^{(c)} \mathcal{S}_{b}^{(c)} > 1$ is always preserved; moreover, the 
saturation $\mathcal{S}_{a}^{(c)} \mathcal{S}_{b}^{(c)} = 1$ describes minimum uncertainty states. In this table, we show all the possible links among RS
uncertainty principles and $\mathcal{S}$-inequalities, with $\mathscr{V}_{\mathrm{J}_{a} \mathrm{J}_{b}}$ restricted to the closed interval $\lbk - 
\sqrt{\mathscr{V}_{\mathrm{J}_{a}} \mathscr{V}_{\mathrm{J}_{b}}}, \sqrt{\mathscr{V}_{\mathrm{J}_{a}} \mathscr{V}_{\mathrm{J}_{b}}} \, \rbk$; moreover, the
superscript $(c)$ of the product $\mathcal{S}_{a}^{(c)} \mathcal{S}_{b}^{(c)}$ denotes the angular-momentum component resulting from the commutation relation
between ${\bf J}_{a}$ and ${\bf J}_{b}$.}
{\begin{tabular}{@{}ccc@{}} 
\toprule
RS uncertainty principles & $\mathcal{R}$-denominators & $\mathcal{S}$-inequalities \\
$\{ {\bf J}_{x},{\bf J}_{y},{\bf J}_{z} \} \in \mathcal{H}_{\dim}$ & $\mathcal{R}_{abc} \neq 0 \; \forall \ro \in \mathcal{H}_{\dim}$ & 
$\mathcal{S}_{a}^{(c)} \mathcal{S}_{b}^{(c)} \coloneq \mathscr{V}_{\mathrm{J}_{a}} \mathscr{V}_{\mathrm{J}_{b}} / \mathcal{R}_{abc}^{2}$ \\ 
\colrule
$\mathscr{V}_{\mathrm{J}_{x}} \mathscr{V}_{\mathrm{J}_{y}} - \lpar \mathscr{V}_{\mathrm{J}_{x} \mathrm{J}_{y}} \rpar^{2} \geq \frac{1}{4} | \lg {\bf J}_{z} 
\rg |^{2}$ & $\mathcal{R}_{xyz} \coloneq \lbk ( \mathscr{V}_{\mathrm{J}_{x} \mathrm{J}_{y}} )^{2} + \frac{1}{4} | \lg {\bf J}_{z} \rg |^{2} \rbk^{\half}$ &
$\mathcal{S}_{x}^{(z)} \mathcal{S}_{y}^{(z)} \geq 1$ \\
$\mathscr{V}_{\mathrm{J}_{x}} \mathscr{V}_{\mathrm{J}_{z}} - \lpar \mathscr{V}_{\mathrm{J}_{x} \mathrm{J}_{z}} \rpar^{2} \geq \frac{1}{4} | \lg {\bf J}_{y} 
\rg |^{2}$ & $\mathcal{R}_{xzy} \coloneq \lbk ( \mathscr{V}_{\mathrm{J}_{x} \mathrm{J}_{z}} )^{2} + \frac{1}{4} | \lg {\bf J}_{y} \rg |^{2} \rbk^{\half}$ &
$\mathcal{S}_{x}^{(y)} \mathcal{S}_{z}^{(y)} \geq 1$ \\
$\mathscr{V}_{\mathrm{J}_{y}} \mathscr{V}_{\mathrm{J}_{z}} - \lpar \mathscr{V}_{\mathrm{J}_{y} \mathrm{J}_{z}} \rpar^{2} \geq \frac{1}{4} | \lg {\bf J}_{x} 
\rg |^{2}$ & $\mathcal{R}_{yzx} \coloneq \lbk ( \mathscr{V}_{\mathrm{J}_{y} \mathrm{J}_{z}} )^{2} + \frac{1}{4} | \lg {\bf J}_{x} \rg |^{2} \rbk^{\half}$ &
$\mathcal{S}_{y}^{(x)} \mathcal{S}_{z}^{(x)} \geq 1$ \\
\botrule
\end{tabular}}
\label{tab1}
\end{table}
%%%%%%%%%%%%%%%%%%%%%%%%%%%%%%%%%%%%%%%%%%%%%%%%%%%%%%%%%%%%%%%%%%%%%%%%%%%%%%%%%%%%%%%%%%%%%%%%%%%%%%%%%%%%%%%%%%%%%%%%%%%%%%%%%%%%%%%%%%%%%%%%%%%%%%%%%%

In order to illustrate the theoretical framework developed in this work, let us now consider the $\mathcal{S}$-inequality $\mathcal{S}_{y}^{(x)} 
\mathcal{S}_{z}^{(x)} \geq 1$ and the $\mathcal{E}$-inequalities $\{ \mathcal{E}_{y},\mathcal{E}_{z} \}$ described in Appendix B by means of Eq. (\ref{b3}).
In particular, this procedure allows not only to establish a direct link between squeezing and entanglement effects for the modified LMG model, but also
to investigate how these effects are affected by the transverse magnetic field $h$ and anisotropy parameter $\gam$ with $N=20$ fixed. For instance, 
Fig. \ref{fig1}(a,c) represents the plots of $\mathcal{S}_{y}^{(x)}(t)$ (dot-dashed line) and $\mathcal{S}_{z}^{(x)}(t)$ (solid line) versus $t \in [0,50]$
for different values of $(h,\gam)$: (a) $(-0.1,0.2)$ and (c) $(-0.13,0.1)$. It is interesting to observe at a first glance how the squeezing effect is sensitive 
to small variations in the parameters $|h|$ and $\gam$. Further numerical computations corroborate this sensitivity and allow us to produce the following proper 
description about such an evidence as time goes on: for $\gam > |h|$, both the parameters $\mathcal{S}_{y}^{(x)}(t)$ and $\mathcal{S}_{z}^{(x)}(t)$ exhibit squeezing 
effect and oscillatory behaviour due to the prevalence of quantum correlation effects (here introduced by the anisotropy parameter $\gam$) on the transverse 
magnetic field $h$; however, if one considers $\gam < |h|$, the squeezing effect is strongly reduced (in fact, it always survives for small values of time) in 
both the parameters, which reveals the relative strong influence of $h$ on the quantum correlation effects. With this in mind, let us now consider Fig. \ref{fig1}(b,d) 
where the plots of $\mathcal{E}_{y}(t)$ (dot-dashed line) and $\mathcal{E}_{z}(t)$ (solid line) are exhibited for the same values of $(h,\gam)$ used, respectively, 
in \ref{fig1}(a,c). The `almost perfect match' between squeezing and entanglement effects demonstrate the previous assertions and indeed reinforce the fundamental sequence
{\it correlation} $\mapsto$ {\it entanglement} $\mapsto$ {\it squeezing} of quantum effects.

%%%%%%%%%%%%%%%%%%%%%%%%%%%%%%%%%%%%%%%%%%%%%%%%%%%%%%%%%%%%%%%%%%%%%%%%%%%%%%%%%%%%%%%%%%%%%%%%%%%%%%%%%%%%%%%%%%%%%%%%%%%%%%%%%%%%%%%%%%%%%%%%%%%%%%%%%%
\bfig[!t]
\centering
\begin{minipage}[b]{0.45\linewidth}
\includegraphics[width=\textwidth]{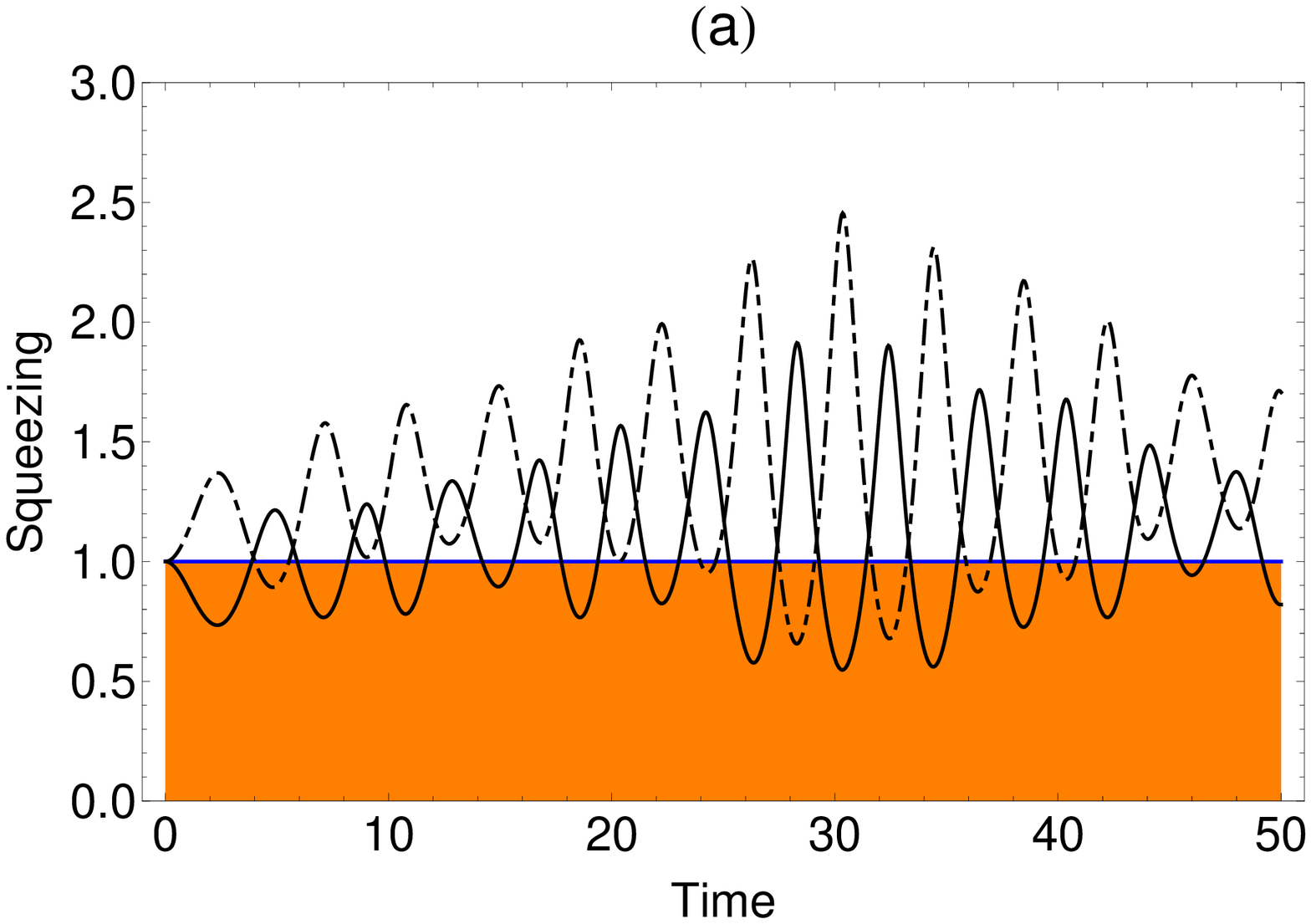}
\end{minipage} \hfill
\begin{minipage}[b]{0.45\linewidth}
\includegraphics[width=\textwidth]{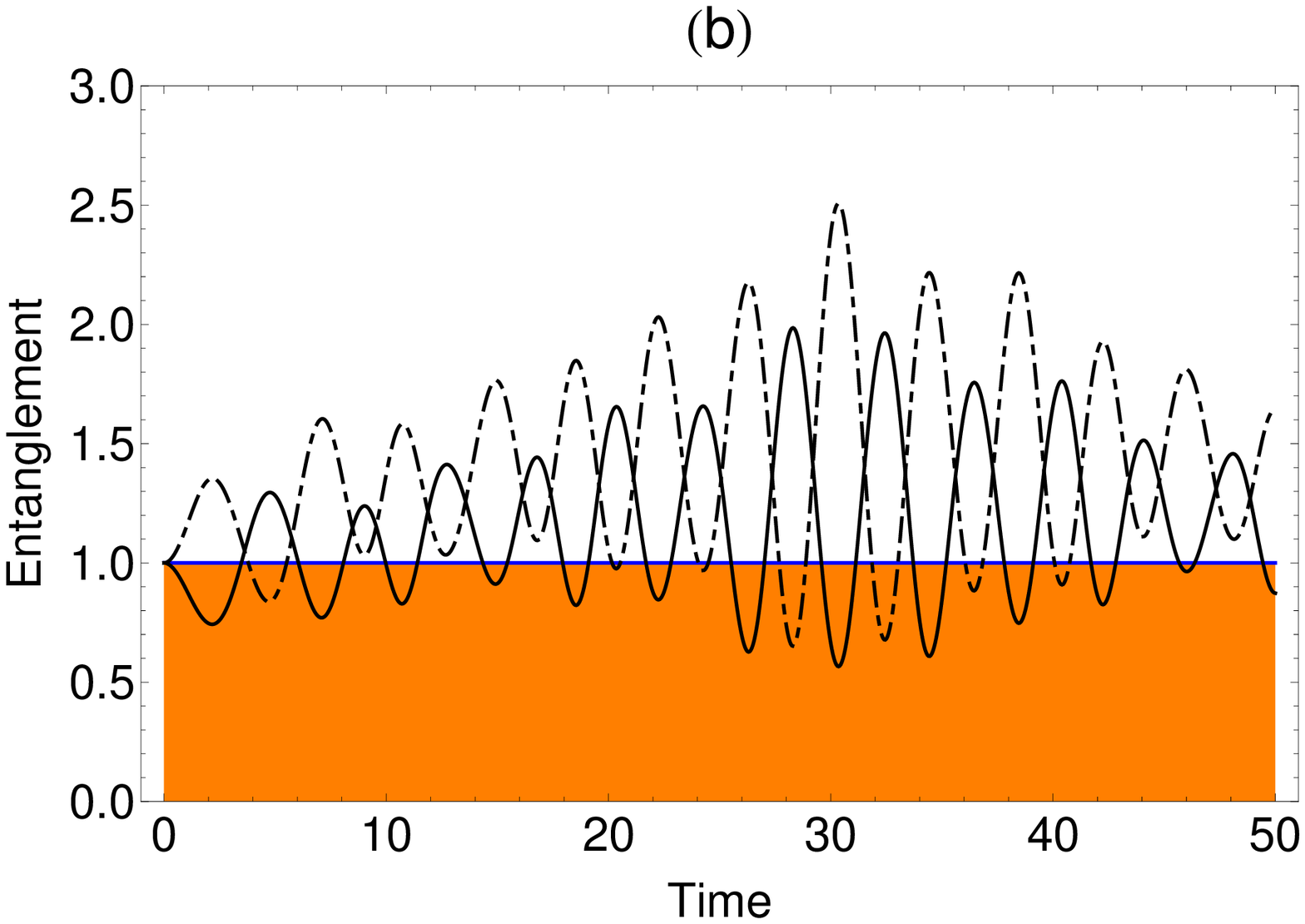}
\end{minipage} \hfill
\begin{minipage}[b]{0.45\linewidth}
\includegraphics[width=\textwidth]{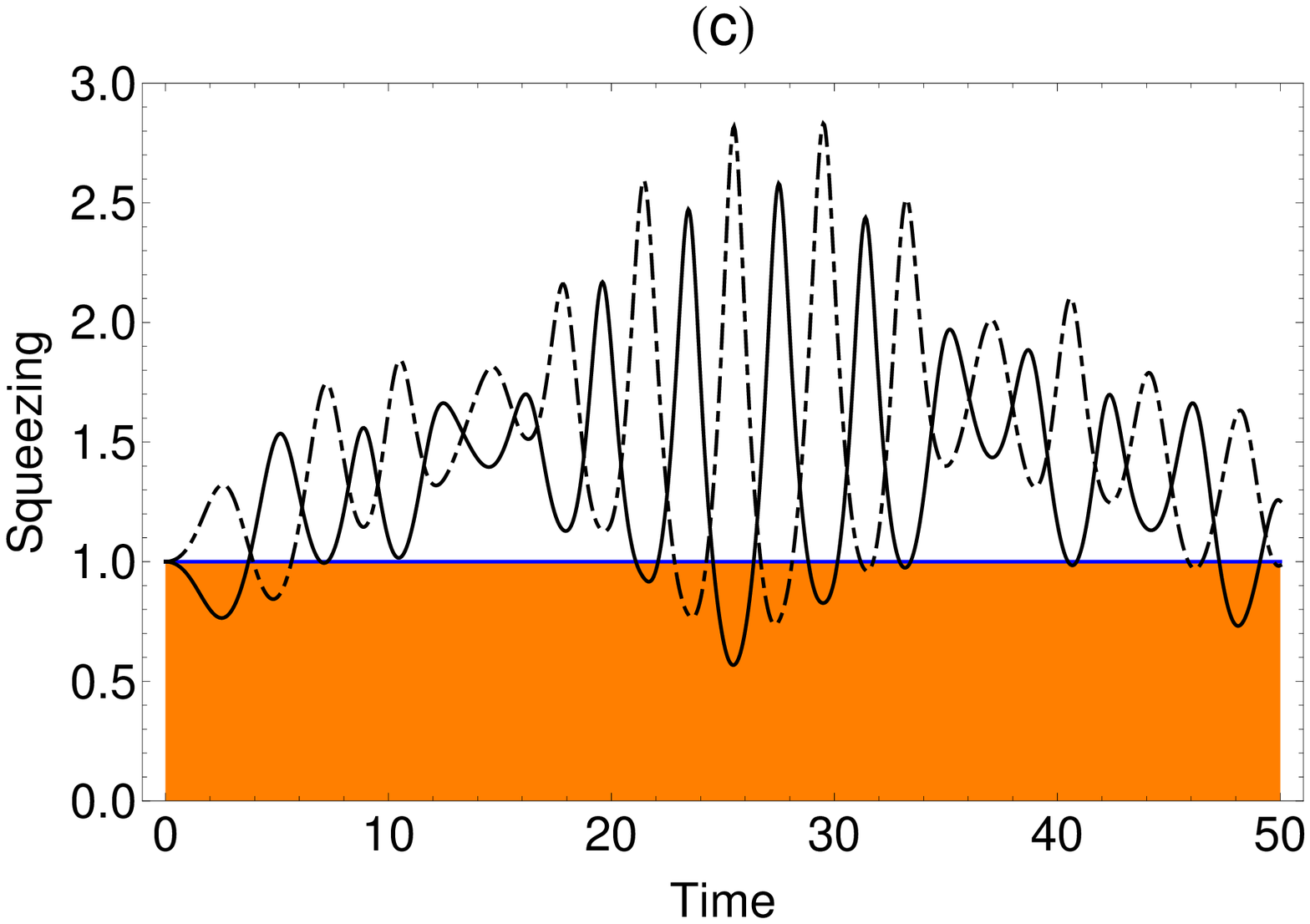}
\end{minipage} \hfill
\begin{minipage}[b]{0.45\linewidth}
\includegraphics[width=\textwidth]{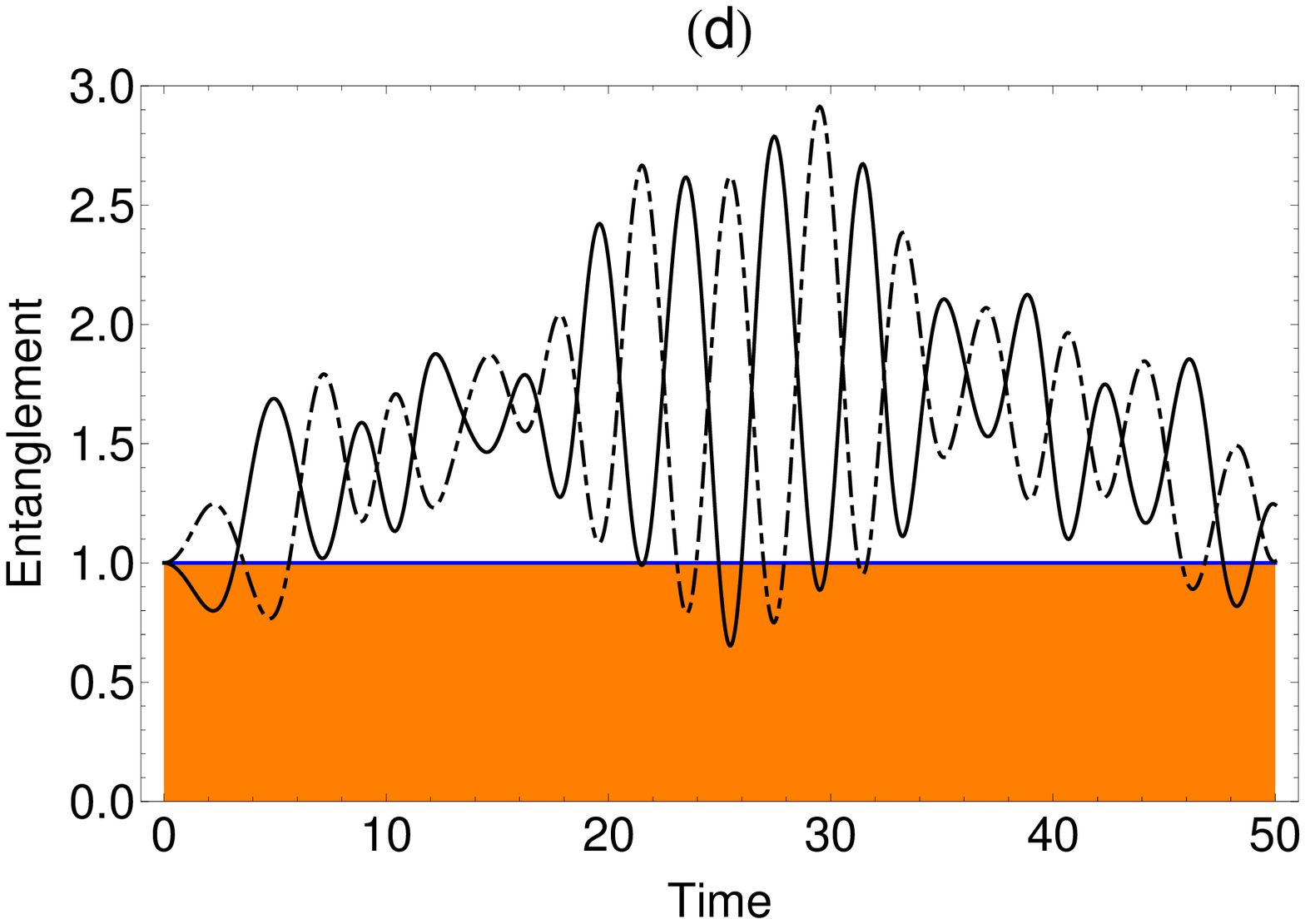}
\end{minipage}
%%%%%%%%%%%%%%%%%%%%%%%%%%%%%%%%%%%%%%%%%%%%%%%%%%%%%%%%%%%%%%%%%%%%%%%%%%%%%%%%%%%%%%%%%%%%%%%%%%%%%%%%%%%%%%%%%%%%%%%%%%%%%%%%%%%%%%%%%%%%%%%%%%%%%%%%%%
\caption{The `almost perfect match' between squeezing and entanglement effects is clearly visible in this selection of pictures, where the parameters (a,c) 
$\mathcal{S}_{y}^{(x)}(t)$ (dot-dashed line) and $\mathcal{S}_{z}^{(x)}(t)$ (solid line) as well as (b,d) $\mathcal{E}_{y}(t)$ (dot-dashed line) and
$\mathcal{E}_{z}(t)$ (solid line) are plotted as functions of time and placed side by side for immediate comparision. Furthermore, this set of plots corresponds
to two different values of $(h,\gam)$, namely, (a,b) $(-0.1,0.2)$ and (c,d) $(-0.13,0.1)$, which leads us to investigate how the quantum correlations are 
affected by the transverse magnetic field $h$ and anisotropy parameter $\gam$ via $(2j+1)^{2}$-dimensional discrete phase space inherent to the modified 
LMG model.}
\label{fig1} 
\efig
%%%%%%%%%%%%%%%%%%%%%%%%%%%%%%%%%%%%%%%%%%%%%%%%%%%%%%%%%%%%%%%%%%%%%%%%%%%%%%%%%%%%%%%%%%%%%%%%%%%%%%%%%%%%%%%%%%%%%%%%%%%%%%%%%%%%%%%%%%%%%%%%%%%%%%%%%%
At this moment, we pay special attention to the results obtained by T\'{o}th and co-workers\cite{SS1} where a full set of generalized spin-squeezing inequalities 
for entanglement detection was derived and studied in detail. For such task, let us formally establish an entanglement criterion for separable states through a 
set of inequalities involving the mean values $\lbr \lg {\bf J}_{a} \rg, \lg {\bf J}_{a}^{2} \rg \rbr_{a=x,y,z}$ that summarizes those aforementioned results.
\begin{description}
\item[Entanglement criterion.] Let $\ro$ describe an $N$-particle density operator, as well as $\lbr \lg {\bf J}_{x} \rg, \lg {\bf J}_{y} \rg, \lg {\bf J}_{z} \rg, 
\lg {\bf J}_{x}^{2} \rg, \lg {\bf J}_{y}^{2} \rg, \lg {\bf J}_{z}^{2} \rg \rbr$ characterize the mean values of collective angular-momentum operators associated with
the physical system of interest which, by hypothesis, are previously known. For separable states represented by the convex sum $\ro = \sum_{k} p_{k} \ro_{k}^{(1)} 
\otimes \ro_{k}^{(2)} \otimes \ldots \otimes \ro_{k}^{(N-1)} \otimes \ro_{k}^{(N)}$, where $p_{k}$ features the probability distribution for a given $k$, one verifies that
\brr
\lb{e19}
\lg {\bf J}_{x}^{2} \rg + \lg {\bf J}_{y}^{2} \rg + \lg {\bf J}_{z}^{2} \rg & \leq & \frac{1}{4} N(N+2) \\
\lb{e20}
\frac{1}{2} (N+2) \lpar \mathscr{V}_{\mathrm{J}_{x}} + \mathscr{V}_{\mathrm{J}_{y}} + \mathscr{V}_{\mathrm{J}_{z}} \rpar & \geq & \frac{1}{4} N(N+2) \\
\lb{e21}
\frac{1}{2} (N+2) \lbk \lg {\bf J}_{a}^{2} \rg + \lg {\bf J}_{b}^{2} \rg - (N-1) \mathscr{V}_{\mathrm{J}_{c}} \rbk & \leq & \frac{1}{4} N(N+2) \\
\lb{e22}
(N-1) \lpar \mathscr{V}_{\mathrm{J}_{a}} + \mathscr{V}_{\mathrm{J}_{b}} \rpar - \lg {\bf J}_{c}^{2} \rg + N & \geq & \frac{1}{4} N(N+2)
\err
are always fulfilled, $\frac{1}{4} N(N+2)$ being considered a lower/upper bound in all cases, with $\{ a,b,c \}$ labelling all the possible permutations of $\{ x,y,z \}$. 
Hence, the violation of any inequality (\ref{e20})-(\ref{e22}) leads, in this framework, to entangled states since Eq. (\ref{e19}) is valid for all quantum states --- 
indeed, such inequalities define a polytope in the three-dimensional $( \lg {\bf J}_{x}^{2} \rg, \lg {\bf J}_{y}^{2} \rg, \lg {\bf J}_{z}^{2} \rg )$-space, where the 
separable states lie inside this geometric representation. It is important to stress that physical systems with particle-number fluctuations (\ie, the particle number $N$ 
is not constant, such as in BEC experiments) are discarded in this criterion.\cite{SS2}
\end{description}

\begin{remark}
Let us now apply these inequalities to the modified LMG model with the main aim of corroborating the previous results and presenting alternative forms for
detection of spin-squeezing and entanglement effects. Numerical computations show that Eqs. (\ref{e21}) and (\ref{e22}) are not satisfied for determined
values of time, which leads us to choose one of them whose particular time evolution combines with those exhibited by $\mathcal{S}$ and $\mathcal{E}$-parameters
--- see Fig. \ref{fig1}(a)-(d). So, let us define the parameter
\be
\lb{e23}
E_{a} \coloneq \frac{(N-1) \mathscr{V}_{\mathrm{J}_{a}}}{\lg {\bf J}_{b}^{2} \rg + \lg {\bf J}_{c}^{2} \rg - \frac{N}{2}} \geq 1 \qquad (a,b,c=x,y,z)
\ee
based on Eq. (\ref{e21}), where the violation $E_{a} < 1$ directly implies in the quantum effects under investigation. Note that, compared with Eq. (\ref{b3}),
$E_{a}$ has subtle differences: it requires the sum of mean values related to the angular-momentum-squared operators (see denominator of both expressions), 
while the previous one involves only squared mean values of the angular-momentum operators. Figure \ref{fig2} shows the plots of $E_{y}(t)$ (dot-dashed
line) and $E_{z}(t)$ (solid line) versus $t \in [0,50]$ for the same set of values associated with the transverse magnetic field and anisotropy parameters
used in the previous figure, \ie, (a) $(-0.1,0.2)$ and (b) $(-0.13,0.1)$. It is important to stress that the functional characteristics of the involved 
parameters in the study of spin-squeezing and entanglement effects, as well as the associated entanglement dynamics to the modified LMG model, represent two 
essential factors that could explain (since the time evolution also depends on the initial state chosen for the physical system), in part, the similarities 
of behaviour related to the $\mathcal{S}$, $\mathcal{E}$, and $E$-parameters viewed in Figs. \ref{fig1} and \ref{fig2}. In summary, the entanglement criterion 
conceived by T\'{o}th {\it et al}.\cite{SS1} reinforces the `almost perfect match' between spin-squeezing and entanglement effects, which leads us to investigate, 
in this particular case, the quantum correlation rules via discrete Wigner and Husimi functions.\footnote{For completeness sake, we present a case study for 
$h=0$ (absence of transverse magnetic field) in appendix C, whose numerical computations lead us to discuss the real necessity of demanding the respective 
validity domains of the entanglement criteria used in this work.}
\end{remark}
%
%%%%%%%%%%%%%%%%%%%%%%%%%%%%%%%%%%%%%%%%%%%%%%%%%%%%%%%%%%%%%%%%%%%%%%%%%%%%%%%%%%%%%%%%%%%%%%%%%%%%%%%%%%%%%%%%%%%%%%%%%%%%%%%%%%%%%%%%%%%%%%%%%%%%%%%%
\bfig[!t]
\centering
\begin{minipage}[b]{0.45\linewidth}
\includegraphics[width=\textwidth]{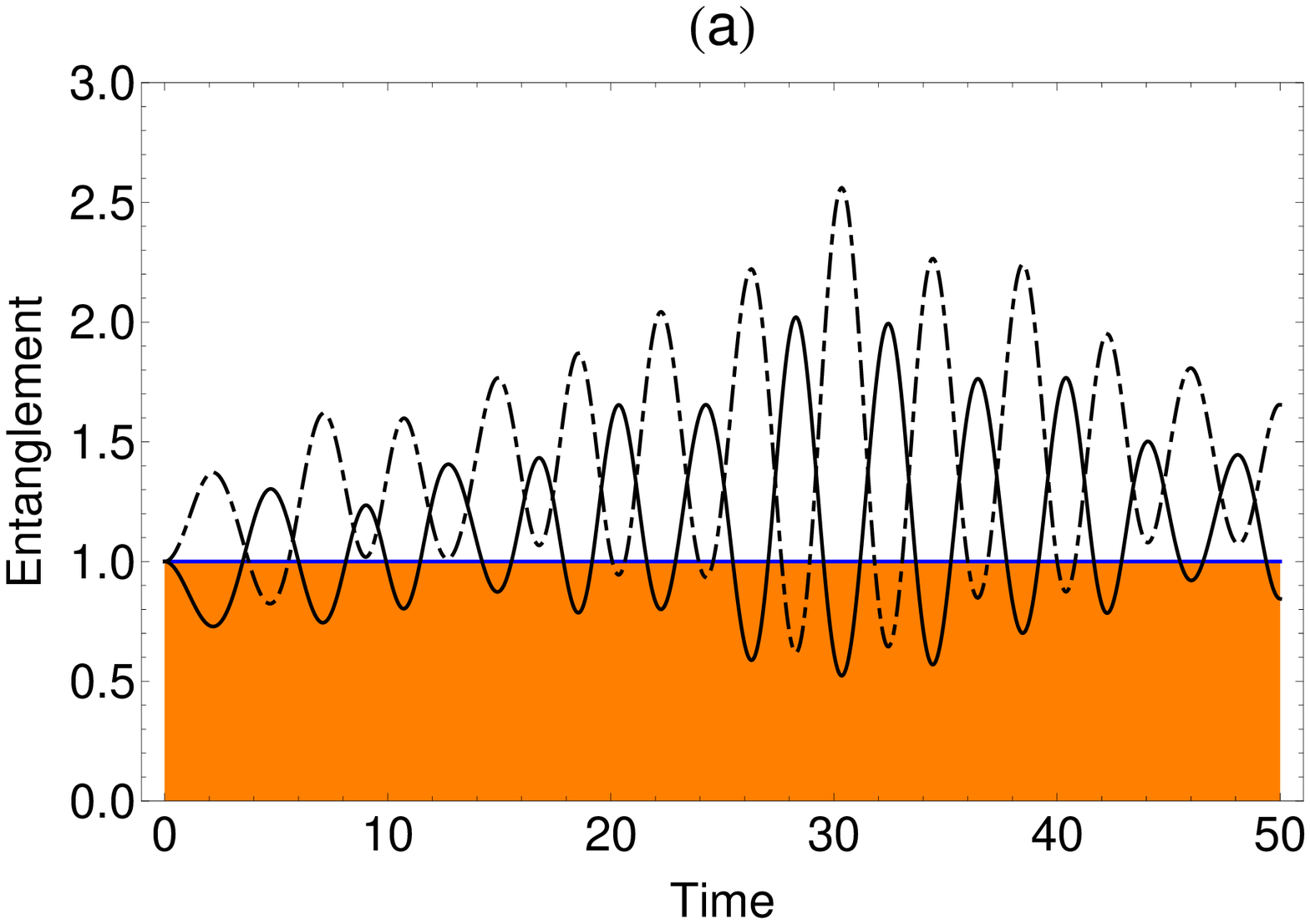}
\end{minipage} \hfill
\begin{minipage}[b]{0.45\linewidth}
\includegraphics[width=\textwidth]{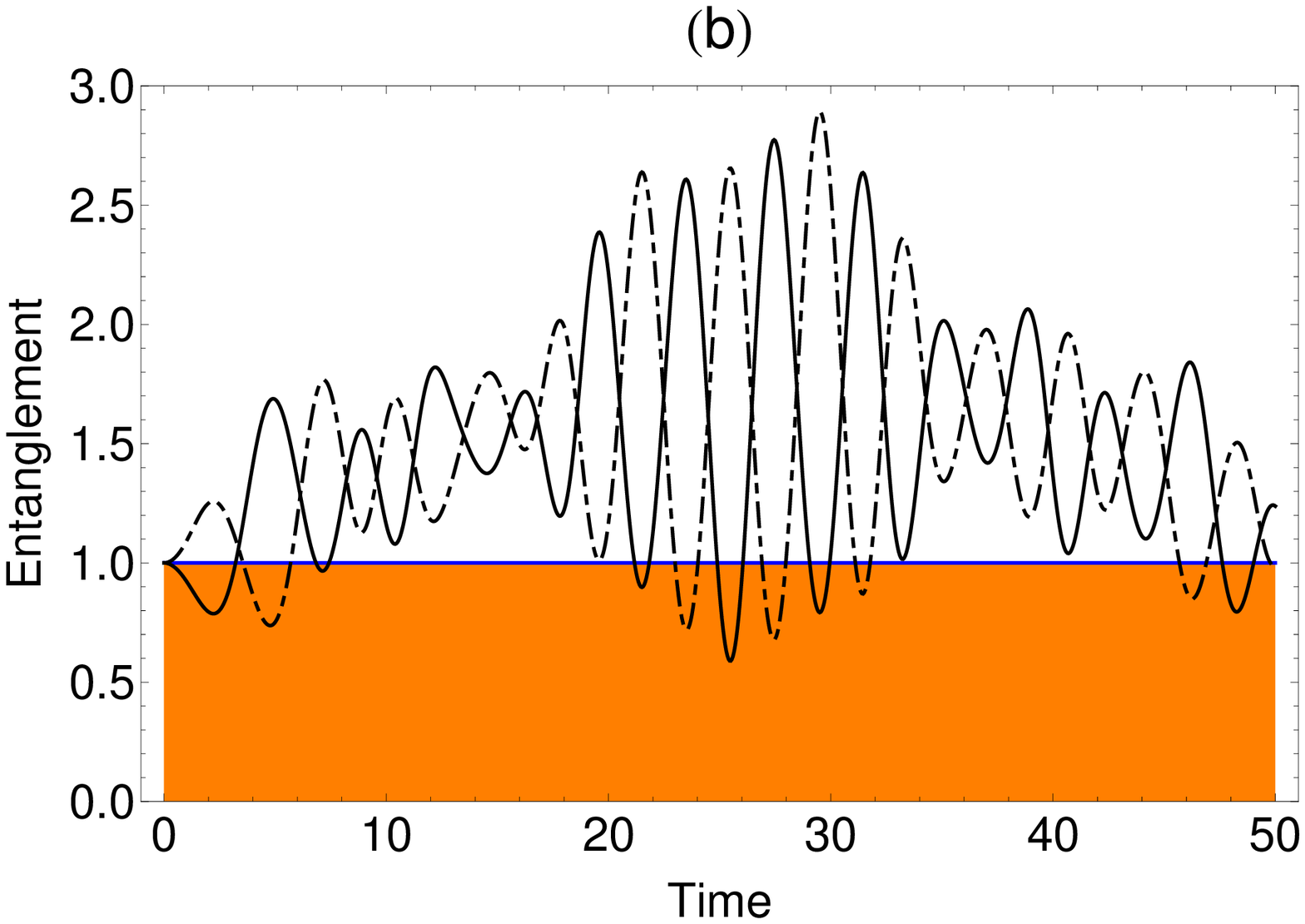}
\end{minipage}
%%%%%%%%%%%%%%%%%%%%%%%%%%%%%%%%%%%%%%%%%%%%%%%%%%%%%%%%%%%%%%%%%%%%%%%%%%%%%%%%%%%%%%%%%%%%%%%%%%%%%%%%%%%%%%%%%%%%%%%%%%%%%%%%%%%%%%%%%%%%%%%%%%%%%%%%
\caption{Plots of $E_{y}(t)$ (dot-dashed line) and $E_{z}(t)$ (solid line) as a function of time for the same values of $(h,\gam)$ used in the previous
figure, that is, (a) $(-0.1,0.2)$ and (b) $(-0.13,0.1)$. Here, two factors immediately emerge from our considerations on these pictures: (i) their 
similarities with the previous cases (emphasizing, mainly, the strong influence of $h$ on the entanglement effects), and consequently, (ii) the important 
role of quantum correlations in such measurements. It is worth mentioning that the particle number $N$ used in the numerical calculations remains the same
(\ie, $N=20$), which preserves important basic quantum effects (contrary to the thermodynamic limit).}
\label{fig2} 
\efig
%%%%%%%%%%%%%%%%%%%%%%%%%%%%%%%%%%%%%%%%%%%%%%%%%%%%%%%%%%%%%%%%%%%%%%%%%%%%%%%%%%%%%%%%%%%%%%%%%%%%%%%%%%%%%%%%%%%%%%%%%%%%%%%%%%%%%%%%%%%%%%%%%%%%%%%%

Henceforth, let us mention some few words on the entanglement dynamics of the Hamiltonian operator ${\bf H}^{\prime}$ (Schr\"{o}dinger picture): it basically
involves the action of collective spin operators responsible for introducing the quantum effects attributed to two-body correlations --- here mediated 
by the anisotropy parameter $\gam$ --- upon a nonseparable initial state, that is, the collective spin-coherent state (\ref{e5}) with $\th = \frac{\pi}{2}$
and $\varphi = 0$ fixed. In this sense, since the discrete Wigner function reflects the action of the time-evolution operator ${\bf U}(t) \coloneq \exp 
\lpar - \frac{\nc}{\hbar} {\bf H}^{\prime} t \rpar$ upon the initial state $| \frac{\pi}{2},0 \rg$ in a $(2j+1)^{2}$-dimensional discrete phase space,
it seems natural to investigate its behaviour for those specific values of time where effectively occur the spin-squeezing and entanglement effects.

%%%%%%%%%%%%%%%%%%%%%%%%%%%%%%%%%%%%%%%%%%%%%%%%%%%%%%%%%%%%%%%%%%%%%%%%%%%%%%%%%%%%%%%%%%%%%%%%%%%%%%%%%%%%%%%%%%%%%%%%%%%%%%%%%%%%%%%%%%%%%%%%%%%%%%%%%
\subsection{Finite-dimensional discrete phase spaces: a case study of quantum correlations via Wigner and Husimi functions}
%%%%%%%%%%%%%%%%%%%%%%%%%%%%%%%%%%%%%%%%%%%%%%%%%%%%%%%%%%%%%%%%%%%%%%%%%%%%%%%%%%%%%%%%%%%%%%%%%%%%%%%%%%%%%%%%%%%%%%%%%%%%%%%%%%%%%%%%%%%%%%%%%%%%%%%%

Initially, let us consider the time evolution of $\mathscr{W}_{\frac{\pi}{2},0}(\mu,\nu;t)$ upon a $21^{2}$-dimensional phase space labeled by the pair $(\mu,\nu)$ 
of dimensionless discrete angular-momentum and angle variables,\cite{GR1} where the particular operator ${\bf H}^{\prime} = - h {\bf J}_{z} - \frac{1}{N} 
\lpar {\bf J}_{x}^{2} + \gam {\bf J}_{y}^{2} \rpar$ will take place at this description. As usual, the parameters employed in the numerical computations 
are the same as those mentioned in figure \ref{fig1}(b), \ie, $N=20$ and $(h,\gam)=(-0.1,0.2)$. Furthermore, we choose six representative time values of 
$\mathcal{E}_{z}(t)$ which reflect the first local minimum and maximum points exhibited in figure \ref{fig1}(b). In this way, figure \ref{fig3} shows the 
contour plots of $\mathscr{W}_{\frac{\pi}{2},0}(\mu,\nu;t)$ versus $(\mu,\nu) \in [-10,10]$ where, in particular, (a) $t=0$ (initial Wigner function), 
(b) $t=2.15$ (first local minimum point), (c) $t=4.75$ (first local maximum point), (d) $t=7.10$ (second local minimum point), (e) $t=9.05$ (second local 
maximum point), and (f) $t=9.95$ (corresponds to $\mathcal{E}_{z}(9.95) \approx 1$). In all these pictures, it is interesting to observe how quantum 
correlations associated with two-body interaction term modify the initial correlations present in the collective spin-coherent state $| \frac{\pi}{2},0 \rg$;
besides, note that both $\mu$ and $\nu$ were conveniently shifted in the snapshots. Next, let us describe certain specific points inherent to the time evolution 
of the discrete Wigner function.
%%%%%%%%%%%%%%%%%%%%%%%%%%%%%%%%%%%%%%%%%%%%%%%%%%%%%%%%%%%%%%%%%%%%%%%%%%%%%%%%%%%%%%%%%%%%%%%%%%%%%%%%%%%%%%%%%%%%%%%%%%%%%%%%%%%%%%%%%%%%%%%%%%%%%%%%%%
\bfig[!t]
\centering
\begin{minipage}[b]{0.3\linewidth}
\includegraphics[width=\textwidth]{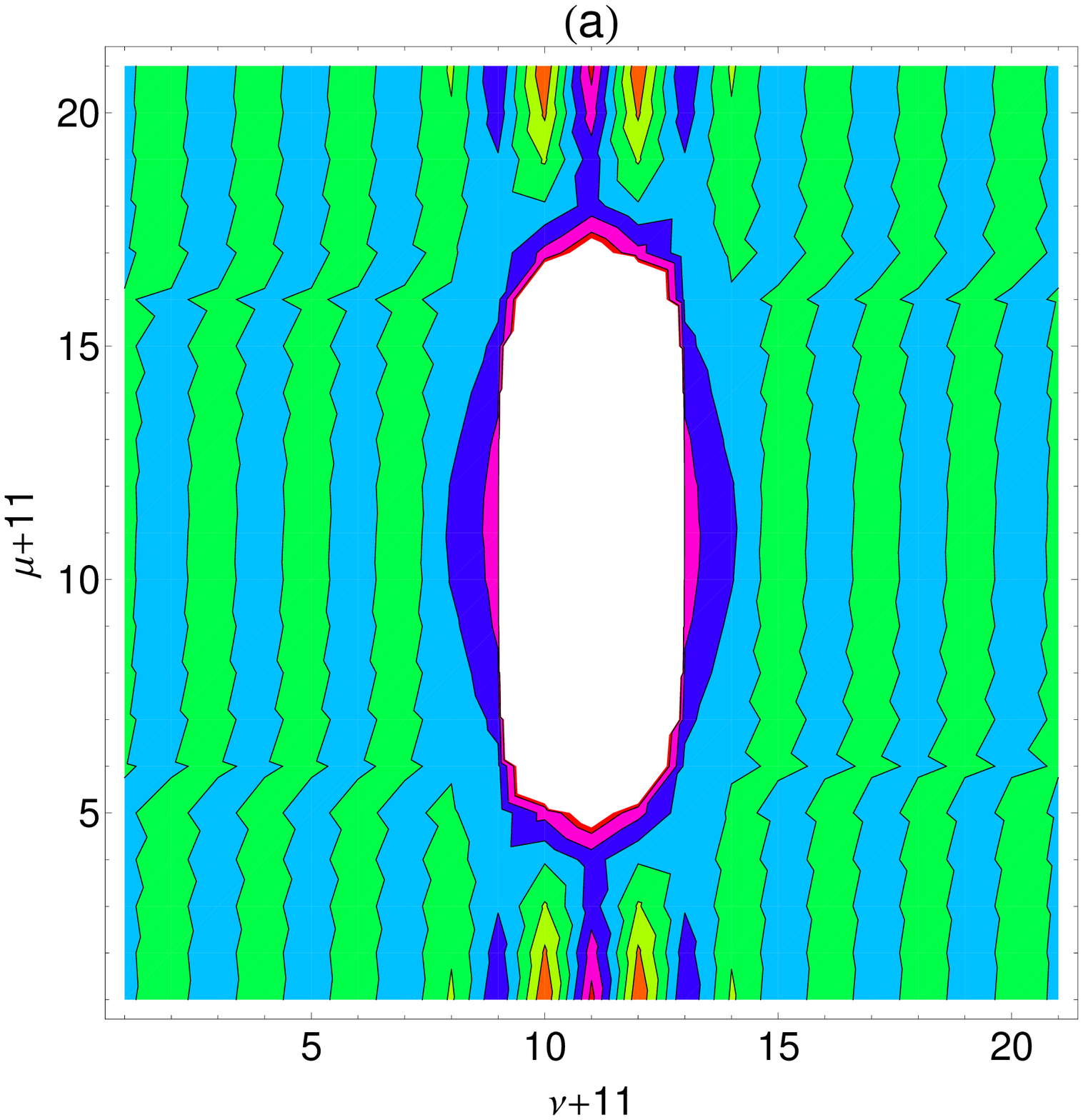}
\end{minipage} \hfill
\begin{minipage}[b]{0.3\linewidth}
\includegraphics[width=\textwidth]{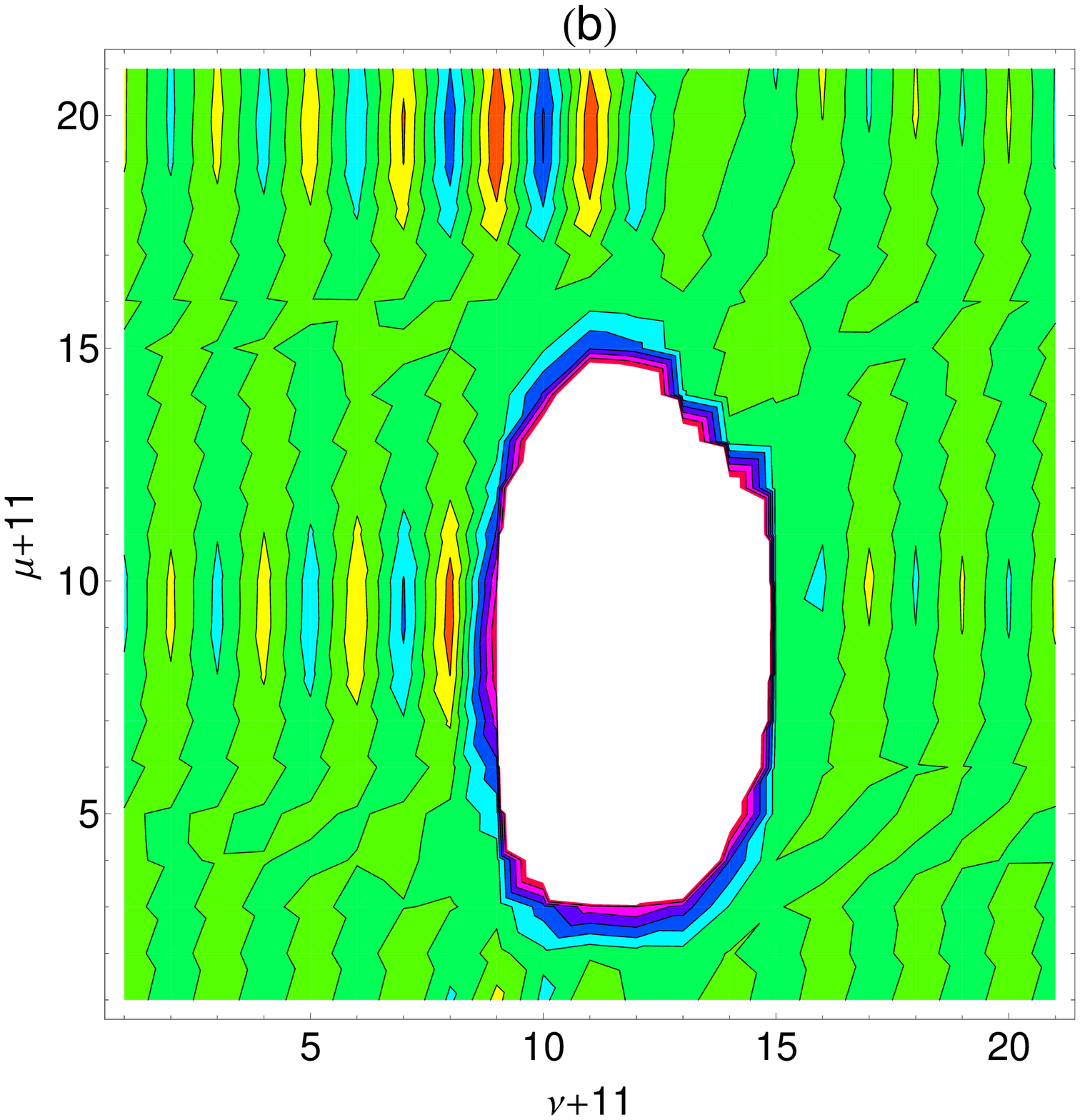}
\end{minipage} \hfill
\begin{minipage}[b]{0.3\linewidth}
\includegraphics[width=\textwidth]{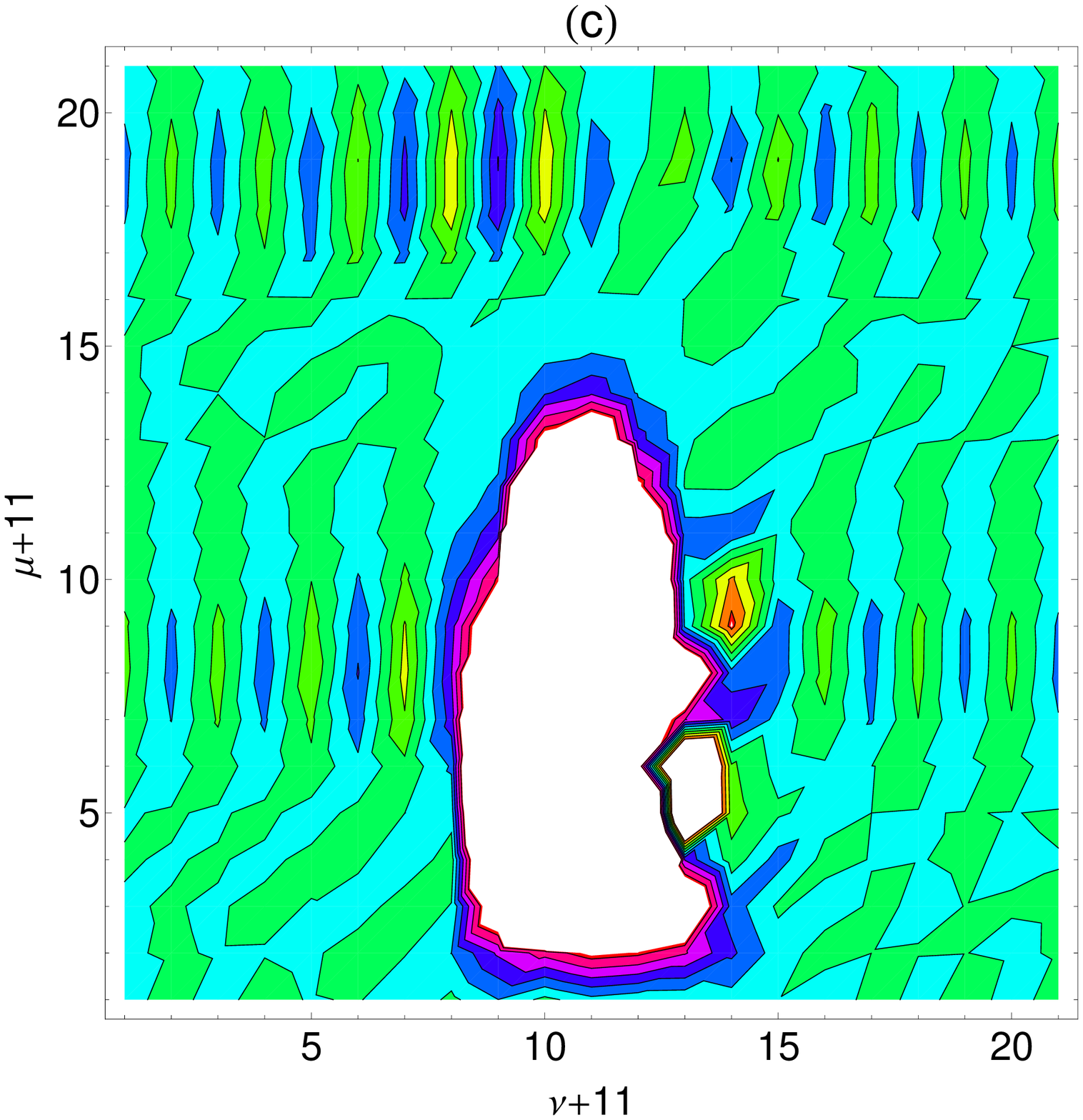}
\end{minipage} \hfill
\begin{minipage}[b]{0.3\linewidth}
\includegraphics[width=\textwidth]{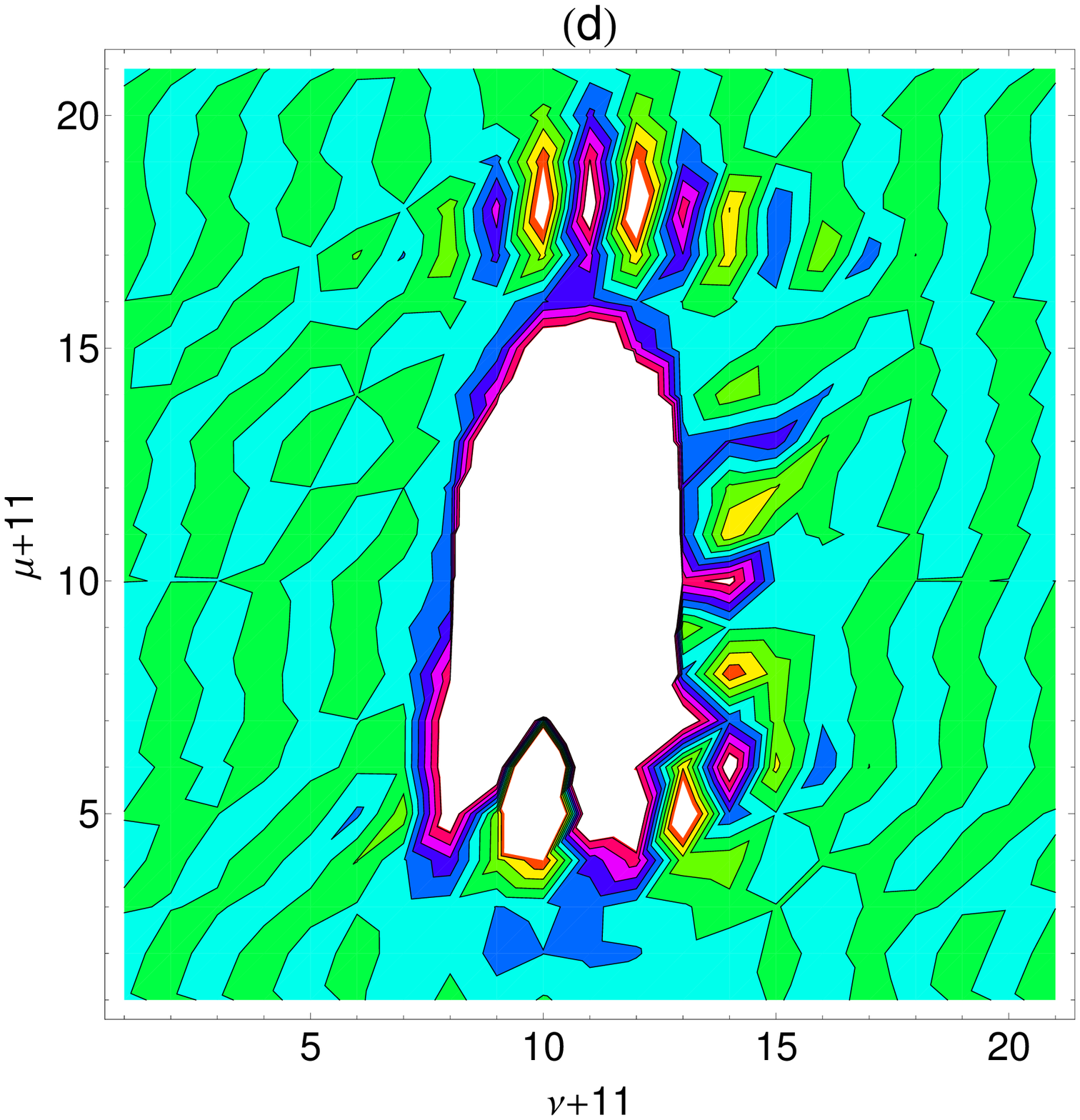}
\end{minipage} \hfill
\begin{minipage}[b]{0.3\linewidth}
\includegraphics[width=\textwidth]{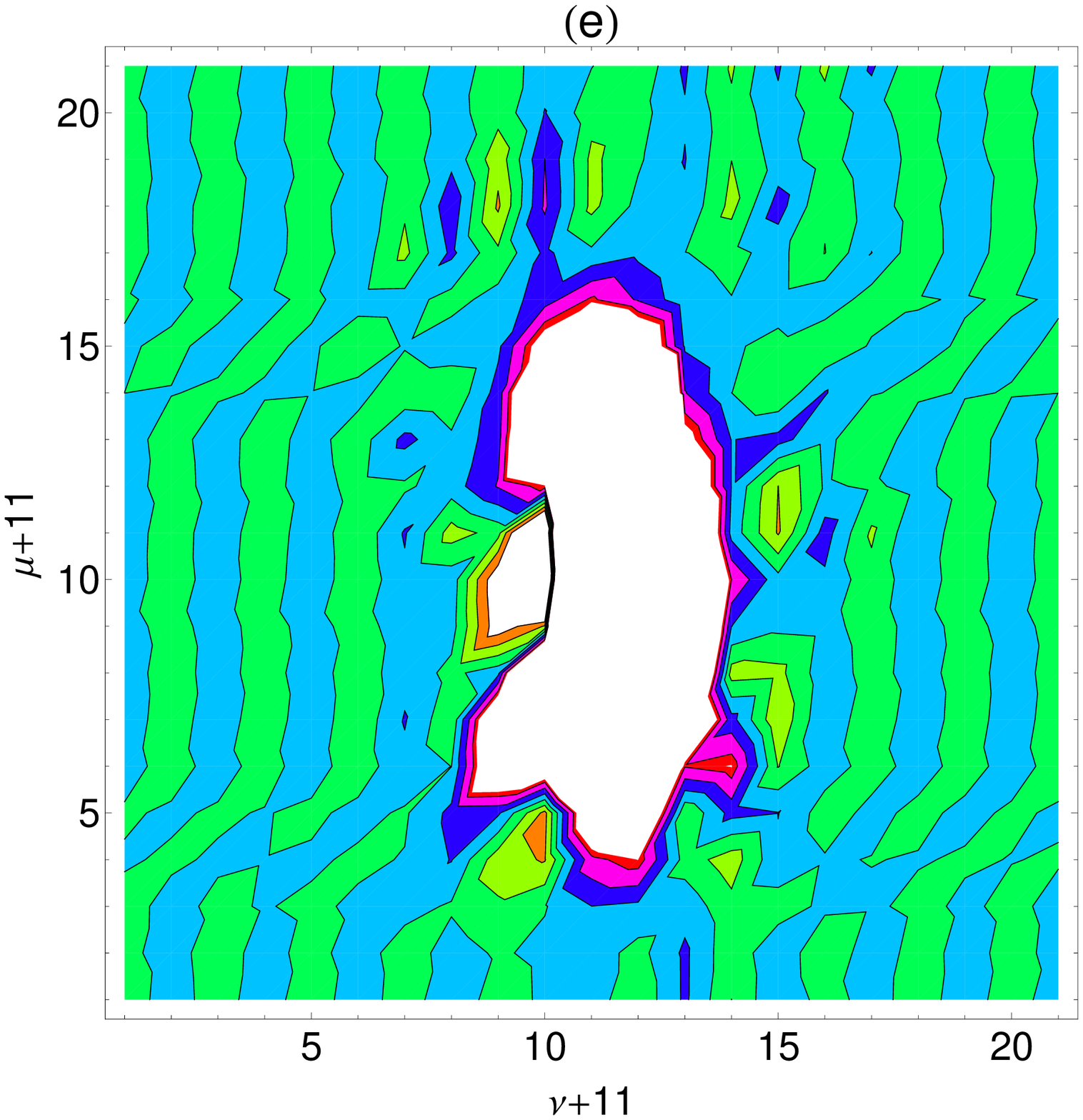}
\end{minipage} \hfill
\begin{minipage}[b]{0.3\linewidth}
\includegraphics[width=\textwidth]{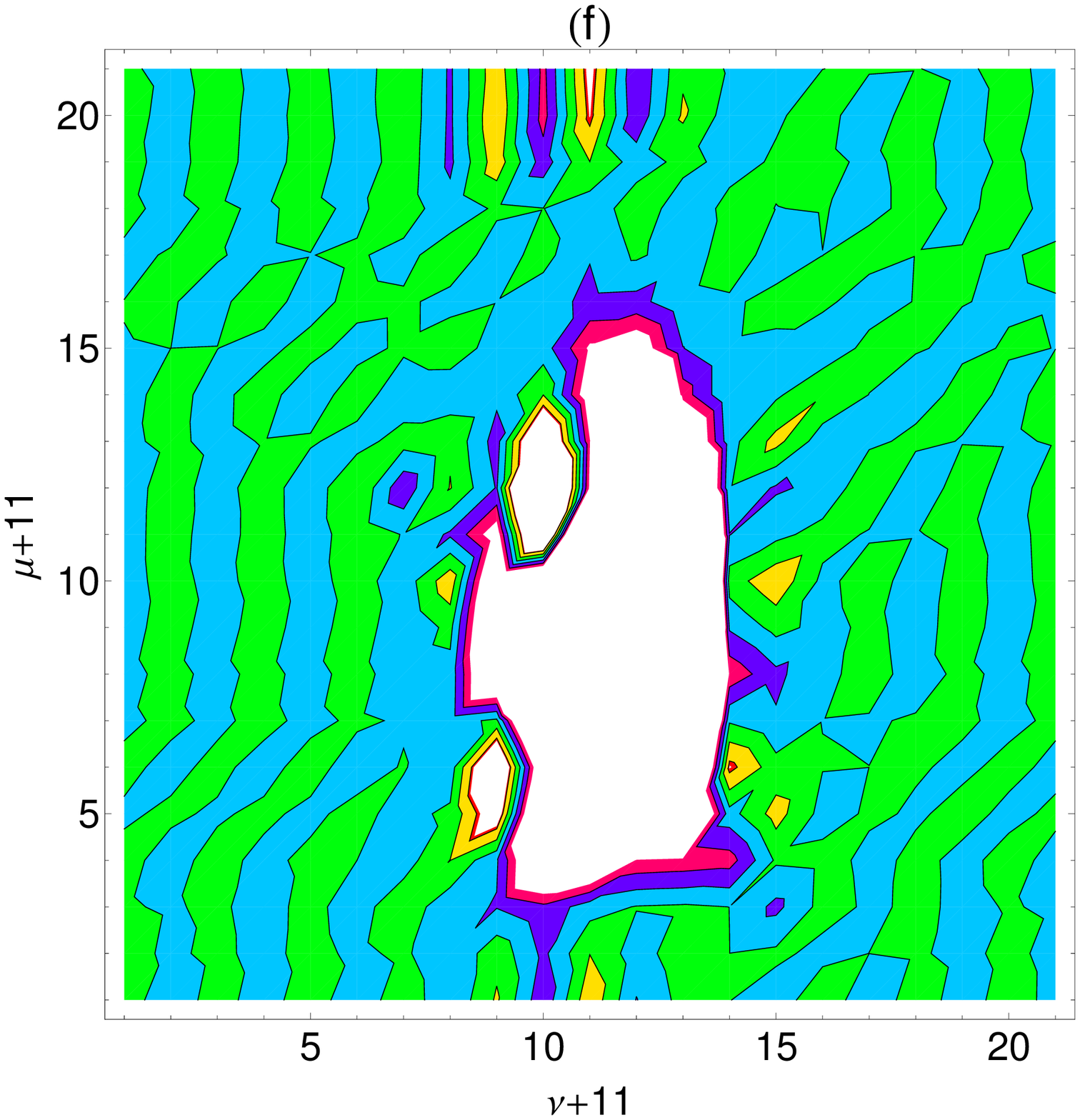}
\end{minipage}
%%%%%%%%%%%%%%%%%%%%%%%%%%%%%%%%%%%%%%%%%%%%%%%%%%%%%%%%%%%%%%%%%%%%%%%%%%%%%%%%%%%%%%%%%%%%%%%%%%%%%%%%%%%%%%%%%%%%%%%%%%%%%%%%%%%%%%%%%%%%%%%%%%%%%%%%%%
\caption{(Color online) Time evolution of $\mathscr{W}_{\frac{\pi}{2},0}(\mu,\nu;t)$ for the modified LMG model with $N=20$ and $(h,\gam)=(-0.1,0.2)$ fixed, 
where the labels $\mu$ and $\nu$ characterize, respectively, the dimensionless angular-momentum and angle pair. In particular, these pictures show how the
two-body interaction term present in the Hamiltonian operator ${\bf H}^{\prime}$ affects the initial Wigner distribution $\mathscr{W}_{\frac{\pi}{2},0}(\mu,\nu;0)$ 
(via contour plots) for determined values of time. We have adopted in our numerical computations the values (a) $t=0$, (b) $t=2.15$, (c) $t=4.75$, (d) $t=7.10$,
(e) $t=9.05$, and (f) $t=9.95$, which illustrate a representative but not complete evolution of the discrete Wigner function. Moreover, these values correspond
to the first local minimum and maximum points of $\mathcal{E}_{z}(t)$ --- see solid curve in figure \ref{fig1}(b) --- while (f) represents the situation 
$\mathcal{E}_{z}(9.95) \approx 1$. Although this phase space appears to the eyes as continuous (due to the large number of points associated with $N=20$), it is
important to stress that such a particular phase space is (by construction) genuinely discrete.}
\label{fig3} 
\efig
%%%%%%%%%%%%%%%%%%%%%%%%%%%%%%%%%%%%%%%%%%%%%%%%%%%%%%%%%%%%%%%%%%%%%%%%%%%%%%%%%%%%%%%%%%%%%%%%%%%%%%%%%%%%%%%%%%%%%%%%%%%%%%%%%%%%%%%%%%%%%%%%%%%%%%%%%%
%
\begin{itemize}
\item {\bf 4(a).} The function $\mathscr{W}_{\frac{\pi}{2},0}(\mu,\nu;0)$ exhibits a rotational symmetry with negative values located at the proximities
of $(-10,0)$ and $(10,0)$ (see small orange zones); as well as different widths in the respective $\mu$ and $\nu$-directions. This important fact suggests, 
if one considers both the directions, that marginal distributions related to the angular momentum and angle variables can be somehow relevant and necessary
in the analysis of quantum correlations. The approximate format of an ellipse --- white zone in the middle part of the contour plot --- represents a region 
where there exists raised probability peaks which feature, by their turn, the initial correlations associated with $| \frac{\pi}{2},0 \rg$.

\item {\bf 4(b).} This first case of time evolution corresponds to $\mathscr{W}_{\frac{\pi}{2},0}(\mu,\nu;t=2.15)$ and exemplifies the first local minimum
point of $\mathcal{E}_{z}(t)$ (see solid curve in figure \ref{fig1}(b) when $t=2.15$). The entanglement effects (associated with the two-body interaction, and
synonymous of quantum correlation) introduce additional symmetries that modify the statistical weights of the initial quantum state $| \frac{\pi}{2},0 \rg$, 
originating, in this way, new interference patterns which should explain the changed widths of the discrete Wigner distribution function in $t=2.15$, as well 
as the appearance of new zones where negative probabilities occur. Note that $\mathscr{W}_{\frac{\pi}{2},0}(\mu,\nu;t)$ exhibits a motion towards the frontier
of the discrete phase space labeled by $\mu=-10$, the transverse magnetic field $h$ being the agent responsible for such behaviour. Numerical computations
indeed corroborate this assertion and allow us to show that, for $h=0$, the Wigner function remains frozen in the discrete phase space, having its widths
modified as time goes by. 

\item {\bf 4(c).} This particular case depicts $\mathscr{W}_{\frac{\pi}{2},0}(\mu,\nu;t=4.75)$ and represents the first local maximum point of $\mathcal{E}_{z}(t)$. 
Here, $\mathscr{W}_{\frac{\pi}{2},0}(\mu,\nu;t)$ begins its motion towards the centre of such a discrete phase space, where the interference patterns associated
with the components of ${\bf U}(t=4.75) | \frac{\pi}{2},0 \rg$ have specifically yielded the portrait verified in this situation. The small white `island' located
in the right hand side of this picture and quite near to the raised probability peaks (main white zone) depicts a small region with negative probabilities.

\item {\bf 4(d).} The second minimum point of $\mathcal{E}_{z}(t)$ has as discrete phase-space representative the contour plot of $\mathscr{W}_{\frac{\pi}{2},0}
(\mu,\nu;t=7.10)$. These unique interference patterns (once ${\bf H}^{\prime}$ describes for $h \neq 0$ a nonperiodic system) with small `islands' of positive 
and negative probabilities represent the accumulated effects attributed to the transverse magnetic field and anisotropy parameter. Moreover, let us briefly 
mention that ${\bf U}(t=7.10) | \frac{\pi}{2},0 \rg$ has a coincidence probability (also known as fidelity and/or time correlation) with the original state close 
to $89\%$, while \ref{fig3}(b) results in $55\%$.

\item {\bf 4(e).} Note that $\mathscr{W}_{\frac{\pi}{2},0}(\mu,\nu;t=9.05)$ corresponds to the second local maximum point of $\mathcal{E}_{z}(t)$, \ie, we are
describing a situation where the entanglement does not occur. For the sake of comparison, the coincidence probabilities associated with \ref{fig3}(c) and \ref{fig3}(e) 
assume the respective percentages of $48\%$ and $78\%$. This difference can be justified, in principle, by means of a simple inspection between both the pictures: 
while \ref{fig3}(e) is located near the centre of discrete phase space, \ref{fig3}(c) stays in the proximity of $\mu=-10$. However, if one considers the previous 
case \ref{fig3}(d) for subsequent comparison, different contributions should be taken into account in the analysis.

\item {\bf 4(f).} This last case depicts $\mathscr{W}_{\frac{\pi}{2},0}(\mu,\nu;t=9.95)$ for $\mathcal{E}_{z}(9.95) \approx 1$. It is interesting to observe
how the nonperiodic dynamics related to the Hamiltonian operator ${\bf H}^{\prime}$ affects any reconstruction process of $\mathscr{W}_{\frac{\pi}{2},0}
(\mu,\nu;0)$: some numerical estimates of $\mathfrak{F}(t) \coloneq \tr [ \ro(0) \ro(t) ]$ for $t=9.95$ results in $51\%$ of time correlation.
\end{itemize}

Now, let us show an essential formal result that explores the connection between discrete Husimi and Wigner distribution functions,\cite{Haroche}
\be
\lb{e24}
\mathscr{H}_{\rho}(\mu,\nu;t) = \frac{1}{2j+1} \sum_{\mu^{\prime},\nu^{\prime} = -j}^{j} E(\mu,\nu | \mu^{\prime},\nu^{\prime}) \, 
\mathscr{W}_{\rho}(\mu^{\prime},\nu^{\prime};t) .
\ee
Here, $E(\mu,\nu | \mu^{\prime},\nu^{\prime})$ defines a smoothing process characterized by the discrete phase-space function\cite{Ruzzi}
\bd
E(\mu,\nu | \mu^{\prime},\nu^{\prime}) = \frac{1}{2j+1} \sum_{\eta,\xi=-j}^{j} \exp \lbr \frac{2 \pi \nc}{2j+1} \lbk \eta (\mu^{\prime} - \mu) + \xi
(\nu^{\prime} - \nu) \rbk \rbr \mathscr{K}(\eta,\xi) 
\ed
that closely resembles the role of a usual Gaussian function in the continuous phase space, since $\mathscr{K}(\eta,\xi) \coloneq \mathscr{M}(\eta,\xi) /
\mathscr{M}(0,0)$ is here responsible for the sum of products of Jacobi $\vartheta$-functions evaluated at integer arguments, namely,
\brr
\mathscr{M}(\eta,\xi) &=& \frac{\sqrt{\fa}}{2} \bigl\{ \vartheta_{3}(\fa \eta | \nc \fa) \lbk \vartheta_{3}(\fa \xi | \nc \fa) + \exp (\nc \pi \eta) 
\vartheta_{4}(\fa \xi | \nc \fa) \rbk \nn \\ 
& & + \exp (\nc \pi \xi) \vartheta_{4}(\fa \eta | \nc \fa) \lbk \vartheta_{3}(\fa \xi | \nc \fa) + \exp [ \nc \pi (\eta + 2j+1) ] 
\vartheta_{4}(\fa \xi | \nc \fa) \rbk \bigr\} \nn
\err
where $\fa = (4j+2)^{-1}$ --- in particular, see Appendix A of Ref. \refcite{MR} for definitions and technical details on Jacobi $\vartheta$-functions. 
So, in order to illustrate how the smoothing process acts on the summand of Eq. (\ref{e24}), let us consider $\mathscr{W}_{\frac{\pi}{2},0}(\mu,\nu;t)$
and also the same set of parameters used in the previous figure. In addition, it should be noticed that $\mathscr{H}_{\frac{\pi}{2},0}(\mu,\nu;t)$ is strictly 
positive as well as limited to the closed interval $[0,1]$ for any $t \geq 0$; consequently, all the negative values and non-regular patterns described in figure
\ref{fig3} will be washed out by $E(\mu,\nu | \mu^{\prime},\nu^{\prime})$. Figure \ref{fig4} exhibits the contour plots of $\mathscr{H}_{\frac{\pi}{2},0}(\mu,\nu;t)$,
where such effects are effectively checked along the pictures \ref{fig4}(a-f). In these cases, the width changes observed for the $(\mu,\nu)$-directions are related 
to the two-body interaction term (here mediated by means of $N$ and $\gam$), while the motion towards frontier located at $\mu=-10$ is associated with the 
transverse magnetic field $h$. Since the widths associated with the angular-momentum and angle distributions are also affected by those effects, it seems convenient
at this moment to employ the Wehrl-type entropy functionals\cite{Mizrahi} for understanding how the correlations between the discrete variables $\mu$ and 
$\nu$ of a finite-dimensional phase space change for $t \geq 0$.
%%%%%%%%%%%%%%%%%%%%%%%%%%%%%%%%%%%%%%%%%%%%%%%%%%%%%%%%%%%%%%%%%%%%%%%%%%%%%%%%%%%%%%%%%%%%%%%%%%%%%%%%%%%%%%%%%%%%%%%%%%%%%%%%%%%%%%%%%%%%%%%%%%%%%%%%%%
\bfig[!t]
\centering
\begin{minipage}[b]{0.3\linewidth}
\includegraphics[width=\textwidth]{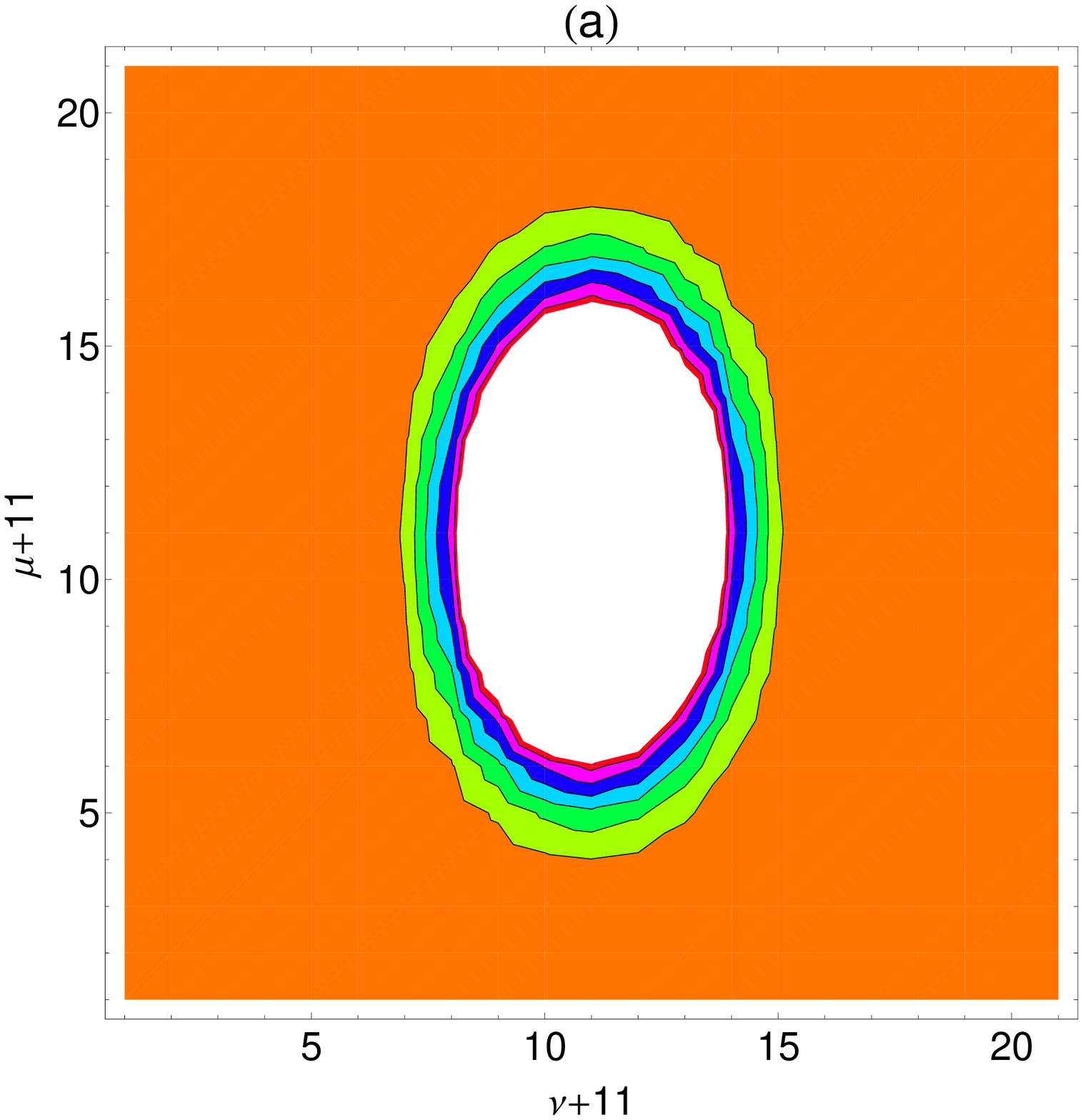}
\end{minipage} \hfill
\begin{minipage}[b]{0.3\linewidth}
\includegraphics[width=\textwidth]{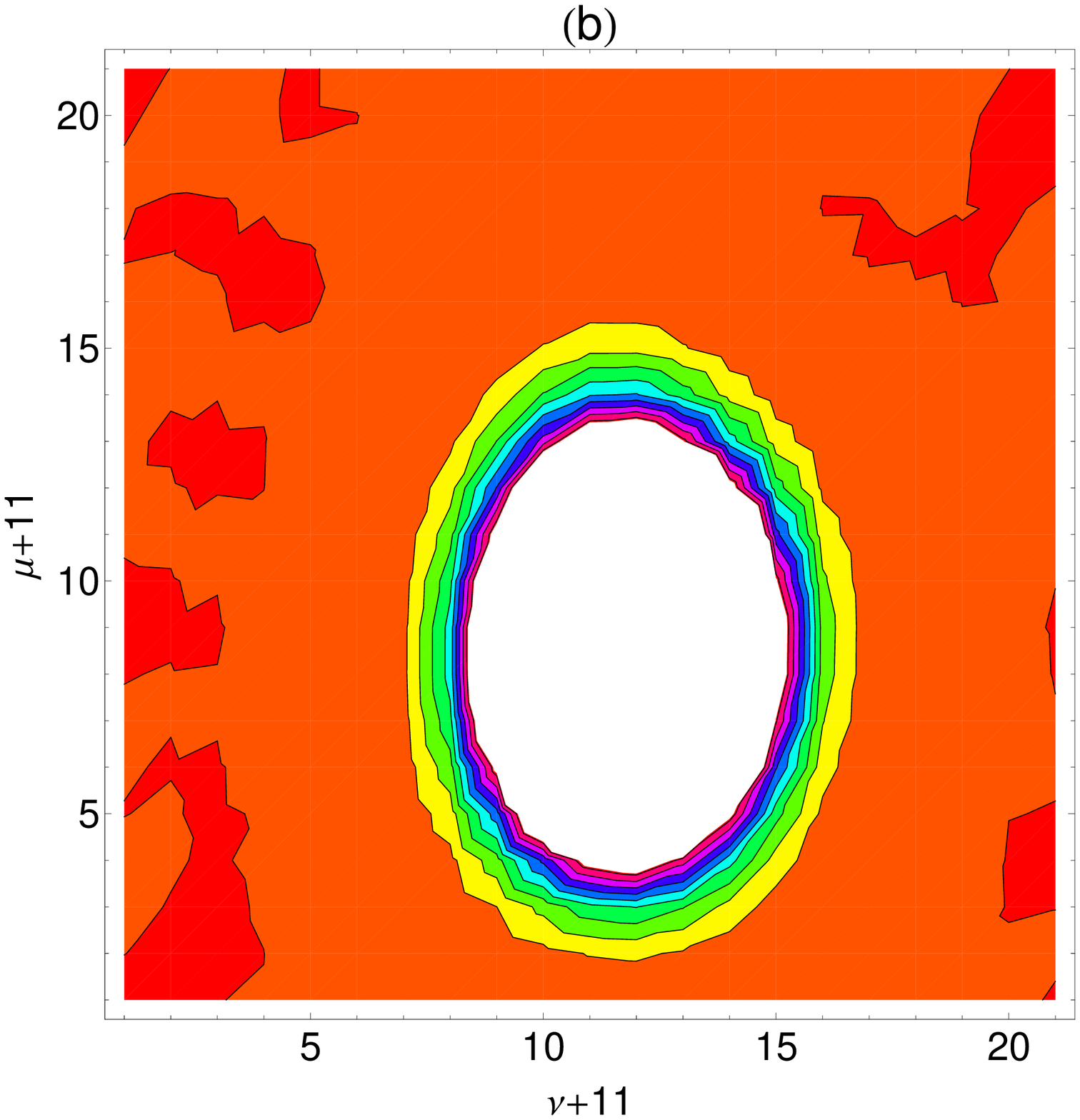}
\end{minipage} \hfill
\begin{minipage}[b]{0.3\linewidth}
\includegraphics[width=\textwidth]{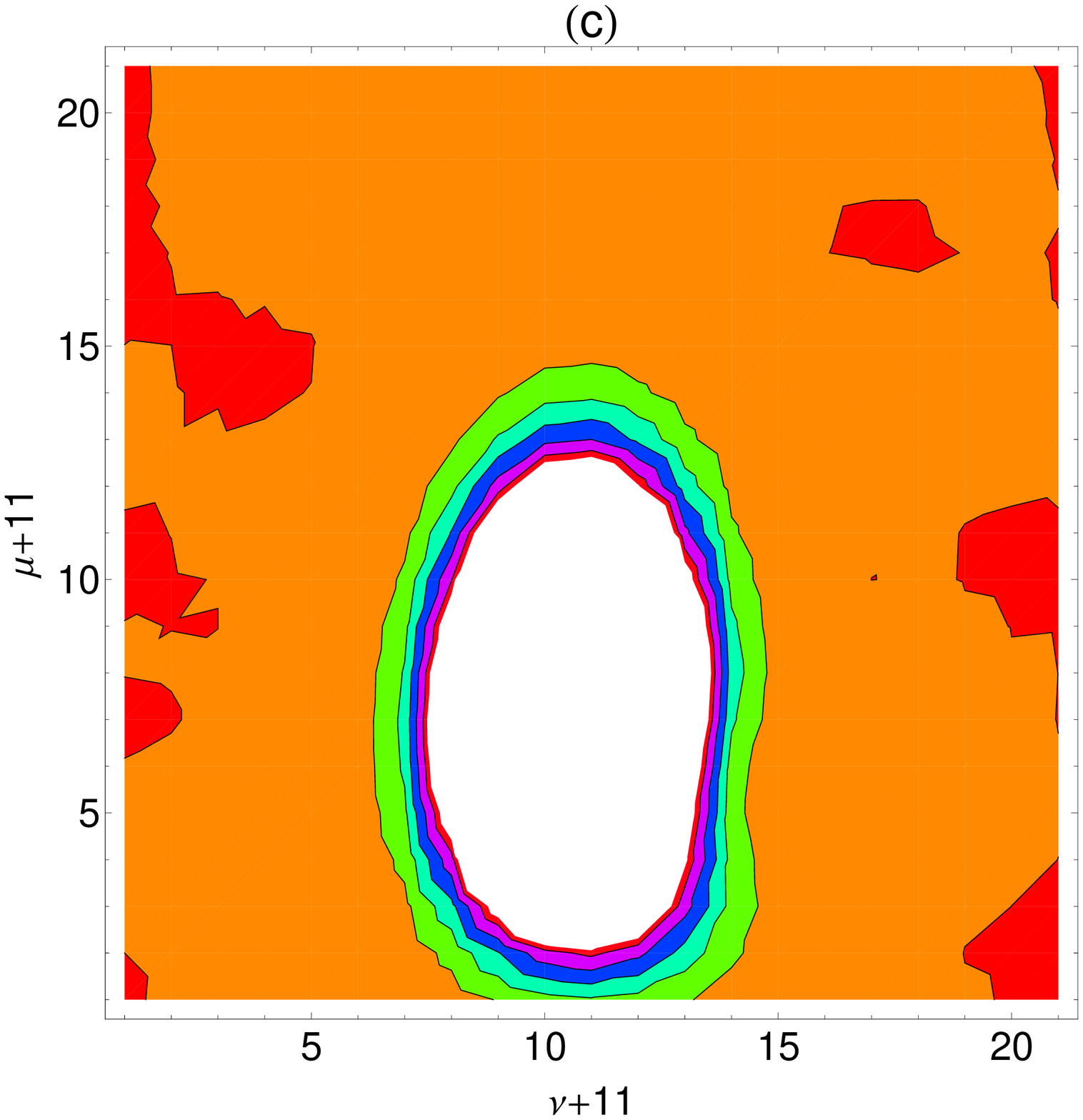}
\end{minipage} \hfill
\begin{minipage}[b]{0.3\linewidth}
\includegraphics[width=\textwidth]{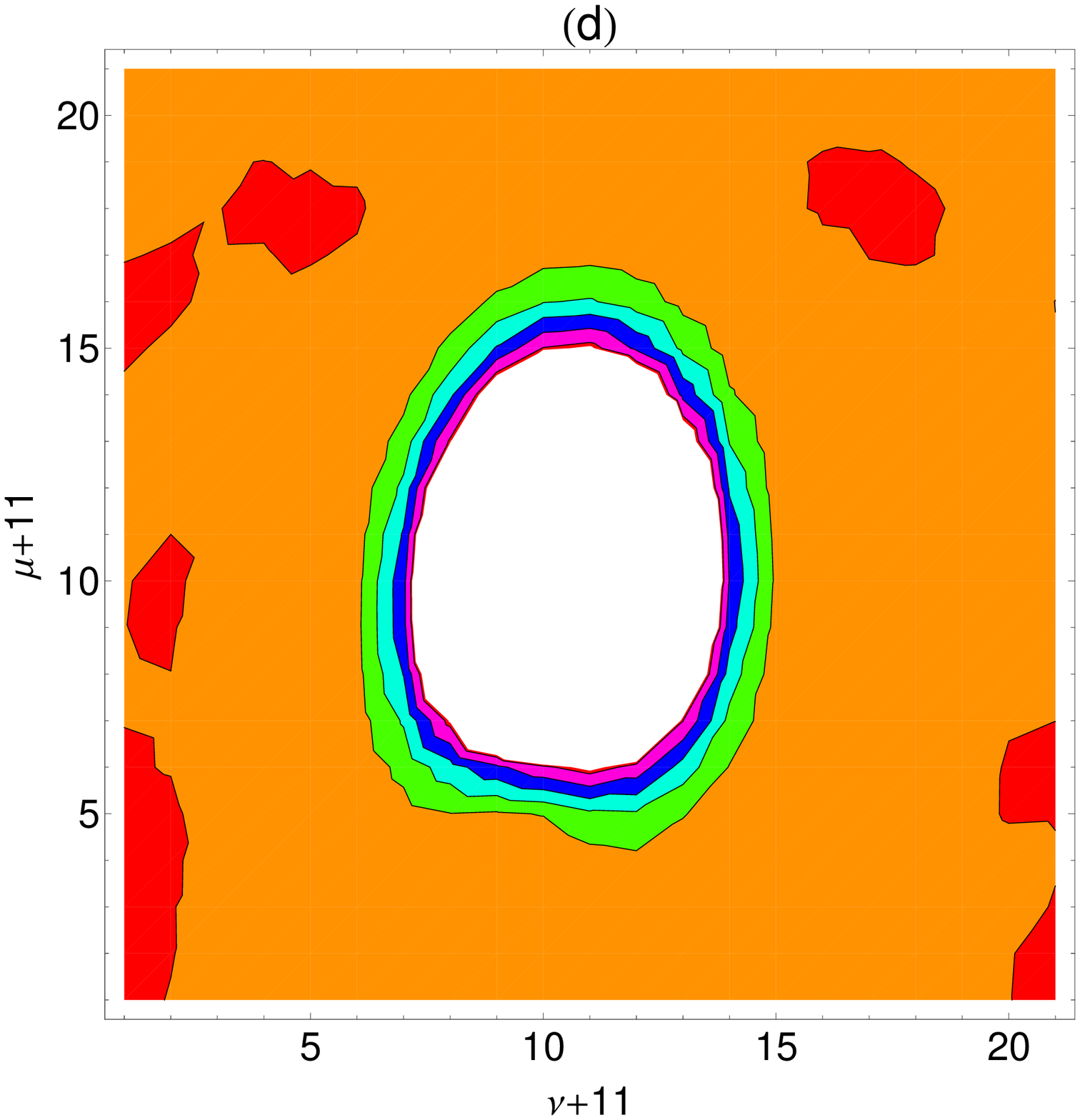}
\end{minipage} \hfill
\begin{minipage}[b]{0.3\linewidth}
\includegraphics[width=\textwidth]{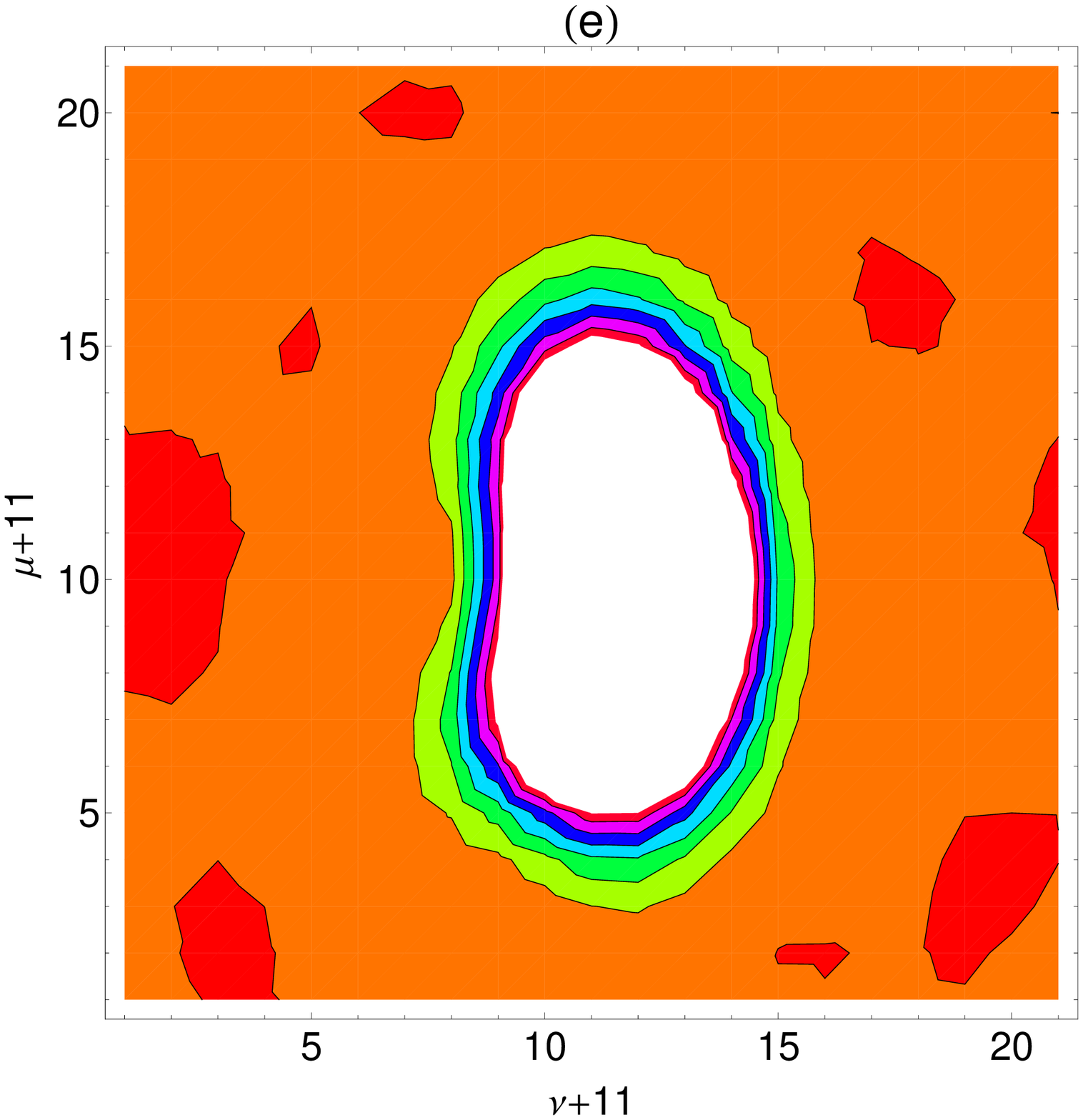}
\end{minipage} \hfill
\begin{minipage}[b]{0.3\linewidth}
\includegraphics[width=\textwidth]{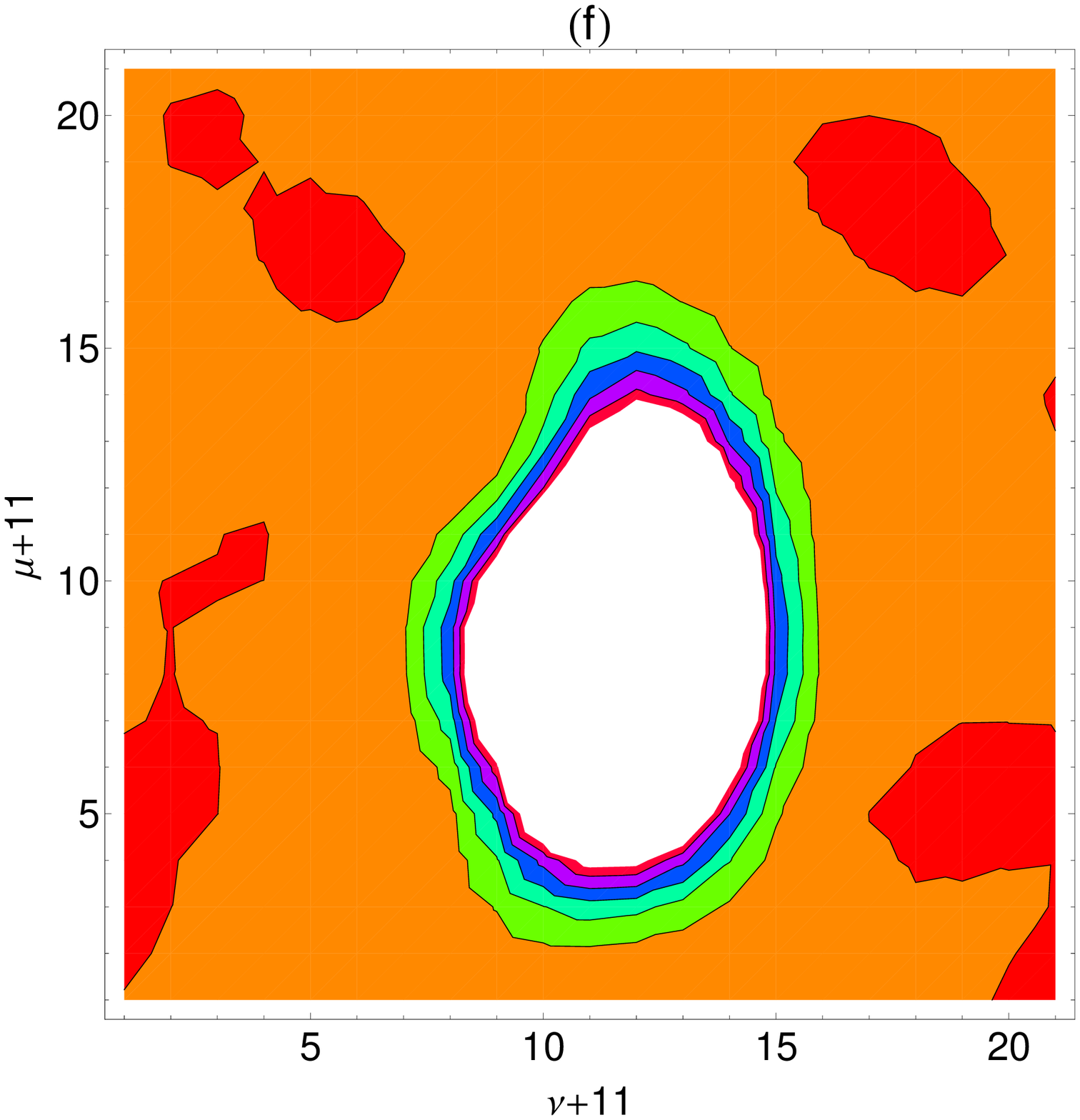}
\end{minipage}
%%%%%%%%%%%%%%%%%%%%%%%%%%%%%%%%%%%%%%%%%%%%%%%%%%%%%%%%%%%%%%%%%%%%%%%%%%%%%%%%%%%%%%%%%%%%%%%%%%%%%%%%%%%%%%%%%%%%%%%%%%%%%%%%%%%%%%%%%%%%%%%%%%%%%%%%%%
\caption{(Color online) Time evolution of $\mathscr{H}_{\frac{\pi}{2},0}(\mu,\nu;t)$ for the same set of parameters employed in the previous figure. In what
concerns the discrete Husimi function, it must be stressed that such a function is particularly obtained from $\mathscr{W}_{\frac{\pi}{2},0}(\mu,\nu;t)$ 
through a smoothing process characterized by a discrete phase-space function that closely resembles the role of the Gaussian function in the continuous 
phase space. This process essentially explains the disappearance of all irregular regions and negative values exhibited by the discrete Wigner distribution, 
which implies in a well-behaved function along its time evolution and limited to the closed interval $[0,1]$. Similarly to the previous case, it is worth
mentioning that such a particular phase space is genuinely discrete (by {\it ab initio} construction), even though it appears to the eyes as contionuous.}
\label{fig4} 
\efig
%%%%%%%%%%%%%%%%%%%%%%%%%%%%%%%%%%%%%%%%%%%%%%%%%%%%%%%%%%%%%%%%%%%%%%%%%%%%%%%%%%%%%%%%%%%%%%%%%%%%%%%%%%%%%%%%%%%%%%%%%%%%%%%%%%%%%%%%%%%%%%%%%%%%%%%%%%

To this end, let us now consider the mutual correlation functional $\mathrm{I}[\mathrm{H};t]$ defined as follows:\cite{MRG} $\mathrm{I}[\mathrm{H};t] 
\coloneq \mathrm{E}[\mathrm{Q};t] + \mathrm{E}[\mathrm{R};t] - \mathrm{E}[\mathrm{H};t] \geq 0$. In this particular definition,
\be
\lb{e25}
\mathrm{E}[\mathrm{H};t] \equiv - \frac{1}{2j+1} \sum_{\mu,\nu = -j}^{j} \mathscr{H}_{\rho}(\mu,\nu;t) \ln \lbk \mathscr{H}_{\rho}(\mu,\nu;t) \rbk
\ee
corresponds to the time-dependent joint entropy functional here expressed in terms of the discrete Husimi function (\ref{e24}), while
\be
\lb{e26}
\mathrm{E}[\mathrm{Q};t] \equiv - \frac{1}{\sqrt{2j+1}} \sum_{\mu = -j}^{j} \mathscr{Q}_{\rho}(\mu;t) \ln \lbk \mathscr{Q}_{\rho}(\mu;t) \rbk 
\ee
and
\be
\lb{e27}
\mathrm{E}[\mathrm{R};t] \equiv - \frac{1}{\sqrt{2j+1}} \sum_{\nu = -j}^{j} \mathscr{R}_{\rho}(\nu;t) \ln \lbk \mathscr{R}_{\rho}(\nu;t) \rbk
\ee
represent the marginal entropies --- which are directly related, by their turn, to the respective marginal distribution functions (for technical details, 
see Ref. \refcite{Ruzzi})
\bd
\mathscr{Q}_{\rho}(\mu;t) = \frac{1}{\sqrt{2j+1}} \sum_{\nu = -j}^{j} \mathscr{H}_{\rho}(\mu,\nu;t) \quad \mbox{and} \quad
\mathscr{R}_{\rho}(\nu;t) = \frac{1}{\sqrt{2j+1}} \sum_{\mu = -j}^{j} \mathscr{H}_{\rho}(\mu,\nu;t) .
\ed
Note that $\mathrm{E}[\mathrm{H};t]$, $\mathrm{E}[\mathrm{Q};t]$ and $\mathrm{E}[\mathrm{R};t]$ constitute some basic mathematical elements for 
describing functional correlations between $\mu$ and $\nu$, and consequently, the width changes associated with the discrete Husimi function
$\mathscr{H}_{\rho}(\mu,\nu;t)$. Within several important properties inherent to the entropy functionals, the Araki-Lieb inequality\cite{Lieb}
$\left| \mathrm{E}[\mathrm{Q};t] - \mathrm{E}[\mathrm{R};t] \right| \leq \mathrm{E}[\mathrm{H};t] \leq \mathrm{E}[\mathrm{Q};t] + \mathrm{E}[\mathrm{R};t]$
has a central role in such a description: for instance, it leads us to investigate how the dynamic correlations (introduced by means of the Hamiltonian
operator ${\bf H}^{\prime}$) affect the underlying correlations of the initial state $\ro(t_{0})$ --- here mapped onto $\mathscr{H}_{\rho}(\mu,\nu;t_{0})$.
So, if one computes the entropy functionals from $\mathscr{H}_{\frac{\pi}{2},0}(\mu,\nu;t)$, some interesting results can be promptly achieved. 

%%%%%%%%%%%%%%%%%%%%%%%%%%%%%%%%%%%%%%%%%%%%%%%%%%%%%%%%%%%%%%%%%%%%%%%%%%%%%%%%%%%%%%%%%%%%%%%%%%%%%%%%%%%%%%%%%%%%%%%%%%%%%%%%%%%%%%%%%%%%%%%%%%%%%%%%
\bfig[!t]
\centering
\begin{minipage}[b]{0.45\linewidth}
\includegraphics[width=\textwidth]{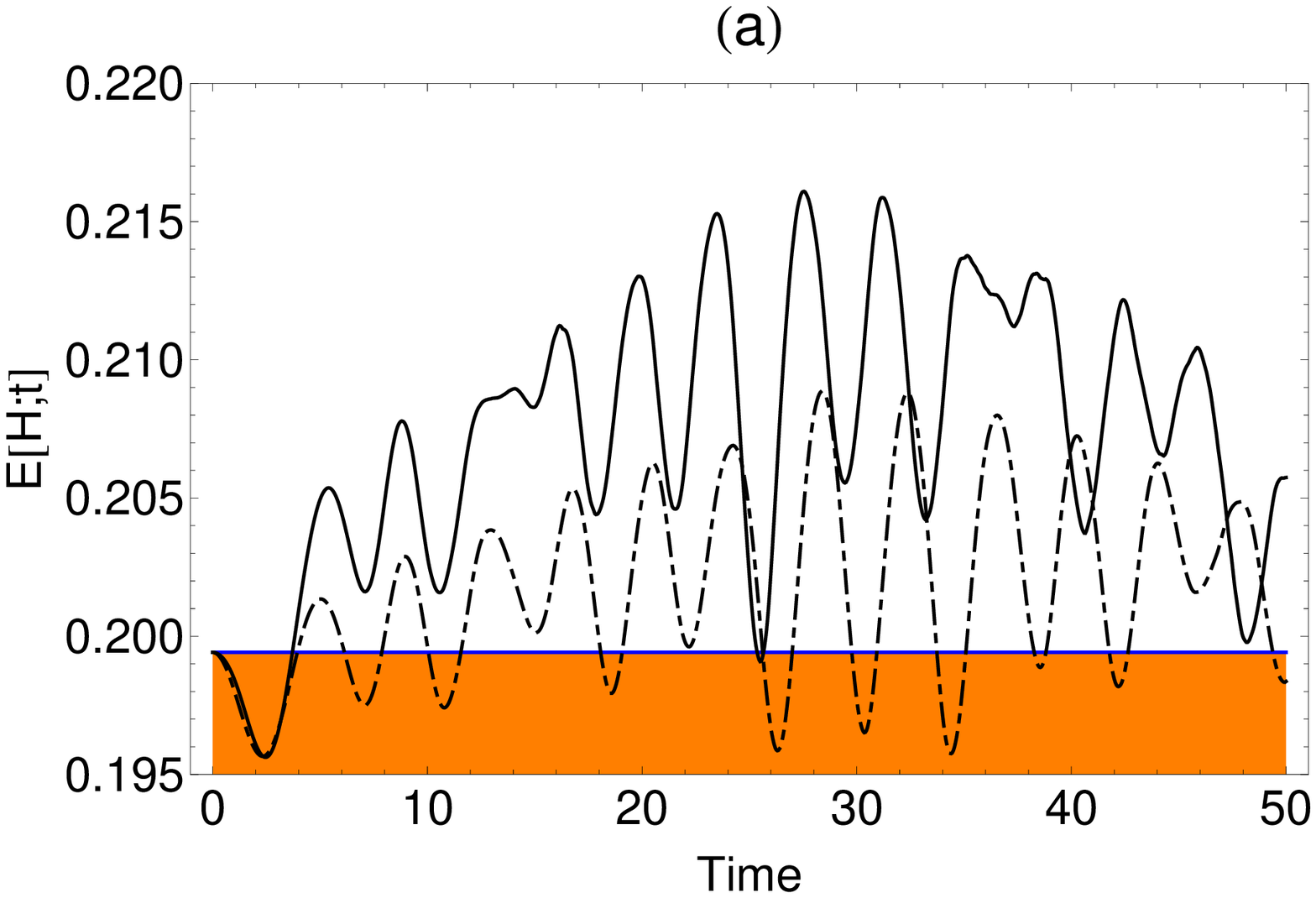}
\end{minipage} \hfill
\begin{minipage}[b]{0.45\linewidth}
\includegraphics[width=\textwidth]{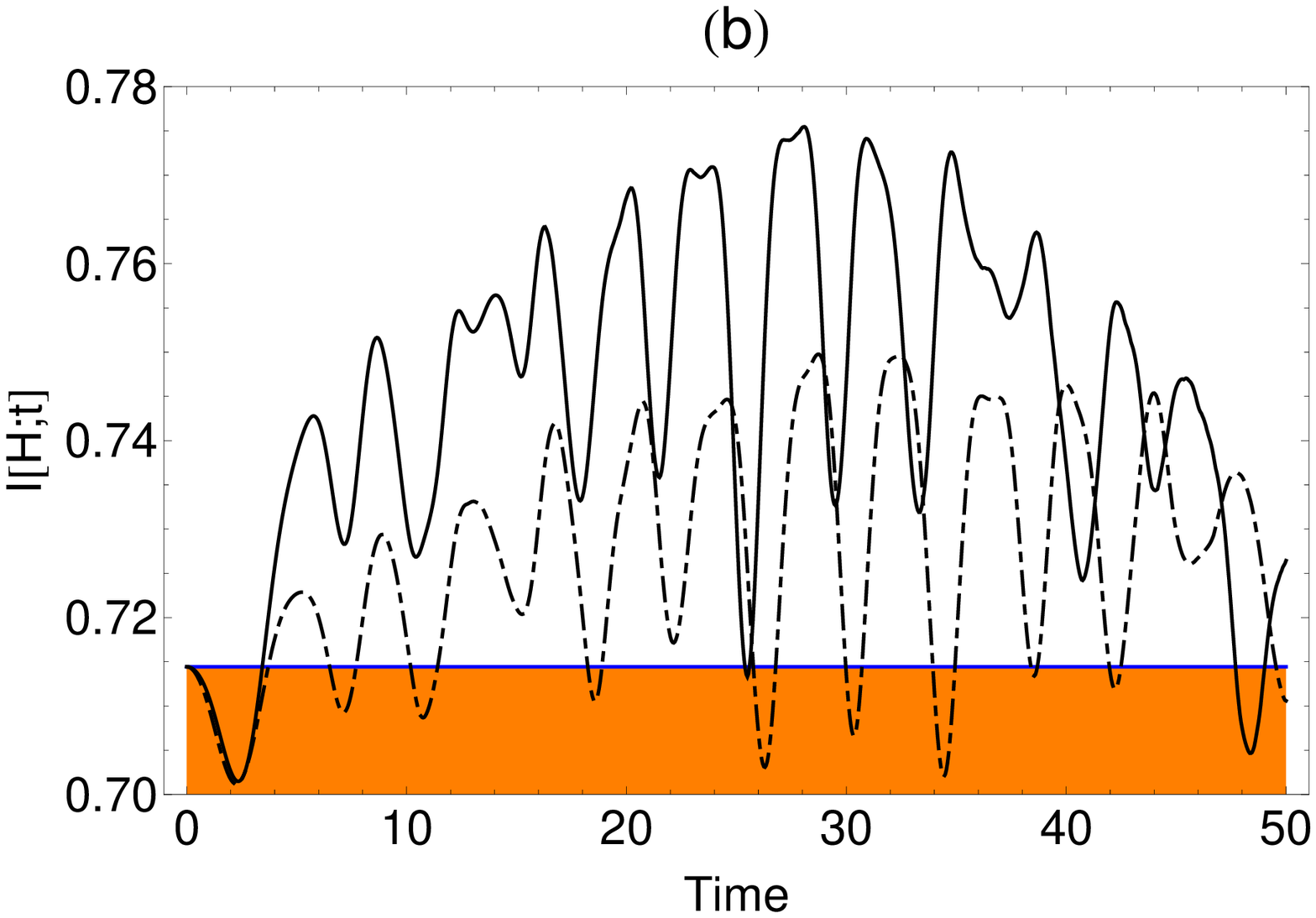}
\end{minipage}
%%%%%%%%%%%%%%%%%%%%%%%%%%%%%%%%%%%%%%%%%%%%%%%%%%%%%%%%%%%%%%%%%%%%%%%%%%%%%%%%%%%%%%%%%%%%%%%%%%%%%%%%%%%%%%%%%%%%%%%%%%%%%%%%%%%%%%%%%%%%%%%%%%%%%%%%
\caption{Plots of (a) $\mathrm{E}[\mathrm{H};t]$ and (b) $\mathrm{I}[\mathrm{H};t]$ versus $t \in [0,50]$ for two different values of transverse magnetic
field $h$ and anisotropy parameter $\gam$ with $N=20$ fixed, namely, the solid curves in both pictures correspond to $(-0.1,0.2)$, while the dot-dashed
curves are related to $(-0.13,0.1)$. In each case, the orange region describes particular situations of $\mathscr{H}_{\frac{\pi}{2},0}(\mu,\nu;t)$ where the
entropy functionals are restricted to (a) $\mathrm{E}[\mathrm{H};t] \leq \mathrm{E}[\mathrm{H};0] \approx 0.1994$ and (b) $\mathrm{I} [\mathrm{H};t] \leq 
\mathrm{I}[\mathrm{H};0] \approx 0.7144$. It is important to stress that $\mathrm{E}[\mathrm{H};t]$ and $\mathrm{I}[\mathrm{H};t]$ are sensitive to small 
variations of $h$.}
\label{fig5} 
\efig
%%%%%%%%%%%%%%%%%%%%%%%%%%%%%%%%%%%%%%%%%%%%%%%%%%%%%%%%%%%%%%%%%%%%%%%%%%%%%%%%%%%%%%%%%%%%%%%%%%%%%%%%%%%%%%%%%%%%%%%%%%%%%%%%%%%%%%%%%%%%%%%%%%%%%%%%
Figure \ref{fig5} illustrates the time evolution of (a) $\mathrm{E}[\mathrm{H};t]$ and (b) $\mathrm{I}[\mathrm{H};t]$ as a function of $t \in [0,50]$ for 
the same values of $(h,\gam)$ used in figure \ref{fig1}. Then, let us initially focus on $(-0.1,0.2)$ (see solid curves in both the pictures) since this 
case coincides with that used in the contour plots of $\mathscr{H}_{\frac{\pi}{2},0}(\mu,\nu;t)$ for specific times. Those concentric ellipses observed in 
figure \ref{fig4}(a) can be considered as a phase-space signature of the initial state $| \frac{\pi}{2},0 \rg$ at $t=0$, here endorsed by $\mathrm{E}[\mathrm{H};0] 
\approx 0.1994$ and $\mathrm{I}[\mathrm{H};0] \approx 0.7144$, which means that such a state presents an initial correlation between $\mu$ and $\nu$ featured 
by a spread angular-momentum distribution and `localized' angle distribution. Indeed, the sum $\mathrm{E}[\mathrm{Q};0] + \mathrm{E}[\mathrm{R};0] \approx
0.5150$ justifies such an assertion and (as expected) also corroborates the right-hand side of the Araki-Lieb inequality. Already at $t=2.15$, one observes
the displacement of the discrete Husimi function towards the frontier $\mu=-10$ accompanied, in this situation, by a decrease of asymmetry between its 
intrinsic widths in the $(\mu,\nu)$-directions. Since $\mathrm{E}[\mathrm{H};t=2.15] \approx 0.1958$ implies in effective information gain upon the discrete 
angular-momentum and angle collective variables, the extra correlations introduced through the two-body interaction term explain, in principle, the effects 
depicted in pictures \ref{fig1}(a,b). However, the deformed widths viewed in \ref{fig4}(c-f) after the first rebound from that frontier and subsequent return towards 
the centre of the finite-dimensional discrete phase space are responsible for the inscreased values of both entropy functionals when $4.75 \leq t \leq 9.95$.
Let us now briefly discuss the case $(-0.13,0.1)$ here characterized by a significant reduction of the effects related to $\gam$. Although the information
gain happens more frequently in such a case, the spin squeezing and entanglement effects exhibited in pictures \ref{fig1}(c,d) are associated with frequent motions of
$\mathscr{H}_{\frac{\pi}{2},0}(\mu,\nu;t)$ towards the frontier $\mu=-10$, followed by small variations of its widths. 

A last pertinent question then emerges from our considerations about discrete Husimi function and Wehrl-type entropy functionals: ``How the smoothing 
process characterized by Eq. (\ref{e24}) affects the description of correlation strength between the discrete labels $\mu$ and $\nu$ here used to describe 
the finite-dimensional phase space"? To answer this question, let us initially comprehend the role of $\mathscr{K}(\eta,\xi)$ present in the function
$E(\mu,\nu | \mu^{\prime},\nu^{\prime})$: it was basically written as a sum of products of Jacobi $\vartheta$-functions evaluated at integer arguments,
which plays, in such an aforementioned discrete phase space, the role reserved to the Gaussian functions in the continuous case ($N \rightarrow \infty$); 
besides, $\sqrt{4j+2}$ represents its respective width and it assumes a constant value, in this situation, for a given $j$. Hence, any sub-Planck 
structures\cite{Zurek2} or even relevant correlations within this range are smeared in the smoothing process, which directly affects the description of
correlation strength via Wehrl-type entropy functionals. Another important restriction associated with $\mathrm{E}[\mathrm{H};t]$ emerges from the mathematical 
property $\mathrm{S}[\ro(t)] = - \tr [ \ro(t) \ln \ro(t) ] \leq \mathrm{E}[\mathrm{H};t]$, \ie, Eq. (\ref{e25}) consists of an upper bound for the von Neumann
entropy;\cite{Mizrahi} consequently, ``if there are small distance fluctuations or if $\mathscr{H}_{\rho}(\mu,\nu;t)$ is concentrated on small regions of discrete
phase space", then the percent error estimated $\delta = \mathrm{S}/\mathrm{E}$ will be very bad. However, these apparent limitations can be circumvented by:
(i) adopting the prescription of Manfredi and Feix\cite{Manfredi} for quantum entropy based on Wigner functions in continuous phase-space, (ii) reformulating
the function $\mathscr{K}(\eta,\xi)$ in order to modify its respective width,\cite{MRG} and finally (iii) including a parallel study on von Neumann entropy\cite{Vedral}
and quantum discord\cite{Modi} which permits us to increase our knowledge base on the intricate mechanisms of correlation strength related to the spin-squeezing 
and entanglement effects for any spin systems. In many ways the analysis presented in this work of the modified LMG model is complementary to that provided in
Ref.~\refcite{Dusuel}; moreover, in what concerns the discrete phase-space approach and its implications for different physical systems, our results sound
promising at a first glance.

%%%%%%%%%%%%%%%%%%%%%%%%%%%%%%%%%%%%%%%%%%%%%%%%%%%%%%%%%%%%%%%%%%%%%%%%%%%%%%%%%%%%%%%%%%%%%%%%%%%%%%%%%%%%%%%%%%%%%%%%%%%%%%%%%%%%%%%%%%%%%%%%%%%%%%%%%%%%
\section{Conclusions}
%%%%%%%%%%%%%%%%%%%%%%%%%%%%%%%%%%%%%%%%%%%%%%%%%%%%%%%%%%%%%%%%%%%%%%%%%%%%%%%%%%%%%%%%%%%%%%%%%%%%%%%%%%%%%%%%%%%%%%%%%%%%%%%%%%%%%%%%%%%%%%%%%%%%%%%%%%%%

In this non-trivial quantum mechanical scenario of correlations, entanglement and spin-squeezing effects, as well as their connections, any associated theoretical
and/or experimental proposals for measures of the aforementioned effects shall necessarily be accomplished by exhaustive tests of confidence within a wide class 
of analogous physical systems. The mere comprehension of these effects by means of a `specific theoretical/experimental measure applied to a particular physical 
system' does not change its respective status of proposal {\it per se} for standard measure: such transition demands `time, patience, and efforts' to understand
the subtle role of correlations and their intrinsic mechanisms in quantum mechanics. In this paper, we have used a theoretical framework of finite-dimensional
discrete phase space as an alternative approach for the study of the spin-squeezing and entanglement effects. The emphasis on covariances and its fundamental role
in the investigation of spin-squeezing effects is not accidental: additional correlations related to the anticommutation relations of the angular-momentum
generators are now included in the analysis of spin systems. This first effective gain does not imply in the match between the previous effects, since both 
the measures here adopted for describing entanglement exhibit `slightly different functional relations' and also present `speculative features', as expected.

In what concerns the connection between the spin-squeezing and entanglement effects, both time-evolution operator and initial state of a given spin system play
a fundamental role in such process: they describe the short-range and/or long-range correlations of the multipartite system under investigation, which could
explain the similarity degree of the spin-squeezing and entanglement measures studied in this paper, as well as its `almost perfect match'. In this sense, the
modified LMG model fulfils a reasonable set of mathematical and physical prerequisites that lead us to corroborate, by means of numerical computations, the 
aforementioned link, for then establishing, subsequently, its inherent limitations. Besides, these results open new possibilities for future investigations in
different physical systems which encompass underlying $\mathfrak{su}(2)$ structures with distinct physical properties --- in particular, those multipartite systems 
where contributions related to the long-range correlations are indeed necessary in the effective description of entanglement. 

Our particular phase-space approach, nevertheless, also reveals its drawbacks. To begin with, Eq. (\ref{e12}) and its constituting blocks coming from the
numerical recipe described in subsection 3.2 do not satisfy the basic criterion `easy-to-compute'. The large number of sums that appeared in the time evolution of 
discrete Wigner function and mean values obligatorily implies in high computational and operational costs, whose complexities grow as $N$ increases. Moreover,
the modulo $N$ extraction phase here adopted for the discrete variables still follows that mathematical prescription discussed in Ref.~\refcite{DM1}, which
represents a necessary operational cost inherent to the mod($N$)-invariant operator basis ${\bf G}(\mu,\nu)$. However, such apparent limitations can be
circumvented in this context by adopting the theoretical framework exposed in Ref.~\refcite{MR} for the mod($N$)-invariant unitary operator basis $\bop(\mu,\nu)$,
and reformulating the content associated with the time evolution of the discrete Wigner function. This procedure will allow a real computational gain in the
numerical calculations.

Now, let us briefly mention some possibilities for future research that stem from the present paper. As a first example, we recall from Ref.~\refcite{Modi} 
those considerations on the concept of quantum discord and its important link with the subtle boundary between entanglement and classical correlations. 
Given that the Wehrl-type entropy functionals failed in the description of correlations related to the entanglement and spin-squeezing effects, it seems reasonable
to introduce such a measure in this context since quantum discord can lead us to a more efficient analysis on quantum and classical correlations in multipartite
physical systems. Another possible example of research consists in employing that finite-dimensional discrete phase-space framework in order to corroborate the 
Lieb's conjecture for the spin coherent states\cite{Solovej} via quantum dynamics of spin systems. 

%%%%%%%%%%%%%%%%%%%%%%%%%%%%%%%%%%%%%%%%%%%%%%%%%%%%%%%%%%%%%%%%%%%%%%%%%%%%%%%%%%%%%%%%%%%%%%%%%%%%%%%%%%%%%%%%%%%%%%%%%%%%%%%%%%%%%%%%%%%%%%%%%%%%%%%%%%%%
\section*{Acknowledgements}
%%%%%%%%%%%%%%%%%%%%%%%%%%%%%%%%%%%%%%%%%%%%%%%%%%%%%%%%%%%%%%%%%%%%%%%%%%%%%%%%%%%%%%%%%%%%%%%%%%%%%%%%%%%%%%%%%%%%%%%%%%%%%%%%%%%%%%%%%%%%%%%%%%%%%%%%%%%%

The authors thank Maurizio Ruzzi and anonymous referee for providing valuable suggestions on an earlier version of this manuscript. Tiago Debarba is supported 
by Conselho Nacional de Desenvolvimento Cient\'{\i}fico e Tecnol\'{o}gico (CNPq), Brazil. 

%%%%%%%%%%%%%%%%%%%%%%%%%%%%%%%%%%%%%%%%%%%%%%%%%%%%%%%%%%%%%%%%%%%%%%%%%%%%%%%%%%%%%%%%%%%%%%%%%%%%%%%%%%%%%%%%%%%%%%%%%%%%%%%%%%%%%%%%%%%%%%%%%%%%%%%%
\appendix
\section{Variances, covariances, and uncertainty relation}
%%%%%%%%%%%%%%%%%%%%%%%%%%%%%%%%%%%%%%%%%%%%%%%%%%%%%%%%%%%%%%%%%%%%%%%%%%%%%%%%%%%%%%%%%%%%%%%%%%%%%%%%%%%%%%%%%%%%%%%%%%%%%%%%%%%%%%%%%%%%%%%%%%%%%%%%

There are interesting results relating covariance functions and variances associated with certain summations of two non-commuting observables ${\bf X}$ 
and ${\bf Y}$, namely,
\bd
\mathscr{V}_{\opx \pm \opy} = \mathscr{V}_{\opx} + \mathscr{V}_{\opy} \pm 2 \mathscr{V}_{\opx \opy} \quad \mbox{and} \quad  
\mathscr{V}_{\opx \pm \nc \opy} = \mathscr{V}_{\opx} - \mathscr{V}_{\opy} \pm 2 \nc \mathscr{V}_{\opx \opy} .
\ed
Note that such connections can be generalized for a large number of non-commuting observables. For instance, if one considers the set
$\{ {\bf X},{\bf Y},{\bf Z} \}$, it is easy to show that
\brr
\mathscr{V}_{\opx + \opy + \opz} &=& \mathscr{V}_{\opx} + \mathscr{V}_{\opy} + \mathscr{V}_{\opz} + 2 \mathscr{V}_{\opx \opy} + 2 \mathscr{V}_{\opx \opz} + 
2 \mathscr{V}_{\opy \opz} \nn \\
&=& \mathscr{V}_{\opx + \opy} + \mathscr{V}_{\opx + \opz} + \mathscr{V}_{\opy + \opz} - \mathscr{V}_{\opx} - \mathscr{V}_{\opy} - \mathscr{V}_{\opz} \nn
\err
yields the statistical balance equation
\be
\lb{a1}
\mathscr{V}_{\opx + \opy + \opz} + \mathscr{V}_{\opx} + \mathscr{V}_{\opy} + \mathscr{V}_{\opz} = \mathscr{V}_{\opx + \opy} + \mathscr{V}_{\opx + \opz} + 
\mathscr{V}_{\opy + \opz} .
\ee
This particular `conservation law' describes, if applied to the angular-momentum generators, how the statistical fluctuations are distributed among the 
observables $\{ {\bf J}_{x},{\bf J}_{y},{\bf J}_{z} \}$ in the measurement process for a given initial quantum state. Table \ref{tab2} shows some analytical 
expressions obtained for the variances and covariances related to the angular-momentum generators and spin coherent states, which exemplify Eq. (\ref{a1}). 
Following, let us discuss how the unitary transformations modify the signal-to-noise ratio for an ideal experimental situation where the deleterious effects 
of a low-efficiency detection process are absent.\cite{MSV} 

The generator of unitary transformations ${\bf T}(\xi,\om)$ maintains, by definition, the linearity of the angular-momentum operators. Indeed, the 
new set 
\bd
\lbr \overline{{\bf J}}_{a} \coloneq {\bf T}^{\dagger}(\xi,\om) {\bf J}_{a} {\bf T}(\xi,\om) \; : \; \xi \in \mathbb{C} \; \, \mbox{and} \; \, \om \in 
\mathbb{R} \rbr_{a=x,y,z} 
\ed
shows explicitly such a mathematical property through the relations
\brr
\lb{a2}
\overline{{\bf J}}_{x} &=& A_{11} {\bf J}_{x} + A_{12} {\bf J}_{y} + A_{13} {\bf J}_{z} \nn \\
\overline{{\bf J}}_{y} &=& A_{21} {\bf J}_{x} + A_{22} {\bf J}_{y} + A_{23} {\bf J}_{z} \nn \\
\overline{{\bf J}}_{z} &=& A_{31} {\bf J}_{x} + A_{32} {\bf J}_{y} + A_{33} {\bf J}_{z} , 
\err
whose coefficients $\{ A_{ij} \}$ can be immediately determined from the results obtained in subsection 2.1. Since the commutation relation
$[ \overline{{\bf J}}_{a},\overline{{\bf J}}_{b} ] = \nc \eps_{abc} \overline{{\bf J}}_{c}$ keeps its invariable form, some restrictions on 
the aforementioned coefficients should be established in this context, that is,
\brr
A_{22} A_{33} - A_{23} A_{32} &=& A_{11} , \quad A_{13} A_{32} - A_{12} A_{33} = A_{21} , \quad A_{12} A_{23} - A_{13} A_{22} = A_{31} , \nn \\
A_{23} A_{31} - A_{21} A_{33} &=& A_{12} , \quad A_{11} A_{33} - A_{13} A_{31} = A_{22} , \quad A_{13} A_{21} - A_{11} A_{23} = A_{32} , \nn \\
A_{21} A_{32} - A_{22} A_{31} &=& A_{13} , \quad A_{12} A_{31} - A_{11} A_{32} = A_{23} , \quad A_{11} A_{22} - A_{12} A_{21} = A_{33} . \nn
\err
It is worth stressing that these general equations are valid for any unitary transformations which preserve Eq. (\ref{a2}); moreover, after certain 
proper manipulations of such equations, this set can be reduced to
\brr
A_{11} A_{12} + A_{21} A_{22} + A_{31} A_{32} &=& 0 \nn \\
A_{11} A_{13} + A_{21} A_{23} + A_{31} A_{33} &=& 0 \nn \\
A_{12} A_{13} + A_{22} A_{23} + A_{32} A_{33} &=& 0 . \nn
\err
As mentioned in the main part of the text, the identity $\vec{\overline{{\bf J}}}^{2} \equiv \vec{{\bf J}}^{2}$ also preserves the original form of the
total spin operator, which implies in the additional relations
\bd
A_{11}^{2} + A_{21}^{2} + A_{31}^{2} = 1 , \quad A_{12}^{2} + A_{22}^{2} + A_{32}^{2} = 1 , \quad A_{13}^{2} + A_{23}^{2} + A_{33}^{2} = 1.
\ed
Note that $\mathscr{V}_{\overline{\mathrm{J}}_{x}} + \mathscr{V}_{\overline{\mathrm{J}}_{y}} + \mathscr{V}_{\overline{\mathrm{J}}_{z}} \equiv
\mathscr{V}_{\mathrm{J}_{x}} + \mathscr{V}_{\mathrm{J}_{y}} + \mathscr{V}_{\mathrm{J}_{z}} = j$ represents an important by-product in this process
since the sum of such variances remains invariant under the unitary transformation described by Eq. (\ref{a2}).
%%%%%%%%%%%%%%%%%%%%%%%%%%%%%%%%%%%%%%%%%%%%%%%%%%%%%%%%%%%%%%%%%%%%%%%%%%%%%%%%%%%%%%%%%%%%%%%%%%%%%%%%%%%%%%%%%%%%%%%%%%%%%%%%%%%%%%%%%%%%%%%%%%%%%%%%
\begin{table}[t]
\tbl{The explicit results for the variances and covariances --- as shown on the table below --- allow us, in principle, to illustrate both the Eqs. 
(\ref{e6}) and (\ref{a1}). Besides, they also lead us to verify the uncertainty relation $\mathscr{V}_{\mathrm{J}_{a}} \mathscr{V}_{\mathrm{J}_{b}} - 
\lpar \mathscr{V}_{\mathrm{J}_{a} \mathrm{J}_{b}} \rpar^{2} \geq \frac{1}{4} \left| \lg \lbk {\bf J}_{a},{\bf J}_{b} \rbk \rg \right|^{2}$ for any
$a,b = x,y,z$. Note that the saturation is reached in this case for $\th = 0$ and $\pi$, which correspond to the fiducial states $| j,-j \rg$ and
$| j,j \rg$.}
{\begin{tabular}{@{}lll@{}} 
\Hline \\ [-1.8ex] 
Variances and covariances associated with $\{ {\bf J}_{x},{\bf J}_{y},{\bf J}_{z} \}$ for the spin coherent states \\ [0.8ex] 
\hline \\ [-1.8ex]
$\mathscr{V}_{\mathrm{J}_{x}} = (j/2) \lbk 1 - \cos^{2} (\varphi) \sin^{2} (\th) \rbk$ \\ [0.8ex]
$\mathscr{V}_{\mathrm{J}_{y}} = (j/2) \lbk 1 - \sin^{2} (\varphi) \sin^{2} (\th) \rbk$ \\ [0.8ex]
$\mathscr{V}_{\mathrm{J}_{z}} = (j/2) \sin^{2} (\th)$ \\ [0.8ex]
$\mathscr{V}_{\mathrm{J}_{x} \mathrm{J}_{y}} = - (j/4) \sin (2 \varphi) \sin^{2} (\th)$ \\ [0.8ex]
$\mathscr{V}_{\mathrm{J}_{x} \mathrm{J}_{z}} = (j/4) \cos (\varphi) \sin (2 \th)$ \\ [0.8ex]
$\mathscr{V}_{\mathrm{J}_{y} \mathrm{J}_{z}} = (j/4) \sin (\varphi) \sin (2 \th)$ \\ [0.8ex]
$\mathscr{V}_{\mathrm{J}_{x} + \mathrm{J}_{y}} = (j/2) \lbr 2 - \sin^{2} (\th) \lbk 1 + \sin (2 \varphi) \rbk \rbr$ \\ [0.8ex]
$\mathscr{V}_{\mathrm{J}_{x} + \mathrm{J}_{z}} = (j/2) \lbk 1 + \sin^{2} (\varphi) \sin^{2} (\th) + \cos (\varphi) \sin (2 \th) \rbk$ \\ [0.8ex]
$\mathscr{V}_{\mathrm{J}_{y} + \mathrm{J}_{z}} = (j/2) \lbk 1 + \cos^{2} (\varphi) \sin^{2} (\th) + \sin (\varphi) \sin (2 \th) \rbk$ \\ [0.8ex]
$\mathscr{V}_{\mathrm{J}_{x} + \mathrm{J}_{y} + \mathrm{J}_{z}} = (j/2) \lbr 2 - \sin (2 \varphi) \sin^{2} (\th) + [ \cos (\varphi) + \sin (\varphi) ] 
\sin (2 \th) \rbr$ \\ [0.8ex]
\Hline \\ [-1.8ex] 
\multicolumn{1}{@{}l}{See Ref.~\refcite{Inomata} for certain mean values involving some quadratic forms of ${\bf J}_{\pm}$ and ${\bf J}_{z}$.} \\
\end{tabular}}
\label{tab2}
\end{table}
%%%%%%%%%%%%%%%%%%%%%%%%%%%%%%%%%%%%%%%%%%%%%%%%%%%%%%%%%%%%%%%%%%%%%%%%%%%%%%%%%%%%%%%%%%%%%%%%%%%%%%%%%%%%%%%%%%%%%%%%%%%%%%%%%%%%%%%%%%%%%%%%%%%%%%%%

The evaluation of the signal-to-noise ratio (SNR) associated with collective and intrinsinc degrees of freedom is crucial in any experimental data 
analysis: indeed, it provides quantitative information on the measured signal and the noise inherent to the experiment under investigation. Thus, 
let us define SNR through the expression $\mathfrak{R}_{a} \coloneq | \lg {\bf J}_{a} \rg | / \sqrt{\mathscr{V}_{\mathrm{J}_{a}}}$, where each component
of $\vec{{\bf J}}$ presents an important role in the measurement process. However, if one considers experimental situations --- rotations and/or  
time evolutions --- that transform ${\bf J}_{a} \mapsto \overline{{\bf J}}_{a}$, it seems natural to analyse the case $\mathfrak{R}_{a} \mapsto 
\overline{\mathfrak{R}}_{a} \coloneq | \lg \overline{{\bf J}}_{a} \rg | / \sqrt{\mathscr{V}_{\overline{\mathrm{J}}_{a}}}$. In this sense, the variances
\brr
\mathscr{V}_{\overline{\mathrm{J}}_{x}} &=& A_{11}^{2} \mathscr{V}_{\mathrm{J}_{x}} + A_{12}^{2} \mathscr{V}_{\mathrm{J}_{y}} + A_{13}^{2} 
\mathscr{V}_{\mathrm{J}_{z}} \nn \\
& & + 2 \lpar A_{11} A_{12} \mathscr{V}_{\mathrm{J}_{x} \mathrm{J}_{y}} + A_{11} A_{13} \mathscr{V}_{\mathrm{J}_{x} \mathrm{J}_{z}} + A_{12} A_{13} 
\mathscr{V}_{\mathrm{J}_{y} \mathrm{J}_{z}} \rpar \nn \\
\mathscr{V}_{\overline{\mathrm{J}}_{y}} &=& A_{21}^{2} \mathscr{V}_{\mathrm{J}_{x}} + A_{22}^{2} \mathscr{V}_{\mathrm{J}_{y}} + A_{23}^{2} 
\mathscr{V}_{\mathrm{J}_{z}} \nn \\
& & + 2 \lpar A_{21} A_{22} \mathscr{V}_{\mathrm{J}_{x} \mathrm{J}_{y}} + A_{21} A_{23} \mathscr{V}_{\mathrm{J}_{x} \mathrm{J}_{z}} + A_{22} A_{23} 
\mathscr{V}_{\mathrm{J}_{y} \mathrm{J}_{z}} \rpar \nn \\
\mathscr{V}_{\overline{\mathrm{J}}_{z}} &=& A_{31}^{2} \mathscr{V}_{\mathrm{J}_{x}} + A_{32}^{2} \mathscr{V}_{\mathrm{J}_{y}} + A_{33}^{2} 
\mathscr{V}_{\mathrm{J}_{z}} \nn \\
& & + 2 \lpar A_{31} A_{32} \mathscr{V}_{\mathrm{J}_{x} \mathrm{J}_{y}} + A_{31} A_{33} \mathscr{V}_{\mathrm{J}_{x} \mathrm{J}_{z}} + A_{32} A_{33} 
\mathscr{V}_{\mathrm{J}_{y} \mathrm{J}_{z}} \rpar \nn
\err
exhibit an explicit dependence on all the previous variance and covariance functions, which modifies substantially the estimate of $\{ \mathfrak{R}_{a} \}$.
Besides, the uncertainty relation
\be
\lb{a3}
\mathscr{V}_{\overline{\mathrm{J}}_{a}} \mathscr{V}_{\overline{\mathrm{J}}_{b}} - \lpar \mathscr{V}_{\overline{\mathrm{J}}_{a} \overline{\mathrm{J}}_{b}} 
\rpar^{2} \geq \frac{1}{4} \left| \lg [ \overline{{\bf J}}_{a},\overline{{\bf J}}_{b} ] \rg \right|^{2}
\ee
is also modified in this context, since the covariances
\brr
\mathscr{V}_{\overline{\mathrm{J}}_{x} \overline{\mathrm{J}}_{y}} &=& A_{11} A_{21} \mathscr{V}_{\mathrm{J}_{x}} + A_{12} A_{22} \mathscr{V}_{\mathrm{J}_{y}}
+ A_{13} A_{23} \mathscr{V}_{\mathrm{J}_{z}} + ( A_{11} A_{22} + A_{12} A_{21} ) \mathscr{V}_{\mathrm{J}_{x} \mathrm{J}_{y}} \nn \\
& & + ( A_{11} A_{23} + A_{13} A_{21} ) \mathscr{V}_{\mathrm{J}_{x} \mathrm{J}_{z}} + ( A_{12} A_{23} + A_{13} A_{22} ) \mathscr{V}_{\mathrm{J}_{y} 
\mathrm{J}_{z}} \nn \\
\mathscr{V}_{\overline{\mathrm{J}}_{x} \overline{\mathrm{J}}_{z}} &=& A_{11} A_{31} \mathscr{V}_{\mathrm{J}_{x}} + A_{12} A_{32} \mathscr{V}_{\mathrm{J}_{y}}
+ A_{13} A_{33} \mathscr{V}_{\mathrm{J}_{z}} + ( A_{11} A_{32} + A_{12} A_{31} ) \mathscr{V}_{\mathrm{J}_{x} \mathrm{J}_{y}} \nn \\
& & + ( A_{11} A_{33} + A_{13} A_{31} ) \mathscr{V}_{\mathrm{J}_{x} \mathrm{J}_{z}} + ( A_{12} A_{33} + A_{13} A_{32} ) \mathscr{V}_{\mathrm{J}_{y} 
\mathrm{J}_{z}} \nn \\
\mathscr{V}_{\overline{\mathrm{J}}_{y} \overline{\mathrm{J}}_{z}} &=& A_{21} A_{31} \mathscr{V}_{\mathrm{J}_{x}} + A_{22} A_{32} \mathscr{V}_{\mathrm{J}_{y}}
+ A_{23} A_{33} \mathscr{V}_{\mathrm{J}_{z}} + ( A_{21} A_{32} + A_{22} A_{31} ) \mathscr{V}_{\mathrm{J}_{x} \mathrm{J}_{y}} \nn \\
& & + ( A_{21} A_{33} + A_{23} A_{31} ) \mathscr{V}_{\mathrm{J}_{x} \mathrm{J}_{z}} + ( A_{22} A_{33} + A_{23} A_{32} ) \mathscr{V}_{\mathrm{J}_{y} 
\mathrm{J}_{z}} \nn
\err
possess a similar dependence if one compares them with the previous case. Finally, it is worth stressing that such results are extremely relevant in the 
investigative process of squeezing and entanglement effects for spin systems.\cite{SS1,Wine,Hald,Kitagawa}

For instance, Mu\~{n}oz and Klimov\cite{MK} have recently introduced a particular set of discrete displacement generators upon a $2^{n} \otimes 2^{n}$ 
discrete phase space, which allows us to establish a specific family of discrete spin coherent states for $n$-qubit systems with interesting mathematical
and physical properties. Generated by application of the discrete displacement generators to a symmetric fiducial state, such states have isotropic
fluctuations in a tangent plane whose geometric features are well-defined: in general, its direction does not coincide with that chosen for the mean value
$\lg \vec{{\bf J}} \rg$. Besides, these reference states constitute the essential basic elements necessary for investigating the squeezing effects 
associated with the non-symmetric $n$-qubit states (resulting from the application of XOR gates) through the covariances $\mathscr{V}_{\mathrm{J}_{a} 
\mathrm{J}_{b}}$. It is important to emphasize that the number of XOR gates applied to the discrete spin coherent states in order to minimize fluctuations
in the original homogeneous plane strongly depends on the number of qubits. Since $\{ \mathscr{V}_{\mathrm{J}_{a} \mathrm{J}_{b}} \}_{a,b=x,y,z}$ is also 
modified by the action of XOR gates, we believe that our results can be somehow useful for the detailed examination of such squeezing effects.

%%%%%%%%%%%%%%%%%%%%%%%%%%%%%%%%%%%%%%%%%%%%%%%%%%%%%%%%%%%%%%%%%%%%%%%%%%%%%%%%%%%%%%%%%%%%%%%%%%%%%%%%%%%%%%%%%%%%%%%%%%%%%%%%%%%%%%%%%%%%%%%%%%%%%%%%
\section{The Kitagawa-Ueda model}
%%%%%%%%%%%%%%%%%%%%%%%%%%%%%%%%%%%%%%%%%%%%%%%%%%%%%%%%%%%%%%%%%%%%%%%%%%%%%%%%%%%%%%%%%%%%%%%%%%%%%%%%%%%%%%%%%%%%%%%%%%%%%%%%%%%%%%%%%%%%%%%%%%%%%%%%

Let us initiate our investigation on spin squeezing and entanglement through the Kitagawa-Ueda model\cite{Kitagawa} --- here described by the Hamiltonian
${\bf H} = \chi {\bf J}_{z}^{2}$ --- which describes a nonlinear interaction proportional to ${\bf J}_{z}^{2}$. The unitary transformations generated by
the time-evolution operator ${\bf U}(\tau) = \exp \lpar - \nc \tau {\bf J}_{z}^{2} \rpar$ for $\tau = \chi t$ represents a basic set of mathematical tools
that allows us, in principle, to calculate certain important quantities necessary to the investigative process. Indeed, if one considers the unitary 
transformations
\brr
\lb{b1}
{\bf J}_{x}(\tau) &\equiv& {\bf U}^{\dagger}(\tau) {\bf J}_{x} {\bf U}(\tau) = \exp ( \nc \tau ) \lbk {\bf J}_{x} \cos ( 2 \tau {\bf J}_{z} ) - {\bf J}_{y}
\sin ( 2 \tau {\bf J}_{z} ) \rbk \nn \\
{\bf J}_{y}(\tau) &\equiv& {\bf U}^{\dagger}(\tau) {\bf J}_{y} {\bf U}(\tau) = \exp ( \nc \tau ) \lbk {\bf J}_{x} \sin ( 2 \tau {\bf J}_{z} ) + {\bf J}_{y}
\cos ( 2 \tau {\bf J}_{z} ) \rbk \nn \\
{\bf J}_{z}(\tau) &\equiv& {\bf U}^{\dagger}(\tau) {\bf J}_{z} {\bf U}(\tau) = {\bf J}_{z} \; \rightleftharpoons \; \lbk {\bf J}_{z},{\bf U}(\tau) \rbk = 0 ,
\err
it is not so hard to yield the mean values (Heisenberg picture)
\brr
\lb{b2}
\lg {\bf J}_{x}(\tau) \rg_{\th,\varphi} &\equiv& \lg \th,\varphi | {\bf J}_{x}(\tau) | \th,\varphi \rg = \mathscr{A}_{\tau}^{2j-1} j \cos \lbk \varphi -
(2j-1) \delta_{\tau} \rbk \sin (\th) \nn \\ 
\lg {\bf J}_{y}(\tau) \rg_{\th,\varphi} &\equiv& \lg \th,\varphi | {\bf J}_{y}(\tau) | \th,\varphi \rg = \mathscr{A}_{\tau}^{2j-1} j \sin \lbk \varphi -
(2j-1) \delta_{\tau} \rbk \sin (\th) \nn \\
\lg {\bf J}_{z}(\tau) \rg_{\th,\varphi} &\equiv& \lg \th,\varphi | {\bf J}_{z}(\tau) | \th,\varphi \rg = - j \cos (\th)
\err
for the spin coherent states, where $\mathscr{A}_{\tau} = \lbk \cos^{2}(\tau) + \sin^{2}(\tau) \cos^{2}(\th) \rbk^{\half}$ denotes the amplitude function,
and $\delta_{\tau} = \arctan \lbk \tan (\tau) \cos (\th) \rbk$ the phase function. Note that such results lead us to conclude that unitary transformations involving 
nonlinear forms of the angular-momentum generators do not preserve, in general, the linear feature of such operators --- for instance, see Eqs. (\ref{b1}) 
and (\ref{a2}). This particularity is directly associated with the generators of the $\mathfrak{su}(2)$ Lie algebra and does not apply to the squeeze 
operator ${\bf S}(\zeta) \coloneq \exp \lbk \half \lpar \zeta^{\ast} {\bf a}^{2} - \zeta {\bf a}^{\dagger 2} \rpar \rbk$, since ${\bf S}(\zeta)$ is responsible 
for unitary transformations in the Heisenberg-Weyl algebra which preserve the linearity of the boson annihilation and creation operators.\cite{Eberly,Ban}

Another interesting point inherent to the time-evolution operator ${\bf U}(\tau)$ establishes the relation $\lg \vec{{\bf J}}^{2}(\tau) \rg_{\th,\varphi} =
\lg \vec{{\bf J}}^{2}(0) \rg_{\th,\varphi} = j(j+1)$, which preserves the mean value of the total spin operator in $\tau=0$. In fact, this result can be 
interpreted as a direct consequence of the unitary transformations originated from the action of ${\bf U}(\tau)$ on the angular-momentum operators. 
Following, let us discuss some important points related to uncertainty relation, squeezing effects and their links with entanglement, considering the spin 
coherent states as initial states of the physical system.
%%%%%%%%%%%%%%%%%%%%%%%%%%%%%%%%%%%%%%%%%%%%%%%%%%%%%%%%%%%%%%%%%%%%%%%%%%%%%%%%%%%%%%%%%%%%%%%%%%%%%%%%%%%%%%%%%%%%%%%%%%%%%%%%%%%%%%%%%%%%%%%%%%%%%%%%
\begin{table}[th]
\tbl{The time-dependent variance and covariance functions, as shown on the table below for the spin coherent states, constitute an important group of
formal mathematical results related to the Kitagawa-Ueda model that permits us to investigate not only the Robertson-Schr\"{o}dinger uncertainty principle,
but also both the squeezing and entanglement effects associated with the collective angular-momentum operators.}
{\begin{tabular}{@{}lll@{}} 
\Hline \\ [-1.8ex] 
Time-dependent mean values, variances and covariance function for the Kitagawa-Ueda model \\ [0.8ex] 
\hline \\ [-1.8ex]
$\lg {\bf J}_{x}^{2}(\tau) \rg_{\th,\varphi} = j/2 + (j/4) (2j-1) \lbk 1 + \mathscr{A}_{2 \tau}^{2j-2} \cos [2 \varphi - (2j-2) \delta_{2 \tau} ] \rbk
\sin^{2}(\th)$ \\ [0.8ex]
$\lg {\bf J}_{y}^{2}(\tau) \rg_{\th,\varphi} = j/2 + (j/4) (2j-1) \lbk 1 - \mathscr{A}_{2 \tau}^{2j-2} \cos [2 \varphi - (2j-2) \delta_{2 \tau} ] \rbk
\sin^{2}(\th)$ \\ [0.8ex]
$\lg {\bf J}_{z}^{2}(\tau) \rg_{\th,\varphi} \equiv \lg {\bf J}_{z}^{2}(0) \rg_{\th,\varphi} = j^{2} \cos^{2} (\th) + (j/2) \sin^{2} (\th)$ \\ [0.8ex]
$\mathscr{V}_{\mathrm{J}_{x}}(\tau) = j/2 + (j/4) (2j-1) \lbk 1 + \mathscr{A}_{2 \tau}^{2j-2} \cos [2 \varphi - (2j-2) \delta_{2 \tau} ] \rbk
\sin^{2}(\th)$ \\ [0.8ex]
$\qquad \qquad \quad - j^{2} \mathscr{A}_{\tau}^{4j-2} \cos^{2} \lbk \varphi - (2j-1) \delta_{\tau} \rbk \sin^{2}(\th)$ \\ [0.8ex]
$\mathscr{V}_{\mathrm{J}_{y}}(\tau) = j/2 + (j/4) (2j-1) \lbk 1 - \mathscr{A}_{2 \tau}^{2j-2} \cos [2 \varphi - (2j-2) \delta_{2 \tau} ] \rbk
\sin^{2}(\th)$ \\ [0.8ex]
$\qquad \qquad \quad - j^{2} \mathscr{A}_{\tau}^{4j-2} \sin^{2} \lbk \varphi - (2j-1) \delta_{\tau} \rbk \sin^{2}(\th)$ \\ [0.8ex]
$\mathscr{V}_{\mathrm{J}_{z}}(\tau) \equiv \mathscr{V}_{\mathrm{J}_{z}}(0) = (j/2) \sin^{2} (\th)$ \\ [0.8ex]
$\mathscr{V}_{\mathrm{J}_{x} \mathrm{J}_{y}}(\tau) = (j/4) (2j-1) \mathscr{A}_{2 \tau}^{2j-2} \sin [2 \varphi - (2j-2) \delta_{2 \tau} ] 
\sin^{2}(\th)$ \\ [0.8ex]
$\qquad \qquad \quad - (j^{2}/2) \mathscr{A}_{\tau}^{4j-2} \sin [2 \varphi - (4j-2) \delta_{\tau} ] \sin^{2}(\th)$ \\ [0.8ex]
\Hline \\ [-1.8ex] 
\multicolumn{1}{@{}l}{See Refs.~\refcite{SS1} and \refcite{Hald} for some measurement criteria involving such physical quantities.} \\
\end{tabular}}
\label{tab3}
\end{table}
%%%%%%%%%%%%%%%%%%%%%%%%%%%%%%%%%%%%%%%%%%%%%%%%%%%%%%%%%%%%%%%%%%%%%%%%%%%%%%%%%%%%%%%%%%%%%%%%%%%%%%%%%%%%%%%%%%%%%%%%%%%%%%%%%%%%%%%%%%%%%%%%%%%%%%%%
%
\begin{itemize}
\item Initially, let us introduce the parameters $\mathcal{S}_{x} \coloneq \mathscr{V}_{\mathrm{J}_{x}} / \mathcal{R}_{xyz}$ and $\mathcal{S}_{y} \coloneq 
\mathscr{V}_{\mathrm{J}_{y}} / \mathcal{R}_{xyz}$ with $\mathcal{R}_{xyz} \coloneq \lbk ( \mathscr{V}_{\mathrm{J}_{x} \mathrm{J}_{y}} )^{2} + \frac{1}{4} 
| \lg {\bf J}_{z} \rg |^{2} \rbk^{\half}$ (in this definition, $\mathcal{R}_{xyz} \neq 0$ represents a condition {\it sine qua non}), which lead us to 
rewrite Eq. (\ref{e6}) as follows: $\mathcal{S}_{x} \mathcal{S}_{y} \geq 1$. In principle, this simplified form allows us to investigate the squeezing 
effects related to the aforementioned model through exact expressions obtained in Table \ref{tab3} for the time-dependent variance and covariance 
functions. Figure \ref{figb1}(a) shows the plots of $\mathcal{S}_{x}$ (dot-dashed line) and $\mathcal{S}_{y}$ (dashed line) versus $\tau \in [ 0,2 \pi ]$ 
for $j=2$ and $\th = \varphi = \frac{\pi}{4}$ fixed. The hachured area exhibited in the picture describes, in this case, that region where the squeezing 
effects occur for $\mathcal{S}_{x(y)}(\tau) < 1$ (but not both simultaneously). Such a particular evidence of squeezing effect generalizes, through an 
effective way, those results obtained by Kitagawa and Ueda\cite{Kitagawa} in the absence of one-axis twisting mechanism.\footnote{In fact, this mathematical
procedure introduces artificially additional quantum correlations (by means of time-dependent unitary transformations that consist of rotations around the 
$x$-axis) for the angular-momentum operators, which allow to reduce the standard quantum noise related to the spin coherent states down to $\half 
\sqrt[3]{\frac{j}{3}}$. It is important to emphasize that such an estimate does not consider the contributions originated from the covariance function.}   
\item How the squeezing and entanglement effects can be connected? S{\o}rensen {\it et al}\cite{Hald} have proposed an interesting experiment for the 
Bose-Einstein condensates (and attainable with present technology) which allows us, in principle, to answer this intriguing question through a fundamental
effect in quantum mechanics: the many-particle entanglement. Since correlations among spins yield the squeezing effect, it is natural to build a bridge
connecting both the entanglement and spin squeezing effects by means of a solid mathematical framework which allows us to establish certain reliable measurement
criteria.\cite{SS1} In this sense, let us comment some few words on the entanglement criterion adopted in such an experiment: it is basically focussed on
the separability criterion for the $N$-particle density operator involving the variances and squared mean values associated with each orthogonal component
of the collective angular-momentum operators, that is, 
\be
\lb{b3}
\mathcal{E}_{a} \equiv \frac{N \mathscr{V}_{\mathrm{J}_{a}}}{\lg {\bf J}_{b} \rg^{2} + \lg {\bf J}_{c} \rg^{2}} < 1 \qquad (a,b,c=x,y,z) .
\ee
Thus, the parameters $\{ \mathcal{E}_{a} \}$ characterize the atomic entanglement in this context, and quantum states with $\mathcal{E}_{a} < 1$ are
referred to as spin squeezed states.\cite{Hald} For instance, Figure \ref{figb1}(b) shows the plots of $\mathcal{E}_{x}$ (dot-dashed line),
$\mathcal{E}_{y}$ (dashed line), as well as $\mathcal{E}_{z}$ (solid line) versus $\tau \in [ 0, 2\pi ]$ for the same values of $\{ j,\th,\varphi \}$ 
adopted in the squeezing criterion --- see Fig. \ref{figb1}(a). Note that $\mathcal{E}_{x}$ and $\mathcal{E}_{y}$ exhibit values less than one, except
$\mathcal{E}_{z}$, which implies in entanglement effect (see hachured region) for the orthogonal components ${\bf J}_{x}(\tau)$ and ${\bf J}_{y}(\tau)$,
as expected, since $[ {\bf H},{\bf J}_{z} ] = 0$. The similarities between both the figures and the evidence of the squeezing and entanglement effects 
for the same values of $\tau$ justify, in such a case, the above connection {\it per se}. 
\end{itemize}
%%%%%%%%%%%%%%%%%%%%%%%%%%%%%%%%%%%%%%%%%%%%%%%%%%%%%%%%%%%%%%%%%%%%%%%%%%%%%%%%%%%%%%%%%%%%%%%%%%%%%%%%%%%%%%%%%%%%%%%%%%%%%%%%%%%%%%%%%%%%%%%%%%%%%%%%%%%%%%%
\bfig[t]
\centering
\begin{minipage}[b]{0.45\linewidth}
\includegraphics[width=\textwidth]{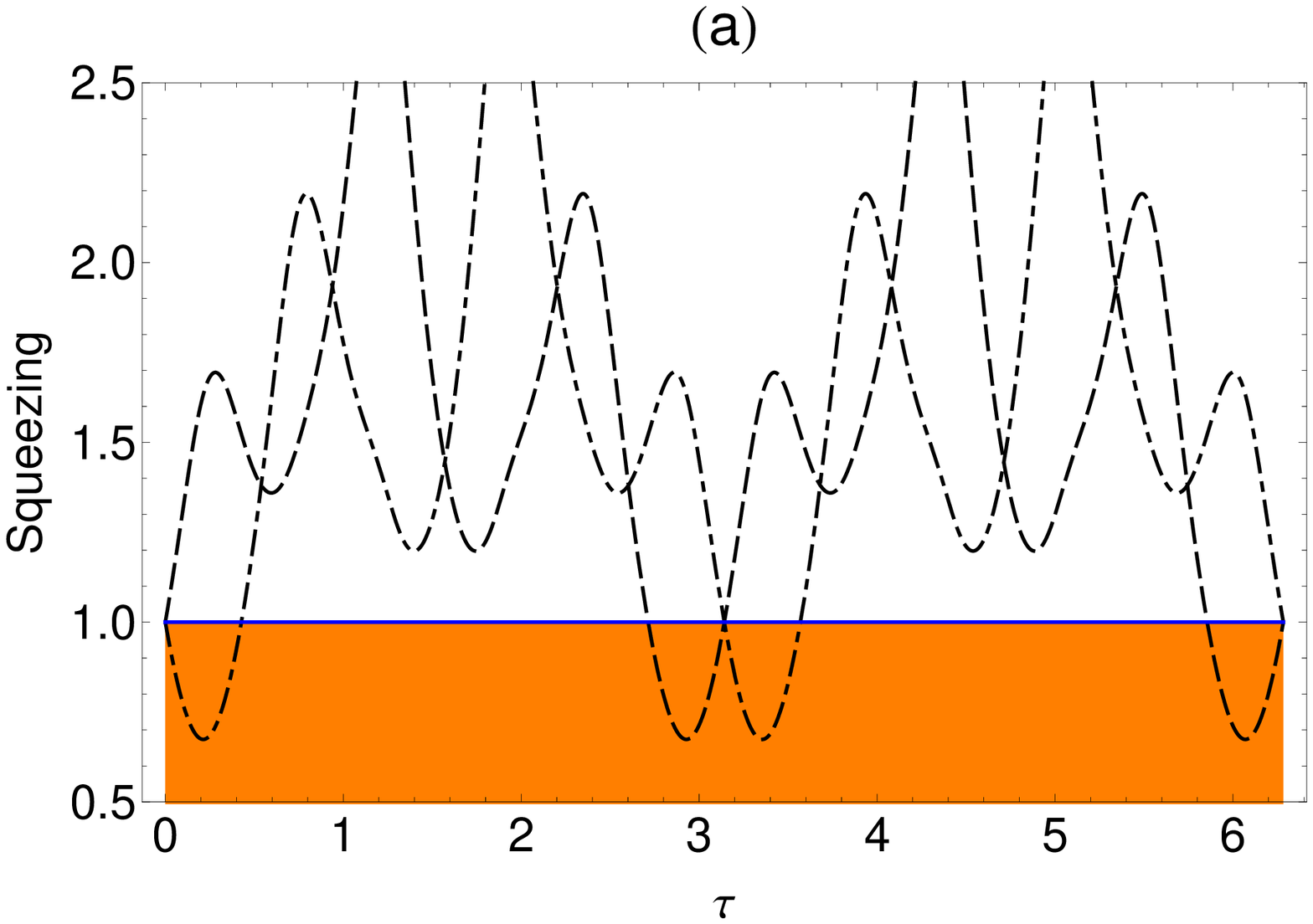}
\end{minipage} \hfill
\begin{minipage}[b]{0.45\linewidth}
\includegraphics[width=\textwidth]{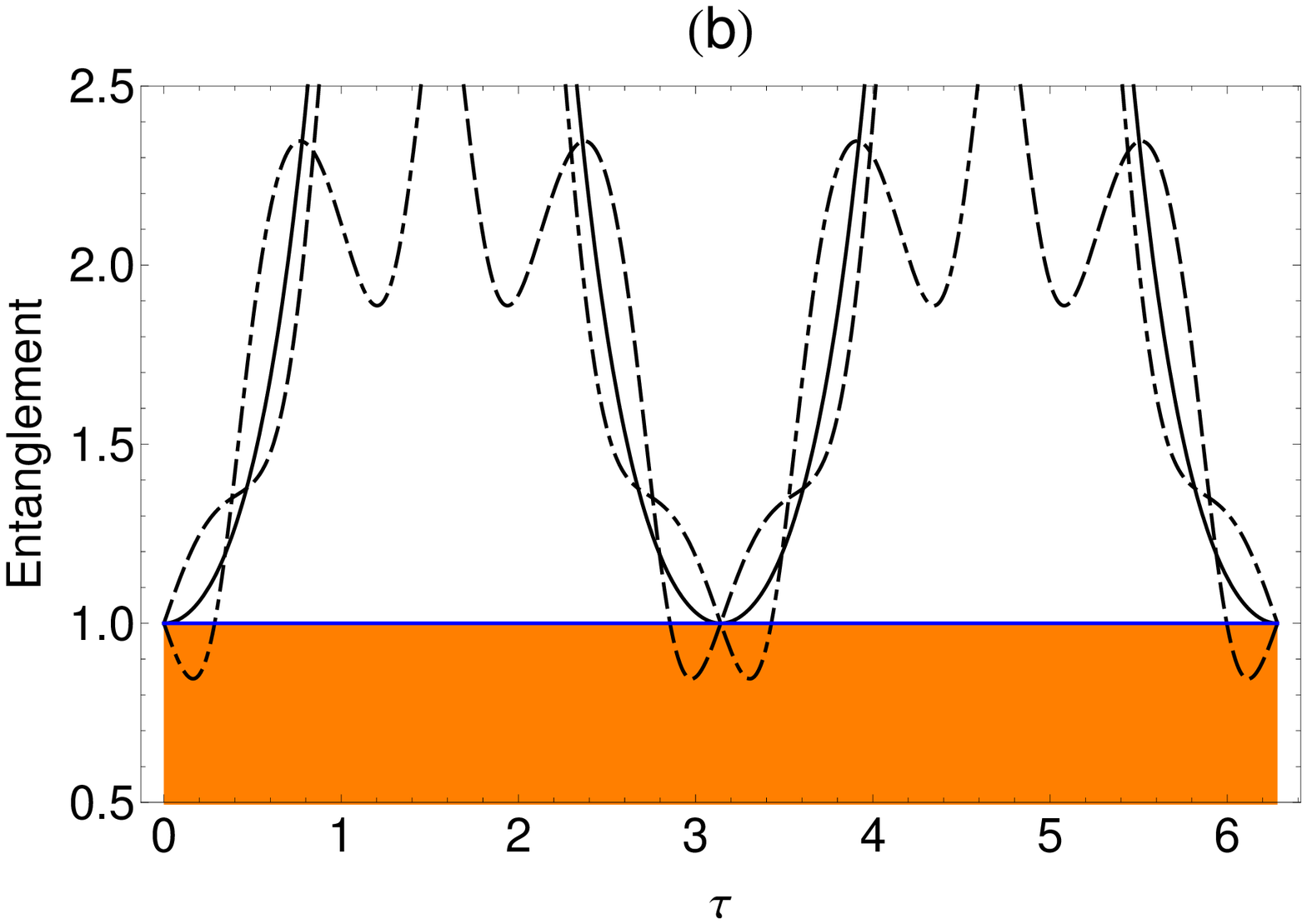}
\end{minipage}
%%%%%%%%%%%%%%%%%%%%%%%%%%%%%%%%%%%%%%%%%%%%%%%%%%%%%%%%%%%%%%%%%%%%%%%%%%%%%%%%%%%%%%%%%%%%%%%%%%%%%%%%%%%%%%%%%%%%%%%%%%%%%%%%%%%%%%%%%%%%%%%%%%%%%%%%%%%%%%%
\caption{(a) represents the plots of $\mathcal{S}_{x}$ (dot-dashed line) and $\mathcal{S}_{y}$ (dashed line) versus $\tau \in [ 0,2\pi ]$ with $j=2$ 
($N=4$ spins) and $\th = \varphi = \frac{\pi}{4}$ fixed. Note that both the parameters $\mathcal{S}_{x}(\tau)$ and $\mathcal{S}_{y}(\tau)$ assume values
less than one in different intervals of $\tau$ (see hachured region), which implies in the squeezing effect for each case obeying the relation
$\mathcal{S}_{x}(\tau) \mathcal{S}_{y}(\tau) \geq 1$ (the saturation is reached for $\tau = m \pi$ with $m \in \mathbb{N}$). (b) corresponds to the plots
of $\mathcal{E}_{x}$ (dot-dashed line), $\mathcal{E}_{y}$ (dashed line), and $\mathcal{E}_{z}$ (solid line) versus $\tau \in [0,2\pi]$, for the same 
values of $\{ j,\th,\varphi \}$ adopted in the previous picture. The hachured area denotes, in this case, that region where the entanglement effect occurs.}
\label{figb1} 
\efig
%%%%%%%%%%%%%%%%%%%%%%%%%%%%%%%%%%%%%%%%%%%%%%%%%%%%%%%%%%%%%%%%%%%%%%%%%%%%%%%%%%%%%%%%%%%%%%%%%%%%%%%%%%%%%%%%%%%%%%%%%%%%%%%%%%%%%%%%%%%%%%%%%%%%%%%%%%%%%%%

%%%%%%%%%%%%%%%%%%%%%%%%%%%%%%%%%%%%%%%%%%%%%%%%%%%%%%%%%%%%%%%%%%%%%%%%%%%%%%%%%%%%%%%%%%%%%%%%%%%%%%%%%%%%%%%%%%%%%%%%%%%%%%%%%%%%%%%%%%%%%%%%%%%%%%%%%%%%%%%
\section{A case study for $h = 0$}
%%%%%%%%%%%%%%%%%%%%%%%%%%%%%%%%%%%%%%%%%%%%%%%%%%%%%%%%%%%%%%%%%%%%%%%%%%%%%%%%%%%%%%%%%%%%%%%%%%%%%%%%%%%%%%%%%%%%%%%%%%%%%%%%%%%%%%%%%%%%%%%%%%%%%%%%%%%%%%%

To what extent the anisotropy parameter $\gam$ can be effectively employed in order to validate the entanglement criteria here adopted in the modified LMG model?
Since the two-body term of the Hamiltonian ${\bf H}^{\prime} = - h {\bf J}_{z} - \frac{1}{N} \lpar {\bf J}_{x}^{2} + \gam {\bf J}_{y}^{2} \rpar$ plays an essential 
role in such a case, we cannot refrain from studying it {\it per se} in a deeper perspective --- in this context, we will assume $h=0$ hereafter. Note that the 
plethora of results exposed and discussed along the text for assigning the existence of entanglement in a multi-spin system can then be directly compared in the 
particular case of the two-body correlations associated with the model at hand. This important procedure will produce an effective range for $\gam$ with $N=20$ 
fixed, via extensive numerical computations of Eqs. (\ref{b3}) and (\ref{e23}), whose physical implications permit, within other features, to answer (in part, at 
least) the aforementioned question.
%%%%%%%%%%%%%%%%%%%%%%%%%%%%%%%%%%%%%%%%%%%%%%%%%%%%%%%%%%%%%%%%%%%%%%%%%%%%%%%%%%%%%%%%%%%%%%%%%%%%%%%%%%%%%%%%%%%%%%%%%%%%%%%%%%%%%%%%%%%%%%%%%%%%%%%%%%
\bfig[t]
\centering
\begin{minipage}[b]{0.45\linewidth}
\includegraphics[width=\textwidth]{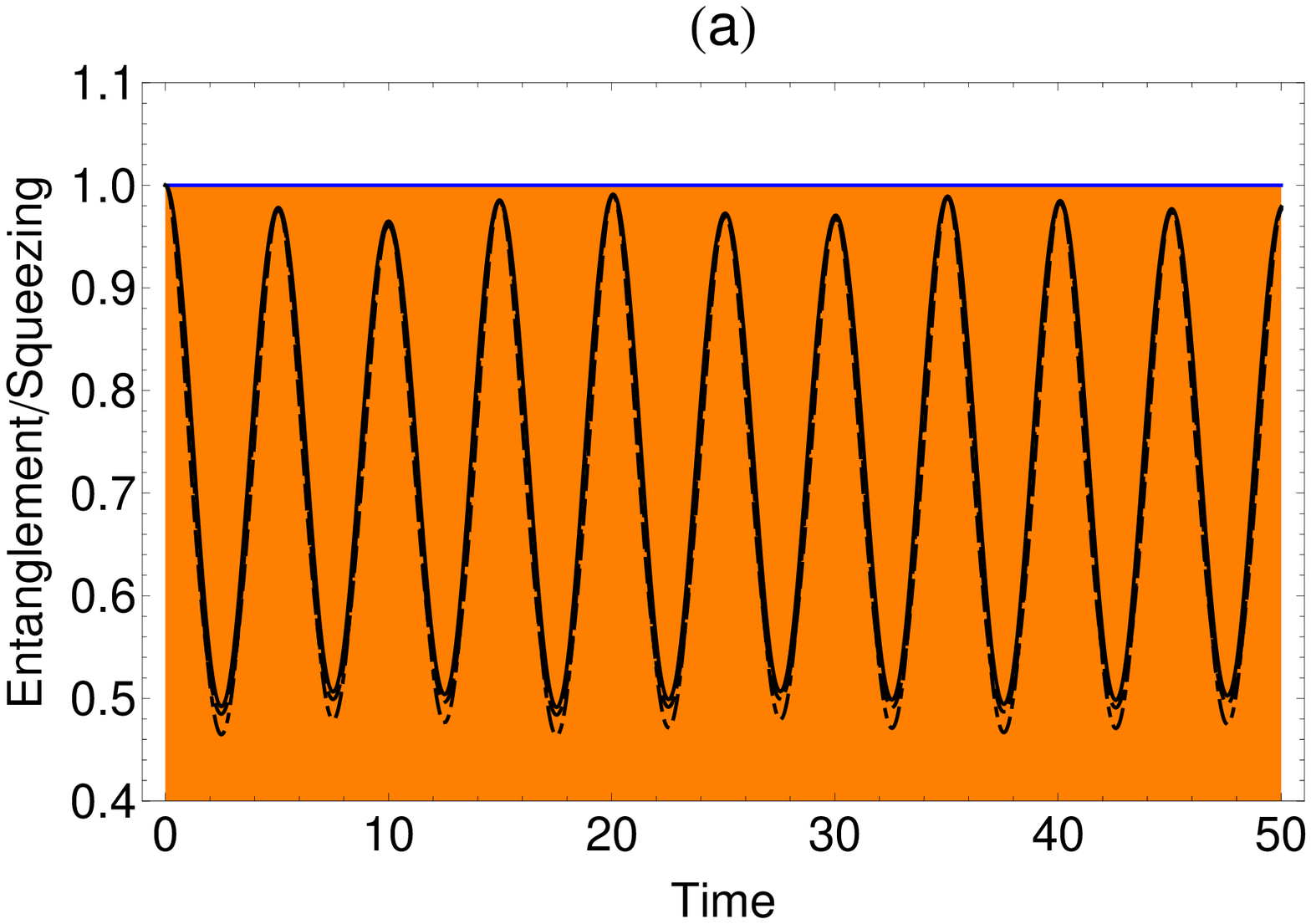}
\end{minipage} \hfill
\begin{minipage}[b]{0.45\linewidth}
\includegraphics[width=\textwidth]{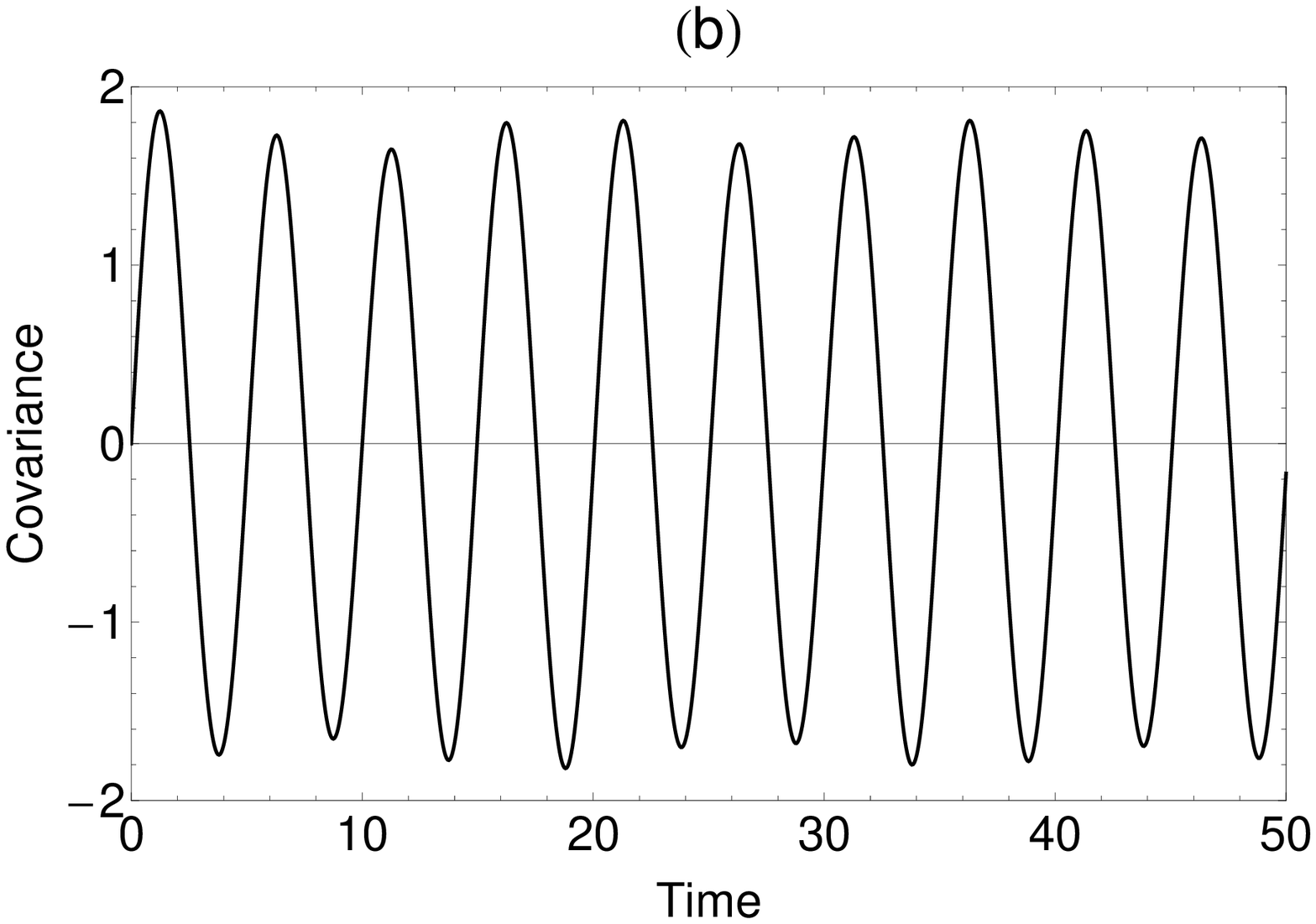}
\end{minipage} \hfill
\begin{minipage}[b]{0.45\linewidth}
\includegraphics[width=\textwidth]{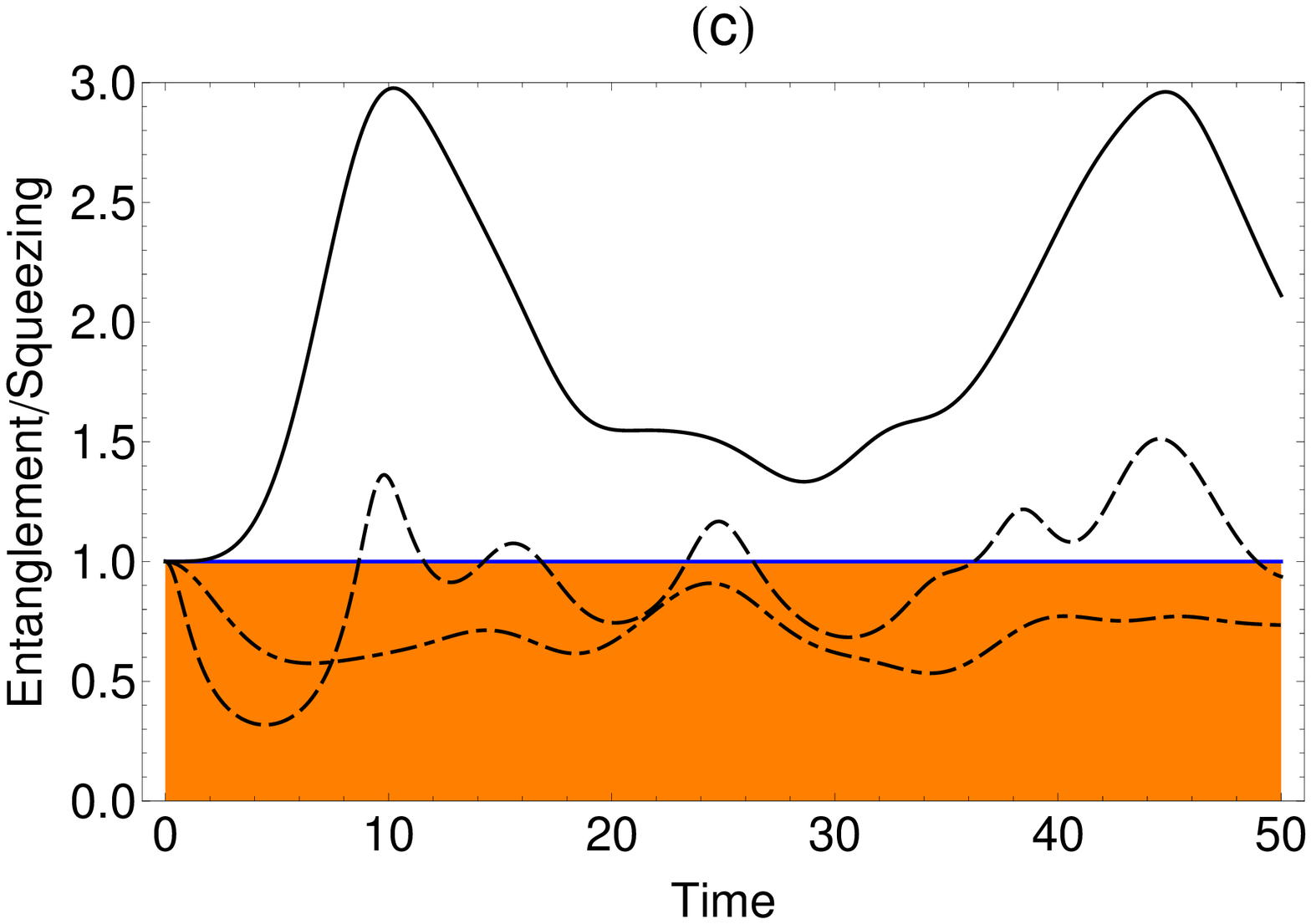}
\end{minipage} \hfill
\begin{minipage}[b]{0.45\linewidth}
\includegraphics[width=\textwidth]{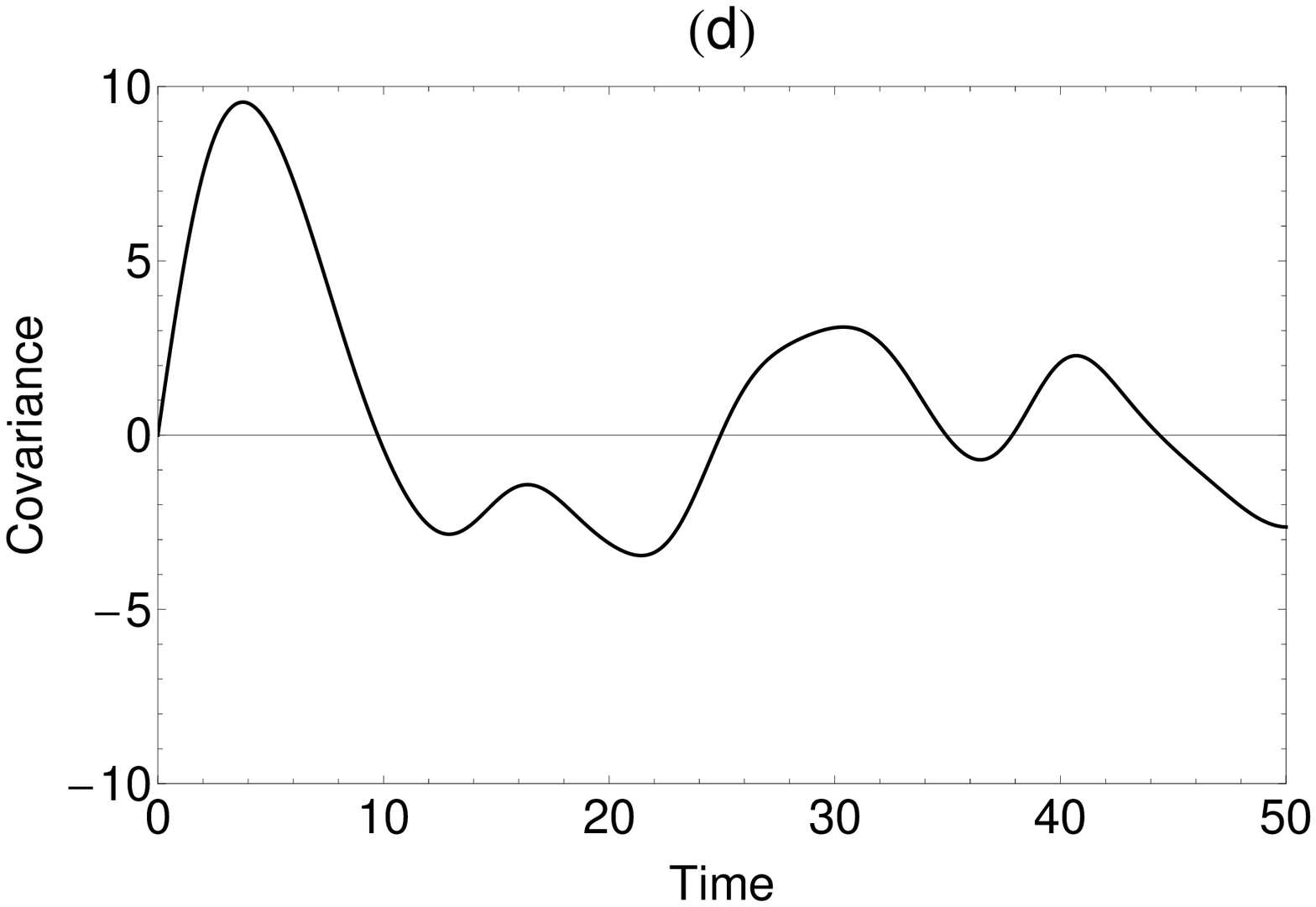}
\end{minipage}
%%%%%%%%%%%%%%%%%%%%%%%%%%%%%%%%%%%%%%%%%%%%%%%%%%%%%%%%%%%%%%%%%%%%%%%%%%%%%%%%%%%%%%%%%%%%%%%%%%%%%%%%%%%%%%%%%%%%%%%%%%%%%%%%%%%%%%%%%%%%%%%%%%%%%%%%%%
\caption{Plots of (a,c) $\mathcal{E}_{z}(t)$ (solid line), $E_{z}(t)$ (dot-dashed line), $\mathcal{S}_{z}^{(x)}(t)$ (dashed line) and (b,d) 
$\mathscr{V}_{\mathrm{J}_{y} \mathrm{J}_{z}}(t)$ versus $t \in [0,50]$ for (a,b) $\gam = 0.5$ and (c,d) $\gam = 0.948$ (critical value), with $N=20$ fixed
and $h=0$ (absence of transverse magnetic field). The transition between the different values of $\gam$, here depicted by pictures (a) and (c), illustrates 
how the squeezing and entanglement criteria (as well as the covariance function) are affected by the anysotropy parameter $\gam$. Such a numerical evidence 
suggests the existence of a validity domain for the aforementioned criteria, where the `almost perfect match' between squeezing and entanglement effects 
should be preserved for any $t \geq 0$.}
\label{figc1} 
\efig
%%%%%%%%%%%%%%%%%%%%%%%%%%%%%%%%%%%%%%%%%%%%%%%%%%%%%%%%%%%%%%%%%%%%%%%%%%%%%%%%%%%%%%%%%%%%%%%%%%%%%%%%%%%%%%%%%%%%%%%%%%%%%%%%%%%%%%%%%%%%%%%%%%%%%%%%%%

Now, we discuss some essential points inherent to the numerical results obtained through the discrete Wigner function (\ref{e13}) evaluated at equally spaced time
intervals (\ie, $0.05$ time units) and $| \gam | \leq 1$. The first one confirms the link between squeezing and entanglement effects verified in section 4 for 
values up to $\gam = 0.5$; however, if one considers the interval $0.5 < \gam < 0.9$, small differences concerning the parameters $\{ \mathcal{S}_{a}^{(x)}(t),
\mathcal{E}_{a}(t), E_{a}(t) \}$ for $a=y,z$ begin to appear, this fact being considered by us as a first numerical evidence about the `limitation' of the 
particular entanglement measures $\mathcal{E}_{a}(t)$ and $E_{a}(t)$ here studied. The second essential point establishes that, for $\gam \approx 0.948$ 
(critical value), $\mathcal{E}_{a}(t)$ does not predict entanglement effects anymore, while $E_{a}(t)$ still does --- in this case, it is worth stressing that
$\mathcal{S}_{a}^{(x)}(t)$ also indicates the presence of the squeezing effect for determined values of time. 

With respect to the proposals of squeezing and entanglement measures discussed in this work, let us draw attention to the presence of the covariance function
$\mathscr{V}_{\mathrm{J}_{y} \mathrm{J}_{z}}(t)$ in the expression for $\mathcal{S}_{a}^{(x)}(t)$ (see table \ref{tab1} and appendix A) which depends explicitly 
on the mean value (\ref{e18}). Our numerical calculations allow to show, in particular, that $\mathscr{V}_{\mathrm{J}_{y} \mathrm{J}_{z}}(t)$ is responsible for 
the occurrence of squeezing effect in the entire domain of $\gam$, since it carries essential additional information, when we take into account the RS inequality 
(instead of the Heisenberg one), associated with the anticommutation relation of the angular-momentum operators $\mathbf{J}_{y}$ and $\mathbf{J}_{z}$. 

Therefore, our results reinforce the real necessity of a deeper discussion on the validity domain related to the entanglement criteria here used to characterize 
the important link between squeezing and entanglement effects.\cite{Kitagawa} Thus, it seems that we cannot assure beforehand the viability of such criteria 
without also especifying their respective validity domains in order to be employed with a consistent degree of confidence.

%%%%%%%%%%%%%%%%%%%%%%%%%%%%%%%%%%%%%%%%%%%%%%%%%%%%%%%%%%%%%%%%%%%%%%%%%%%%%%%%%%%%%%%%%%%%%%%%%%%%%%%%%%%%%%%%%%%%%%%%%%%%%%%%%%%%%%%%%%%%%%%%%%%%%%%%

%%%%%%%%%%%%%%%%%%%%%%%%%%%%%%%%%%%%%%%%%%%%%%%%%%%%%%%%%%%%%%%%%%%%%%%%%%%%%%%%%%%%%%%%%%%%%%%%%%%%%%%%%%%%%%%%%%%%%%%%%%%%%%%%%%%%%%%%%%%%%%%%%%%%%%%%
\end{document}